
%
%
\magnification=\magstep0
\headline={\ifnum\pageno=1\hfil\else\hfil\tenrm--\ \folio\ --\hfil\fi}
\footline={\hfil}
\hsize=6.0truein
\vsize=8.54truein
\hoffset=0.25truein
\voffset=0.25truein
\baselineskip=14pt
%
%
\tolerance=9000
\hyphenpenalty=10000
%
%
%

\font\mbf=cmmib10 \font\mbfs=cmmib10 scaled 833
\font\msybf=cmbsy10 \font\msybfs=cmbsy10 scaled 833

%
%
%

\textfont9=\mbf \scriptfont9=\mbfs \scriptscriptfont9=\mbfs
\def\bmsy{\fam10}
\textfont10=\msybf \scriptfont10=\msybfs \scriptscriptfont10=\msybfs
%
%
\mathchardef\alpha="710B
\mathchardef\beta="710C
\mathchardef\gamma="710D
\mathchardef\delta="710E
\mathchardef\epsilon="710F
\mathchardef\zeta="7110
\mathchardef\eta="7111
\mathchardef\theta="7112
\mathchardef\iota="7113
\mathchardef\kappa="7114
\mathchardef\lambda="7115
\mathchardef\mu="7116
\mathchardef\nu="7117
\mathchardef\xi="7118
\mathchardef\pi="7119
\mathchardef\rho="711A
\mathchardef\sigma="711B
\mathchardef\tau="711C
\mathchardef\upsilon="711D
\mathchardef\phi="711E
\mathchardef\chi="711F
\mathchardef\psi="7120
\mathchardef\omega="7121
\mathchardef\varepsilon="7122
\mathchardef\vartheta="7123
\mathchardef\varpi="7124
\mathchardef\varrho="7125
\mathchardef\varsigma="7126
\mathchardef\varphi="7127
\mathchardef\nabla="7272
\mathchardef\cdot="7201
\def\grad{\nabla} 
%
%
\def\spose#1{\hbox to 0pt{#1\hss}}
\def\lta{\mathrel{\spose{\lower 3pt\hbox{$\mathchar"218$}}
     \raise 2.0pt\hbox{$\mathchar"13C$}}}
\def\gta{\mathrel{\spose{\lower 3pt\hbox{$\mathchar"218$}}
     \raise 2.0pt\hbox{$\mathchar"13E$}}}
%
%

%
%

%
%

%
%
%

\def\cm{{\rm\,cm}}

\def\kpc{{\rm\,kpc}}
\def\mpc{{\rm\,Mpc}}
\def\sec{{\rm\,s}}

\def\kms{{\rm\,km\,s^{-1}}}
\def\kelvin{{\rm\,K}}
\def\erg{{\rm\,erg}}

\def\hz{{\rm\,Hz}}
\def\sr{{\rm\,sr}}
\def\msun{{\,M_\odot}}

\def\etal{et al.$~$}

\def\Sec{\S $~$}

%
%

%
%
\newcount\eqnumber
\eqnumber=1
%
\def\new{{\the\eqnumber}\global\advance\eqnumber by 1}
%
%
\def\ref#1{\advance\eqnumber by -#1 \the\eqnumber
     \advance\eqnumber by #1 }
%
%
\def\last{\advance\eqnumber by -1 {\the\eqnumber}\advance
     \eqnumber by 1}
%
%
\def\eqnam#1{\xdef#1{\the\eqnumber}}
%
%
%
\def\refindent{\par\noindent\hangindent=3pc\hangafter=1 }
\def\aa#1#2#3{\refindent#1, A\&A, #2, #3}
\def\aarev#1#2#3{\refindent#1, A\&A Rev., #2, #3}

\def\aj#1#2#3{\refindent#1, AJ, #2, #3}

\def\apj#1#2#3{\refindent#1, ApJ, #2, #3}
\def\apjsup#1#2#3{\refindent#1, ApJS, #2, #3}

\def\araa#1#2#3{\refindent#1, ARA\&A, #2, #3}

\def\asps#1#2#3{\refindent#1, Astrop. Space Sci., #2, #3}

\def\mnras#1#2#3{\refindent#1, MNRAS, #2, #3}

\def\nature#1#2#3{\refindent#1, Nature, #2, #3}

\def\refbook#1{\refindent#1}

%
%
\def\refrule{\hbox to 3pc{\leaders\hrule depth-2pt height 2.4pt\hfill}}
%
%
%
\def\sect#1 {
  \vskip 1. truecm plus .2cm
  \bigbreak
  \centerline{\bf #1}
  \nobreak
  \bigskip
  \nobreak}
\def\subsec#1#2 {
  \bigbreak
  \centerline{#1.~{\bf #2}}
  \bigskip}
\def\infsec#1#2 {
  \bigbreak
  \line{#1.~{\it #2} \hfill}
  \bigskip}
\newread\epsffilein    
\newif\ifepsffileok    
\newif\ifepsfbbfound   
\newif\ifepsfverbose   
\newdimen\epsfxsize    
\newdimen\epsfysize    
\newdimen\epsftsize    
\newdimen\epsfrsize    
\newdimen\epsftmp      
\newdimen\pspoints     
\pspoints=1bp          
\epsfxsize=0pt         
\epsfysize=0pt         
\def\epsfbox#1{\global\def\epsfllx{72}\global\def\epsflly{72}%
   \global\def\epsfurx{540}\global\def\epsfury{720}%
   \def\lbracket{[}\def\testit{#1}\ifx\testit\lbracket
   \let\next=\epsfgetlitbb\else\let\next=\epsfnormal\fi\next{#1}}%
\def\epsfgetlitbb#1#2 #3 #4 #5]#6{\epsfgrab #2 #3 #4 #5 .\\%
   \epsfsetgraph{#6}}%
\def\epsfnormal#1{\epsfgetbb{#1}\epsfsetgraph{#1}}%
\def\epsfgetbb#1{%
%
%
\openin\epsffilein=#1
\ifeof\epsffilein\errmessage{I couldn't open #1, will ignore it}\else
%
%
   {\epsffileoktrue \chardef\other=12
    \def\do##1{\catcode`##1=\other}\dospecials \catcode`\ =10
    \loop
       \read\epsffilein to \epsffileline
       \ifeof\epsffilein\epsffileokfalse\else
%
%
          \expandafter\epsfaux\epsffileline:. \\%
       \fi
   \ifepsffileok\repeat
   \ifepsfbbfound\else
    \ifepsfverbose\message{No bounding box comment in #1; using defaults}\fi\fi
   }\closein\epsffilein\fi}%
%
%
\def\epsfclipstring{}
\def\epsfsetgraph#1{%
   \epsfrsize=\epsfury\pspoints
   \advance\epsfrsize by-\epsflly\pspoints
   \epsftsize=\epsfurx\pspoints
   \advance\epsftsize by-\epsfllx\pspoints
%
%
   \epsfxsize\epsfsize\epsftsize\epsfrsize
   \ifnum\epsfxsize=0 \ifnum\epsfysize=0
      \epsfxsize=\epsftsize \epsfysize=\epsfrsize
      \epsfrsize=0pt
%
%
     \else\epsftmp=\epsftsize \divide\epsftmp\epsfrsize
       \epsfxsize=\epsfysize \multiply\epsfxsize\epsftmp
       \multiply\epsftmp\epsfrsize \advance\epsftsize-\epsftmp
       \epsftmp=\epsfysize
       \loop \advance\epsftsize\epsftsize \divide\epsftmp 2
       \ifnum\epsftmp>0
          \ifnum\epsftsize<\epsfrsize\else
             \advance\epsftsize-\epsfrsize \advance\epsfxsize\epsftmp \fi
       \repeat
       \epsfrsize=0pt
     \fi
   \else \ifnum\epsfysize=0
     \epsftmp=\epsfrsize \divide\epsftmp\epsftsize
     \epsfysize=\epsfxsize \multiply\epsfysize\epsftmp
     \multiply\epsftmp\epsftsize \advance\epsfrsize-\epsftmp
     \epsftmp=\epsfxsize
     \loop \advance\epsfrsize\epsfrsize \divide\epsftmp 2
     \ifnum\epsftmp>0
        \ifnum\epsfrsize<\epsftsize\else
           \advance\epsfrsize-\epsftsize \advance\epsfysize\epsftmp \fi
     \repeat
     \epsfrsize=0pt
    \else
     \epsfrsize=\epsfysize
    \fi
   \fi
%
%
   \ifepsfverbose\message{#1: width=\the\epsfxsize, height=\the\epsfysize}\fi
   \epsftmp=10\epsfxsize \divide\epsftmp\pspoints
   \vbox to\epsfysize{\vfil\hbox to\epsfxsize{%
      \ifnum\epsfrsize=0\relax
        \includegraphics{#1}%
      \else
        \epsfrsize=10\epsfysize \divide\epsfrsize\pspoints
        \includegraphics{#1}%
      \fi
      \hfil}}%
\global\epsfxsize=0pt\global\epsfysize=0pt}%
%
%
{\catcode`\%=12 \global\let\epsfpercent=
%
%
\long\def\epsfaux#1#2:#3\\{\ifx#1\epsfpercent
   \def\testit{#2}\ifx\testit\epsfbblit
      \epsfgrab #3 . . . \\%
      \epsffileokfalse
      \global\epsfbbfoundtrue
   \fi\else\ifx#1\par\else\epsffileokfalse\fi\fi}%
%
%
\def\epsfempty{}%
\def\epsfgrab #1 #2 #3 #4 #5\\{%
\global\def\epsfllx{#1}\ifx\epsfllx\epsfempty
      \epsfgrab #2 #3 #4 #5 .\\\else
   \global\def\epsflly{#2}%
   \global\def\epsfurx{#3}\global\def\epsfury{#4}\fi}%
%
%
\def\epsfsize#1#2{\epsfxsize}
%
%

\def\lya{{\rm Ly}\alpha}
\def\hi{{H\thinspace I}\ }
\def\hii{{H\thinspace II}\ }

\def\heii{{He\thinspace II}\ }

\def\nhi{N_{HI}}

\def\jhi{J_{\hi}}
\def\jhiu{J_{-21}}
\def\jheii{J_{\heii}}
\def\shi{\sigma_{\hi}(\nu)}
\def\shia{\bar\sigma_{\hi}}
\def\etal{et al.\ }

\def\Sec{\S }

\def\cf{{cf.}\ }

\parskip .15cm plus .1cm
\null\vskip 1.cm

\centerline{\bf THE LYMAN ALPHA FOREST FROM GRAVITATIONAL COLLAPSE}
\centerline{\bf IN THE CDM+$\Lambda$ MODEL}
\bigskip
\bigskip
\centerline{  Jordi Miralda-Escud\'e
\footnote{$^1$}{Institute for Advanced Study, Princeton, NJ 08540},
Renyue Cen
\footnote{$^2$}{Peyton Hall, Princeton University, Princeton, NJ 08544},
Jeremiah P. Ostriker $^2$
and Michael Rauch
\footnote{$^{3,}$}{California Institute of Technology, Pasadena, CA 91125}
\footnote{$^4$}{Hubble Fellow}}
\centerline{ Email: jordi@sns.ias.edu, cen@astro.princeton.edu,}
\centerline{jpo@astro.princeton.edu, mr@astro.caltech.edu }
\vskip 0.5in
\centerline{Received: 1995, ....}

\vskip 1.cm

\sect{ABSTRACT}

  We use an Eulerian hydrodynamic cosmological simulation to model the
$\lya$ forest in a spatially flat,
COBE normalized, cold dark matter model with
$\Omega = 0.4$,
and find that the intergalactic, photoionized gas is predicted to
collapse into sheet-like and filamentary structures which give rise to
absorption lines having similar characteristics as the observed $\lya$
forest. A typical filament is $\sim 1 h^{-1} \mpc$ long with thickness
$\sim 50-100 h^{-1} \kpc$ (in proper units), and baryonic mass
$\sim 10^{10} h^{-1} \msun$.
In comparison our nominal numerical resolution
is $(2.5,9)h^{-1}$kpc (in the two simulations we perform)
with true resolution, perhaps a factor of $2.5$ worse than this;
the nominal mass resolution is $(10^{4.2},10^{5.8})M_\odot$
in gas for the two simulations.
The gas temperature increases with time as structures with larger velocities
collapse gravitationally.

  We show that the predicted distributions of column densities,
b-parameters and equivalent widths of the $\lya$ forest clouds agree
with the observed ones, and that their evolution is consistent with our
model, if the ionizing background has (as expected from our simulations)
an approximately constant intensity between $z=2$ and $z=4$.
The new method of identifying lines as contiguous
regions in the spectrum below a fixed flux threshold is suggested to
analyze the absorption lines, given that the $\lya$ spectra arise from
a continuous density field of neutral hydrogen rather than discrete
clouds. We also predict
the distribution of transmitted flux and its correlation along a
spectrum and on parallel spectra, and the \heii flux decrement as a
function of redshift.
We predict a correlation length of $\sim 80 h^{-1}\kpc$ perpendicular to
the line of sight for features in the Lyman alpha forest.

  In order to reproduce the observed number of lines and average flux
transmission, the baryon content of the clouds may need to be significantly
higher than in previous models because of the low densities and large
volume-filling factors we predict. If the background intensity $\jhi$ is
at least that predicted from the observed quasars, $\Omega_b$ needs to
be as high as $\sim 0.025 h^{-2}$, higher than expected by light element
nucleosynthesis; the model also predicts that most of
the baryons at $z>2$ are in $\lya$ clouds, and that the rate at which
the baryons move to more overdense regions is slow. A large fraction of
the baryons which are not observed at present in galaxies might be
intergalactic gas in the currently collapsing structures, with $T \sim
10^5-10^6 \kelvin$.

\noindent{{\it Subject headings}: intergalactic medium -
quasars: absorption lines - cosmology: large-scale structure of Universe -
hydrodynamics}

\vfill\eject

\sect{1. INTRODUCTION}

  Most of the information available to us by direct observation of
epochs far removed fom the present is in the increasingly detailed
absorption spectra of high redshift quasars. The very numerous lines in
the Lyman alpha forest are believed to arise primarily from neutral
hydrogen along the line of sight, enabling us to study the distribution
of neutral gas over a wide redshift range (from the present to the
highest redshift quasars). By now we can typically observe hundreds of
lines per unit redshift along each line of sight, with each line
yielding an \hi column density $\nhi$, a Doppler width $b$, and a
redshift $z$. Thus, large samples of absorption lines can be obtained by
observing only a relatively small number of quasars. By now fairly good
data is available characterizing the distribution of $b$ and $\nhi$ as a
function of redshift, and the correlations of the lines (e.g.,
Carswell \etal 1991; Rauch \etal 1993; Petitjean \etal 1993;
Schneider \etal 1993; Cristiani \etal 1995; Hu \etal 1995;
Tytler \etal 1995).
In addition, as we shall discuss in this paper, the fluctuations in the
transmitted flux which cannot be separated into individual lines also
contain a large amount of information. The $\lya$ absorption lines also
provide a ``fair'' sample, in the sense that the lines of sight to
distant quasars pass through random parts of the intervening universe.
Thus the usual issues of selection and bias are not
as serious here as when studying other cosmic phenomena.

  Until quite recently there was very little attempt to use this large data
base to constrain theories for the origin of structure. Theories of the
$\lya$ forest were attempts at heuristic modeling rather than {\it ab initio}
theories. One aimed to make a local model for the gas responsible for
the absorption, with sufficient physical detail to understand the
degree of ionization of the gas, its temperature and density, the
processes responsible for confinement in clumps (i.e., the fact that
individual lines are seen), and finally to account for the observed
statistical properties. Some of these models were based on the
hypothesis of a hot intergalactic medium created by early
shock heating,
which could cool into clouds that would be confined by the hot gas
(Sargent \etal 1980; Ostriker \& Ikeuchi 1983; Ikeuchi \& Ostriker
1986). Clouds associated with galaxies were another dominant idea.
Bahcall \& Spitzer (1969) suggested that galaxies were surrounded
by halos of hot gas, which could contain some photoionized clouds
in pressure equilibrium. Another early paper by Arons (1972)
proposed that the clouds could be associated with infalling gas
around forming galaxies. Other recent suggestions include the
possible association of $\lya$ clouds with the outer parts of
\hi disks (Charlton, Salpeter, \& Hogan 1993),
the debris from
merging satellites (Wang 1993; Morris \& van den Bergh 1994),
galactic outflows (Fransson \& Epstein 1982, Wang 1995)
or primordial, linear density fluctuations in the intergalactic medium (Bi
1993).
Rees (1986) and Ikeuchi (1986) suggested that the clouds were related
to subgalactic ``minihalos'' where the photoionized gas
could be maintained in thermal and dynamical equilibrium. A more
detailed study was presented in the classic paper by Bond, Szalay, \&
Silk (1988), where hydrodynamic simulations of spherical halos were
made which suggested that the absorption lines could arise in gas
around the dark matter halos predicted by the cold dark matter model.
This gas could either be infalling and contracting if the halo
was sufficiently massive, or expanding after having been reheated during
reionization. The level of semianalytic modeling possible at the time
this seminal paper was written
did not allow for accurate predictions to be made, which could tell
whether or not the proposed theory was correct. Now, with the advances
of computational capabilities, one can directly model the three
dimensional growth of small scale structure of any {\it ab initio}
model and then shoot lines of sight through the simulation to collect mock
observational spectra which can be compared with real data.

  At this point, we wish to remark that
the models for the formation of structure in the universe from
primordial fluctuations represent an unprecedented step in the history
of science.
Up to now, the physical laws of nature have only attempted
to explain or predict the evolution of a physical system
once a set of initial conditions have been determined
observationally, which can fix the state of the system at an initial
time. The models for the origin of structure specify the statistical
ensemble of the initial conditions for the universe, and they should in
principle fully predict the subsequent evolution of the universe and the
observations we can make, without the need of other observations to fix
any initial conditions. It is only our inability to calculate complex
physical phenomena such as cloud formation from cooled gas, star
formation, supernova explosions, dust formation, etc.
which prevents us from fully realising the predictive potential of
the models and testing them against observations.
At present, we still do not know which model of structure formation
should be correct, and there is also a number of
constants which are free parameters in the models (such as the baryon
to dark matter ratio, the specific entropy of the universe, and the
amplitude of the density fluctuations) and must be determined
by observation.
It is hoped that a more complete
theory in the future will also predict these quantities.

  In order to test such models, we need to identify certain objects and
physical phenomena taking place in the universe which are sufficiently
simple to allow us to predict what should be observed. The observations
that can be most reliably predicted are those of the fluctuations in the
microwave background, since these fluctuations were produced when all
the structures were still in the linear regime (although they are
superposed with other fluctuations arising after non-linear collapse,
such as the Sunyaev-Zeldovich effect). However, most of our observations
of the universe are of objects that have long been in the
non-linear regime at later epochs. The distribution and abundances of
galaxies and X-ray clusters have long been used as a test of the models
of structure formation, since these are the objects most easily observed
near us. At high redshift, galaxies and clusters become progressively
faint, whereas the $\lya$ forest offers us an
excellent probe to directly observe
the evolution of the neutral hydrogen distribution.

In addition to the observational advantages, the $\lya$ forest may
also have the advantage of allowing fairly precise theoretical
predictions of its observable properties.
The $\lya$ absorption spectra allow us to observe gas at lower
densities than any other observational method. In fact, the most
underdense gas in the voids should, as we shall show in this paper,
cause fluctuations in the optical depth between $\lya$ lines that are
observable, and all the neutral gas at intermediate densities, up to the
dense central regions in halos where galaxies should form, is also
observable as absorption lines of increasing column density with
increasing gas density. This gives an enormous advantage over the
study of galaxies, which are indicating the regions where the gas
densities became high enough for star formation to take place, or over
X-ray observations, which are sensitive to objects having high X-ray
surface brightness, corresponding to dense gas in the virialized regions
of galaxy clusters. The processes affecting the low density gas may be
sufficiently simple as to allow detailed modeling and predictability,
and therefore provide powerful tests
for the models of the origin of structure. This is not possible at high
densities, where the gas will typically develop a complicated structure
(such as in the interstellar medium of our galaxy), and star formation
will be taking place.

  The low density gas is affected by gravitational and pressure forces,
which are predicted in detail. In addition, the gas temperature is
influenced by the ionizing radiation, which is responsible for the
heating after reionization and for providing a natural scale, the Jeans
scale, for the smallest objects where the photoionized gas will
collapse. It is this process of reheating that allows us to calculate
the gas distribution with the finite resolution of our numerical code,
even when the smallest structures in the dark matter are unresolved.
However, the presence of the ionizing background introduces two
sources of uncertainty: first, during the epoch of reionization the
intergalactic medium becomes highly inhomogeneous as sources of
ionizing photons start to emit, with individual \hii regions expanding
from the sources and patches of neutral gas left in between (Arons \&
Wingert 1972). After reionization is complete, the photoionized gas
is left at a temperature that depends on the spectrum of the emitted
radiation, and which could spatially fluctuate depending on the epoch
at which each region was reionized (Miralda-Escud\'e \& Rees 1994).
Second, once all the gas is ionized, the discrete number of sources
combined with a large number of absorbers can cause fluctuations in
the intensity of the ionizing background (Zuo 1992a,b; Fardal \& Shull
1993), and correspondingly change the neutral gas density. Inasmuch as
the process of formation of the sources of radiation
(presumably quasars and young stellar populations in galaxies)
is not currently understood, such
effects cannot be predicted, but they can only be calibrated from
observations. In this paper, we shall assume a uniform radiation
background at all epochs, ignoring the effects mentioned above (these
are not likely to be very large, because the expected fluctuations in
intensity and temperature should be small except at very high redshift).

  In addition to the uncertainties related to the ionizing background,
there is also the possibility that kinetic energy is released to the
intergalactic medium from explosions in the gravitationally collapsed
objects (probably arising from active galactic nuclei or supernova
explosions in starbursts).
Other work which we have done (Cen \& Ostriker 1993a) indicates that
such explosive input can alter the properties of the gas considerably
in the vicinity of virialized objects, but is relatively unimportant
in the bulk of the cosmic volume in which most of the QSO absorption
features are found. Thus, in this work we do not include any energetic
input from the collapsed regions; we hope to investigate the effects of
explosions in due course.

  To summarize, the input to our simulations is, in addition to the
model for the initial density fluctuations and the cosmological
parameters adopted, the background radiation field which we evolve
explicitly in time, assuming it is spatially uniform at all times,
and the specific numerical scheme for solving
the hydrodynamic equations and its implementation. Initial results of
this work were presented in Cen \etal (1994; hereafter Paper I); we use
an Eulerian hydrodynamic code and treat a low density, spatially flat
cold dark matter (CDM) model.
More recently, a similar approach has been followed by
Zhang \etal (1995) who use a similar Eulerian code and examine a more
standard $\Omega = 1$ COBE-normalized CDM scenario, and by Hernquist
\etal (1995, hereafter HKWM), who use an SPH (particle based) hydro code
and the same CDM model with a lower normalization, more appropriate to
reproduce the observed abundance of clusters (see also Petitjean \etal
1995, who again take a similar approach but without using a hydro code).
All of these groups, despite the differences in technical approach and
adopted model found surprisingly good agreement between the simulated
and real $\lya$ forest. Differences among the simulations and with
the observations will probably appear as the results are scrutinized
in more detail; however, the overall agreement (which is reinforced by
the results in this paper) makes it quite probable that the physical
environment produced by the codes of widely distributed photoionized
gas in sheets and filaments associated with caustics, outside of
virialized objects, does correspond to the real world.

  In this paper, we return to the simulations of Paper I to provide
greater detail concerning our physical modeling and, more
importantly, to present a much more detailed analysis of our (mock)
observations. The results of Paper I were based on simulations on
a box of $3 h^{-1} \mpc$; here, we shall include results from a larger
simulation in a box of $10 h^{-1} \mpc$, comparing the results to see
the effects of the large-scale power.

  Before beginning the detailed discussion of our results, it may be
useful to say a few words of comparison of the two detailed
hydrodynamic methods mentioned earlier. Roughly speaking, the Eulerian
technique, in its current application, has advantages with regard to
mass resolution and the SPH technique with regard to spatial resolution.
The average gas mass per cell in our large box simulation (with a
length of $10 h^{-1} \mpc$ and $288^3$ cells) is $6\times 10^5 \msun$,
and the cell size is $35 h^{-1} \kpc$ in comoving units; if we take the
actual spatial resolution to be 2.5 cells, this gives $90 h^{-1} \kpc$.
In the SPH simulation of HKWM (with a box of length 11.1
$h^{-1} \mpc$) the mass per gas particle was $1.5 \times 10^8 \msun$, and
the maximal spatial resolution was $10 h^{-1} {\rm comoving} \kpc$
(equivalent to $6.5 h^{-1} \kpc$ Plummer softening). Thus that method was
able to identify and determine the properties of the rare, very high
density features which correspond to the damped $\lya$ and Lyman limit
systems (Katz \etal 1995), which could not be treated with the Eulerian
codes. However, HKWM note (and we concur) that the typical $\lya$ cloud
arises from material having rather modest overdensity in the range
$\rho / \bar\rho = 1 - 10$. If the fluid is overdense by a factor of 10,
the sphere containing 20 particles (roughly the number required to
define local properties in an SPH code) has a diameter of
270 $h^{-1} \kpc$ in the HKWM simulation. Thus, for the bulk of the
small clouds in regions of modest overdensity the Eulerian code has
both greater mass and greater spatial resolution. But, to be
quantitative on the advantages of the particle codes, at an
overdensity $\gta 300$ (appropriate for the high column density
systems) the spatial resolution of the SPH code exceeds that
attainable by the Eulerian code.

  After describing several aspects
of the simulations in \Sec 2, we present the physical characteristics
of the absorbing structures in \Sec 3. The analysis of the simulated
spectra is given in \Sec 4; detailed predictions of our model are
given there for the distribution and correlation properties of the
transmitted flux, and properties of the absorption lines and their
correlations along a single line of sight and on parallel lines of
sight. In \Sec 5 we discuss some aspects of the $\lya$ forest that
are derived from our model, especially the fraction of baryons that
is contained in absorption systems of different column densities.
Finally, our conclusions are summarized in \Sec 6.

\sect{2. THE SIMULATION}

  We use the same Eulerian code that was used for the simulations in
Paper I.
As will be shown below,
many $\lya$ lines originate from
regions enclosed by two semi-planar shocks propagating outwards.
It is therefore essential to model these shocks accurately.
We use a new shock-capturing cosmological hydrodynamic code
based on Harten's Total Variation Diminishing (TVD) hydrodynamic scheme
(Harten 1984), described in Ryu \etal (1993).
The original TVD scheme was improved by adding one additonal
variable (entropy) and
its evolution equation to the conventional hydrodynamic equations.
This improvement allows us to eliminate otherwise large artificial
entropy generation in regions where the gas is not shocked.
In fact, typical, unphysical high temperatures being generated
due to inaccuracy in the conventional hydrodynamic code are
around $10^4-10^5$K when the true temperatures would be much lower.
This affects the formation of $\lya$ clouds
most severely; thus, this modification was absolutely necessary to
use the code for the problem of $\lya$ clouds.
Details on this subtle treatment can be found in Ryu \etal (1993).

  We adopt the same model as in Paper I, a CDM+$\Lambda$ model
with
$\Omega=0.4$,
$\Omega_b=0.0355$ (cf. Walker \etal 1991),
$\lambda=0.6$,
$h=0.65$ and
$\sigma_8=0.79$
(first-year COBE normalization; Efstathiou, Bond, \& White 1992;
Kofman, Gnedin, \& Bahcall 1993.
The second-year COBE data gives a $\sigma_8$ about 15\% higher;
Stompor \& Gorski 1994;
White \& Bunn 1995)
but a very small tilt $(n<1)$ in the spectrum would recover
the initial conditions that we take (for scales $L\le 10h^{-1}$Mpc)
while retaining the CBR normalization (\cf Ostriker \& Steinhardt 1995).
In Paper I, we presented preliminary results from two simulations on
a box of size $3h^{-1}\mpc$: the first one with $N=288^3$ cells and
$144^3$ dark matter particles, which we shall denote as the L3
simulation in this paper, and the second one with $144^3$ cells
and $72^3$ dark matter particles, which will be denoted as the l3
simulation. This paper will examine much more thoroughly a simulation on
a box of $10 h^{-1} \mpc$ with $288^3$ cells, which we call the L10
simulation, and we shall compare the results to the L3 and l3 simulations
on the smaller box. As we shall see, the effects of the large-scale
power that can be included in the L10 simulation are substantially
important, whereas the differences between the L3 and l3 simulations
are small suggesting that the resolution employed is sufficient for
reproducing the $\lya$ clouds.
The cell size in the L10 simulation is $35 h^{-1} \kpc$ (comoving)
corresponding to a baryonic mass of $6.3\times 10^5 M_\odot$.
The power spectrum transfer function is computed
using the method described in Cen, Gnedin, \& Ostriker (1993).

  Atomic processes within a primeval plasma of hydrogen and helium of
composition (76\%,24\%) in mass are all included using the same heating,
cooling, and ionization terms as in Cen (1992).
We calculate self-consistently
the average background photoionizing
radiation field as a function of frequency, assuming the radiation field
is uniform. The evolution
of the radiation field is calculated given the average attenuation in the
simulated box and the emission (both from the gas itself and from the
assumed sources of ionizing photons).
The time-dependent equations for the ionization structure of the gas
are solved by iteration using an implicit method, to avoid the
instabilities that arise in solving stiff equations. In general, the
abundances of different species are close to ionization equilibrium
after most of the gas has been photoionized,
when the universe becomes optically thin and the
intensity of the ionizing background starts rising.
However, before reionization is complete the intensity of ionizing
photons is very low and the ionization timescale correspondingly long,
so ionization equilibrium cannot be assumed.

  At every timestep during the simulation, two-dimensional tables as a
function of density and pressure of the gas are made giving the rates
of ionization and recombination for each species,
as well as the cooling and
heating rate and the radiation emissivity rate, given the spectrum of
the background radiation at that timestep.
Interpolation within this table
is then used to calculate these quantities at every cell, allowing us
to speed up the calculation by a large factor and to solve for the
ionization of the gas using the time-dependent equations.
This approach is an improvement over the assumption of ionization
equilibrium; in particular, it is absolutely necessary to use as the gas
is being reionized. In this way, we ensure that the energy of
reionization (i.e., the average thermal energy given by the
photoionization of every hydrogen atom) is correctly
deposited on the gas. The assumption
of ionization equilibrium would lead to a very low initial temperature
for the intergalactic gas, because the gas is suddenly ionized as the
intensity of the background rises, and it is then heated only at the
rate at which the ions can recombine and be photoionized again.
It needs to be pointed out here that our (invalid)
assumption of a uniform
background intensity has probably also reduced the initial temperature
of the gas, because, as the energy of reionization is deposited in the gas,
the neutral fraction remains high making line cooling very
efficient. In reality, the reionization would occur in expanding HII
regions around discrete sources, so every element of gas is reionized
during a very short time as it is engulfed by an expanding HII region,
and the energy of all the photoionizations can be more efficiently
preserved in the form of thermal energy for the gas.

We model galaxy formation as in Cen \& Ostriker (1992, 1993a,b).
The material turning into collisionless particles as
``galaxies" is assumed to emit ionizing radiation, with two types of spectra:
one characteristic of star formation regions and the other
characteristic of quasars, with efficiencies (i.e., the fraction of
rest-mass energy converted into radiation) of
$e_{UV,*}=5\times 10^{-6}$, and $e_{UV,Q}=6\times 10^{-6}$, respectively.
We adopt the emission spectrum of massive stars from Scalo (1986)
and that of quasars from Edelson and Malkan (1986).
Details of how we identify galaxy formation and
follow the motions of formed galaxies
have been described in Cen \& Ostriker (1993a).
Note that in the simulations which we are using in this paper
no supernova energy input into IGM is included;
we will study this effect in a later paper.

\topinsert
\centerline{
\epsfxsize=4.0in \epsfbox{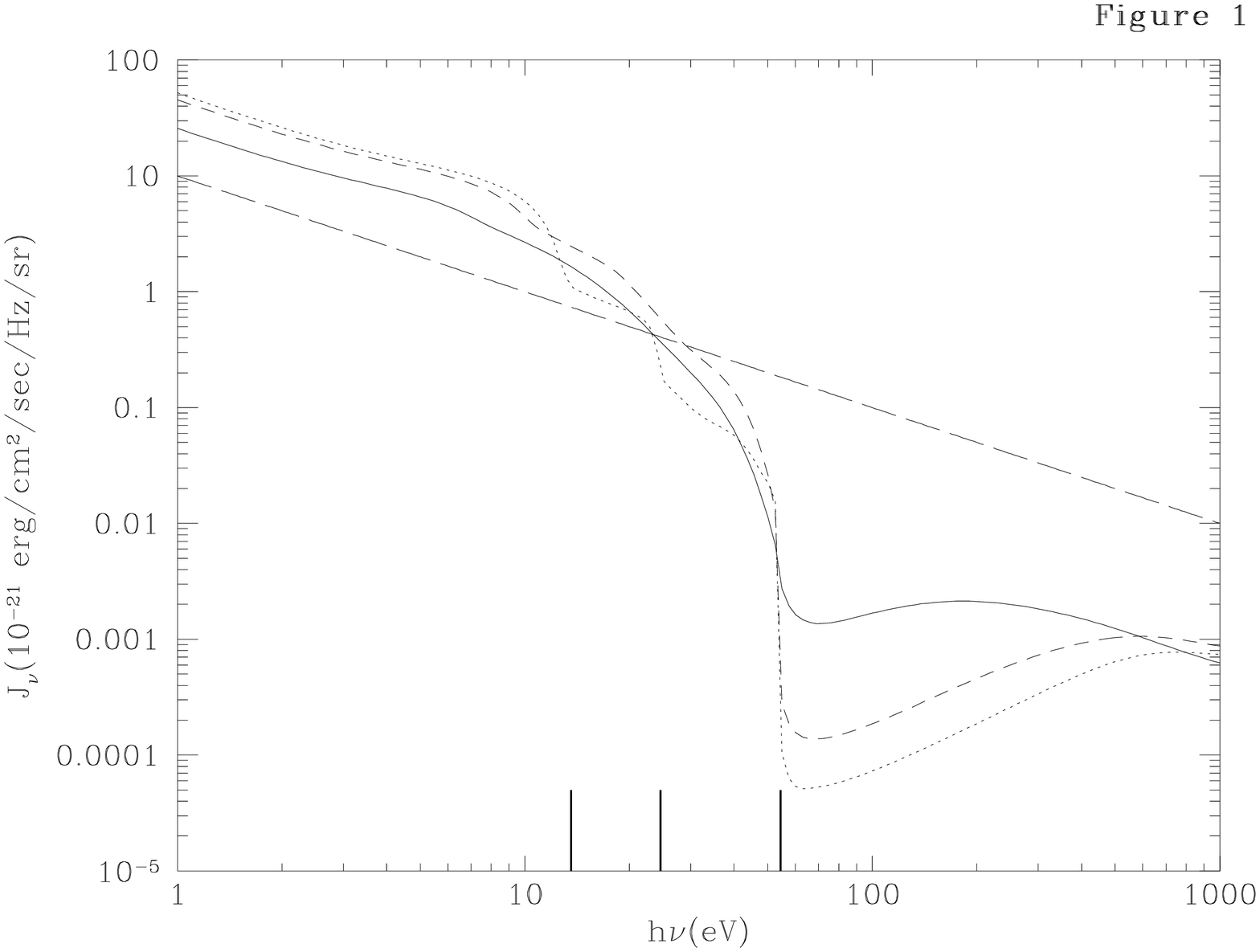}
}

\baselineskip=12truept \leftskip=3truepc \rightskip=3truepc
\noindent {\bf Figure 1:}
The radiation field in the simulation
is shown for three different redshifts z=2 (solid curve),
z=3 (dashed curve) and z=4 (dotted curve).
Also shown for comparison as a long dashed line is
a power-law case with index $-1$ (used in HKWM);
our radiation spectra are softer
due in part to our intrinsic spectra and to the absorption effects.
\endinsert

  The resulting spectrum of the ionizing radiation field is shown in
Figure 1 for the L10 simulation at redshifts $z=2$, 3 and 4, and for
the L3 and l3 simulations at $z=3$. In general, the results in this
paper will be shown at this same set of redshifts for the L10
simulation, and only at $z=3$ for the other simulations, since this
is enough to compare the results and see the effects of the large-scale
power and the resolution. The intensity of the radiation depends mainly
on the fraction of baryons which have been turned to stars in each
simulation. This fraction is given in Table 1. In general, a higher
fraction of baryons are turned into stars in the high-resolution
simulation, but this fraction is also increased by the large-scale power
included in the L10 simulation. We need to keep in mind, however, that
the fraction of baryons in ``stars'' is only an indication of the gas
that has collapsed to very high densities, but much of this gas might
not turn into stars and remain in the form of gaseous disks. This is
in fact probably the case, given the large amount of neutral gas that
is observed in the damped systems.

\topinsert
\vskip 0.2truecm
\centerline{\bf TABLE 1: FRACTION OF BARYONS IN STARS}
\bigskip
\hrule\vskip0.1truecm\hrule
$$\vbox{\tabskip 1em plus 2em minus 5em
\halign to\hsize{\hfil # \hfil & \hfil # \hfil & \hfil # \hfil \cr
Simulation & z & Fraction \cr
\noalign{\medskip\hrule\medskip}
L10 & 2 & 0.086 \cr
L10 & 3 & 0.063 \cr
L10 & 4 & 0.042 \cr
L3 & 3 & 0.071 \cr
l3 & 3 & 0.060 \cr
\noalign{\medskip\hrule\medskip}
}}$$
\vskip 0.3truecm
\endinsert

  HKWM and Katz \etal (1995) adopted a fixed model for the ionizing
background in their simulation, and required an intensity of only
$10^{-22} \erg\cm^{-2}\sec^{-1}\hz^{-1}\sr^{-1}$ at the Lyman limit,
which was necessary
to reproduce the observed number of lines in the $\lya$ forest.
They also adopt a specific spectral form $J_\nu \propto \nu^{-1}$
whereas we adopt a source function $S_\nu$ and, with allowance  for
absorption and cosmological expansion, compute $J_\nu$ as a function
of time. We show for comparison the spectral shape adopted by HKWM
and note that their assumed spectrum is much ``harder"
in the range $13.6 eV < E< 54eV$ relevant for hydrogen and ``softer"
thereafter than is ours.
As we shall see, we will need to adopt a low intensity of the
background, similar to HKWM, and recalculate the neutral fractions for
this lower intensity.
With the intensity that we obtain in the simulation shown
in Figure 1, there would be too few lines in our simulated
spectra (our statement in Paper I that a high intensity could
reproduce the number of observed lines was due to an error of a factor
of three in computing the number of lines per unit redshift, plus the
fact that lines were counted in real space instead of simulated spectra).
In Katz \etal (1995), it was assumed that all
the dense and cooled gas would remain in the form of neutral atomic
material, with no stars being formed. It is interesting to notice that
the fraction of baryons in the damped systems in their simulation is
similar to the fraction of baryons that are turned into stars in ours.

\sect{3. PHYSICAL CONDITIONS OF THE ABSORBING SYSTEMS}

  In this Section, we analyze the nature of the structures formed in the
simulations and we show some representative examples of the absorption
spectra that are produced, and the physical conditions in the objects
yielding absorption lines of different column densities. We shall
concentrate on the simulation in the $10 h^{-1} \mpc$ box (L10), since
the larger size of this simulation allows us to have a more
representative set of collapsed objects and, as we shall see, the
inclusion of the large-scale waves which are absent in the simulations
on the $3 h^{-1} \mpc$ box has highly noticeable effects, whereas the
corresponding loss of resolution in the large box has little effect on
the systems causing the $\lya$ forest.

\topinsert
\centerline{
\hskip -0.5truecm
\epsfxsize=3.2in \epsfbox{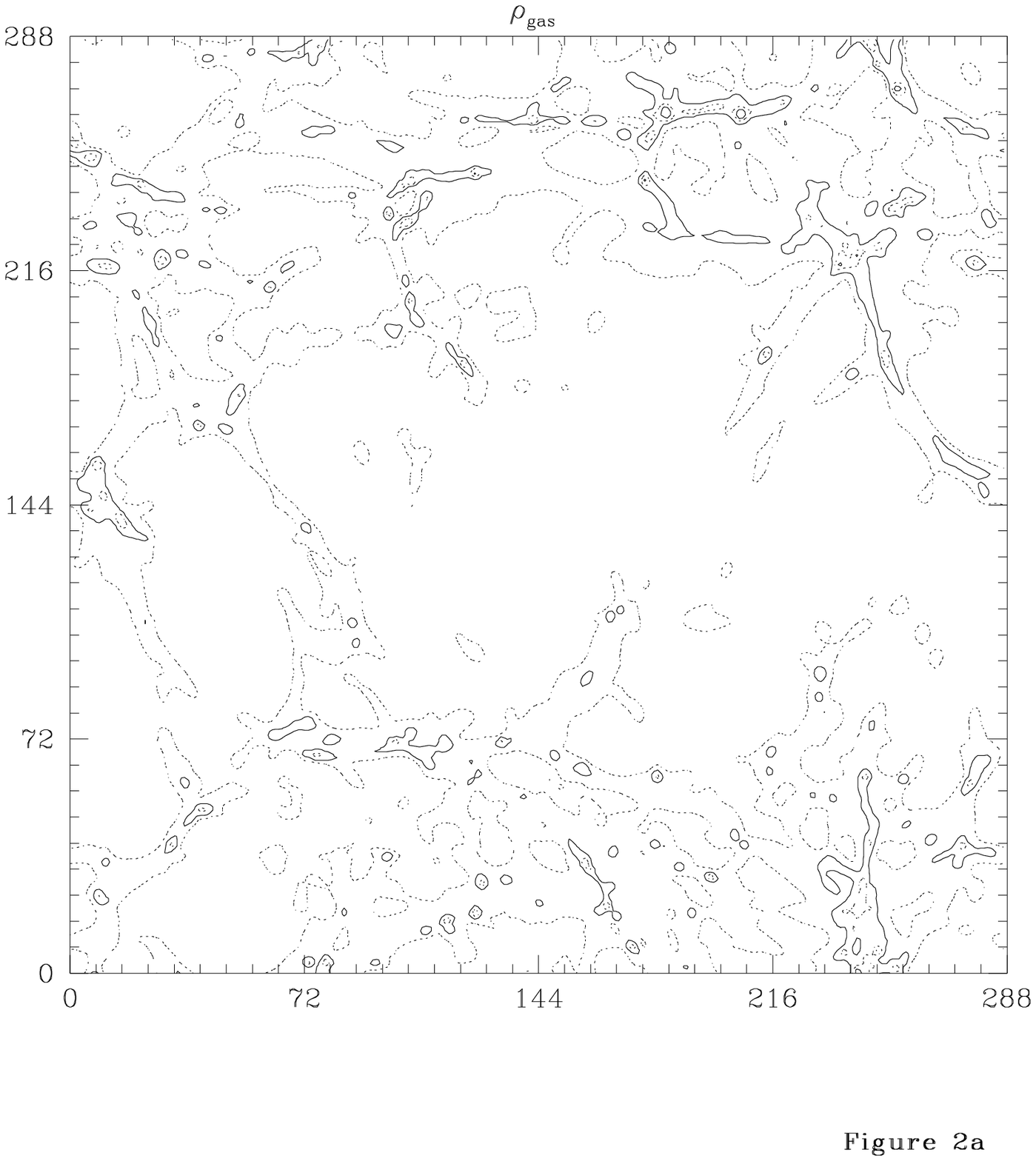} \hskip -0.2truecm
\epsfxsize=3.2in 
}
\centerline{
\epsfxsize=3.2in 
}

\baselineskip=12truept \leftskip=3truepc \rightskip=3truepc
\noindent{\bf Fig. 2:} Slice of the $10 h^{-1} \mpc$ simulation, averaged over
48 cells, showing contours of (a) gas density, where contours are at
$\rho/\bar\rho = 10^{0.5(i-1)}, i=1,2,3,...$, with solid contours for even
$i$ and dotted contours for odd $i$; (b) dark matter density, with the same
contours; and (c) neutral column density, with contours
$10^{12+0.5i} \cm^{-2}$, $i=1$ shown dotted and $i=2,3,...$ shown solid.
The neutral column density is for $\jhi$ as obtained in the simulation
(Table 1), but we use $\jhiu=0.1$ in the rest of the paper to obtain a
higher neutral density and similar $\lya$ absorption as observed.
\endinsert

\subsec{3.1}{General Characteristics of the Absorbing Structures}

  Figures 2(a,b,c) show a slice of the L10 simulation with a thickness
equal to 48 cells.
The projected quantities shown are the gas density,
the dark matter density, and the column density of neutral hydrogen.
The first two quantities are averaged over the projected cells.
Contours are shown for $\rho/\bar\rho = 10^{0.5(i-1)}$ ($i=1,2,3,...$),
where $\rho$ is the density and $\bar\rho$ the average density of gas
and dark matter for Figures 2(a,b), respectively.
The contours are plotted as solid lines for even $i$, and as dotted
lines for odd $i$.
Figure 2c shows the integrated column density of neutral gas.
Contours are shown at column densities
$10^{12+0.5i} \cm^{-2}, ~ i=1$ with the dotted line
and at ~$i=2,3,...$ with solid lines.
As we shall see below, in order to reproduce
the observed number of lines and average depression of the flux due to
the $\lya$ forest, the density of neutral gas in the simulation has to
be multiplied by a factor $\sim 15 - 20$, which can be achieved by
either lowering the intensity of the ionizing background in the
simulation by this factor, or raising the baryon density by the square
root of this factor.

  The distribution of dark matter and gas on large scales follows the
interconnected structure of sheets, filaments and halos that is
produced in any hierarchical gravitational collapse theory. The
gas and dark matter follow each other very well on large scales.
However, on small scales the dark matter is very clumpy, showing the
small structure of small halos that collapsed at an earlier epoch and
are now merging, whereas the gas is much smoother.
In particular,  the central density of spheroidal structures (halos) is
much higher in dark matter than in gas.
This is due to the
gas pressure, which prevents it from collapsing on scales smaller than
the Jeans length, $\lambda_J \equiv c_s(\pi/G\bar\rho)^{1/2}$, where $c_s$ is
the sound speed of the gas (e.g., Peebles 1980, \Sec 16). At z=3, and
for a gas temperature $T=15000 \kelvin$ ($c_s=19\kms$), the Jeans length
in our model (with $\Omega=0.4$ at $z=0$) is $\sim 800 h^{-1} \kpc$ in
comoving scale, which is 23 cells in the simulation. The smallest
structures where the gas will start to collapse will be on this scale,
and the photoionized gas remains smooth over the Jeans length, which
decreases in high density regions as $\rho^{-1/2}$.
There are numerous very small structures (size of
1 to 3 cells in projection)
in the dark matter slice showing a contour level of
$\rho/\bar\rho = 1$.
These structure are partly due to the discreteness
and small number of dark matter particles involved
(a projected cell with density equal to the mean density
contains only six particles).

\topinsert
\vskip -0.6truecm
\centerline{
\hskip -0.5truecm
\epsfxsize=2.9in \epsfbox{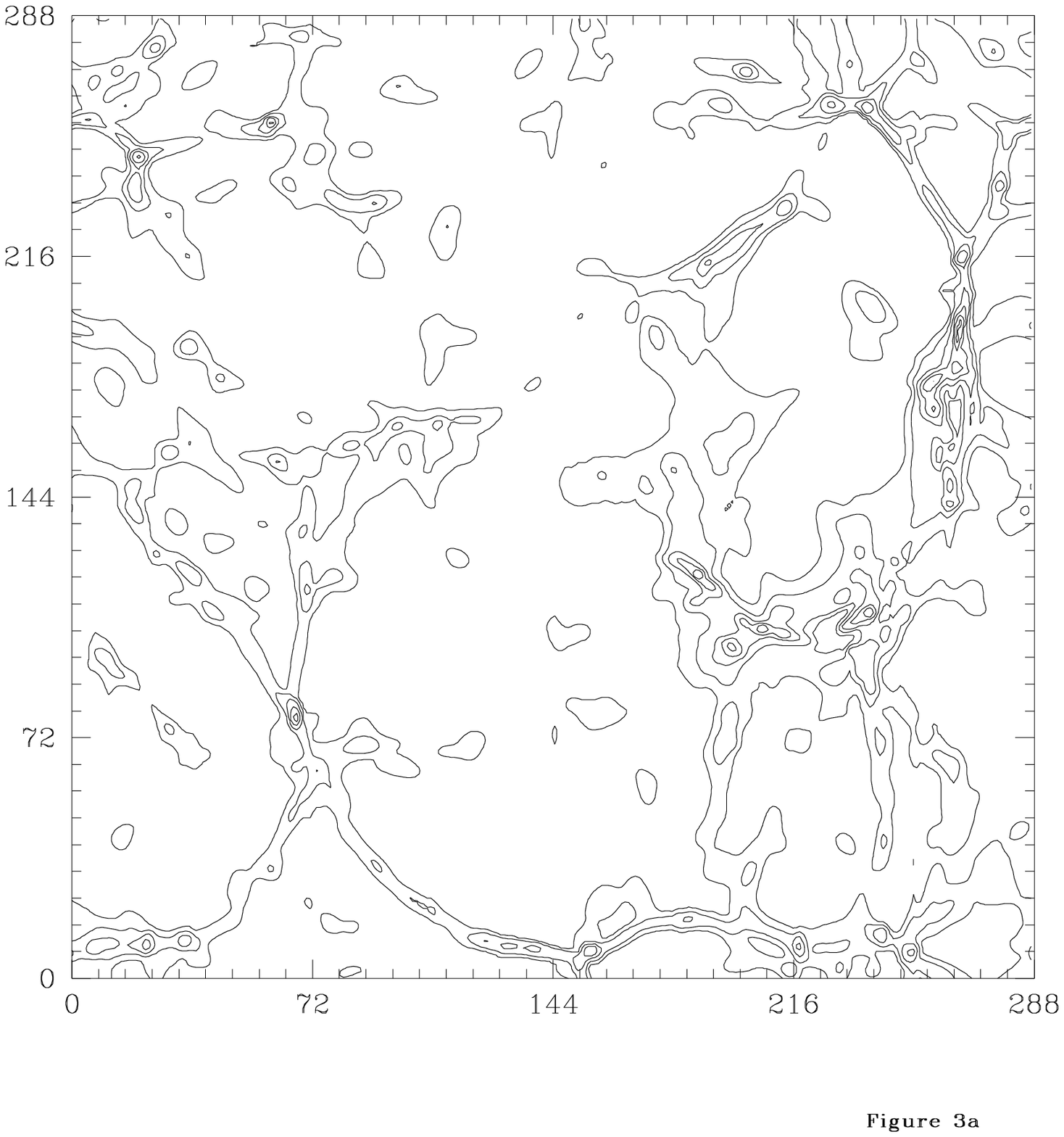}
\epsfxsize=2.9in \epsfbox{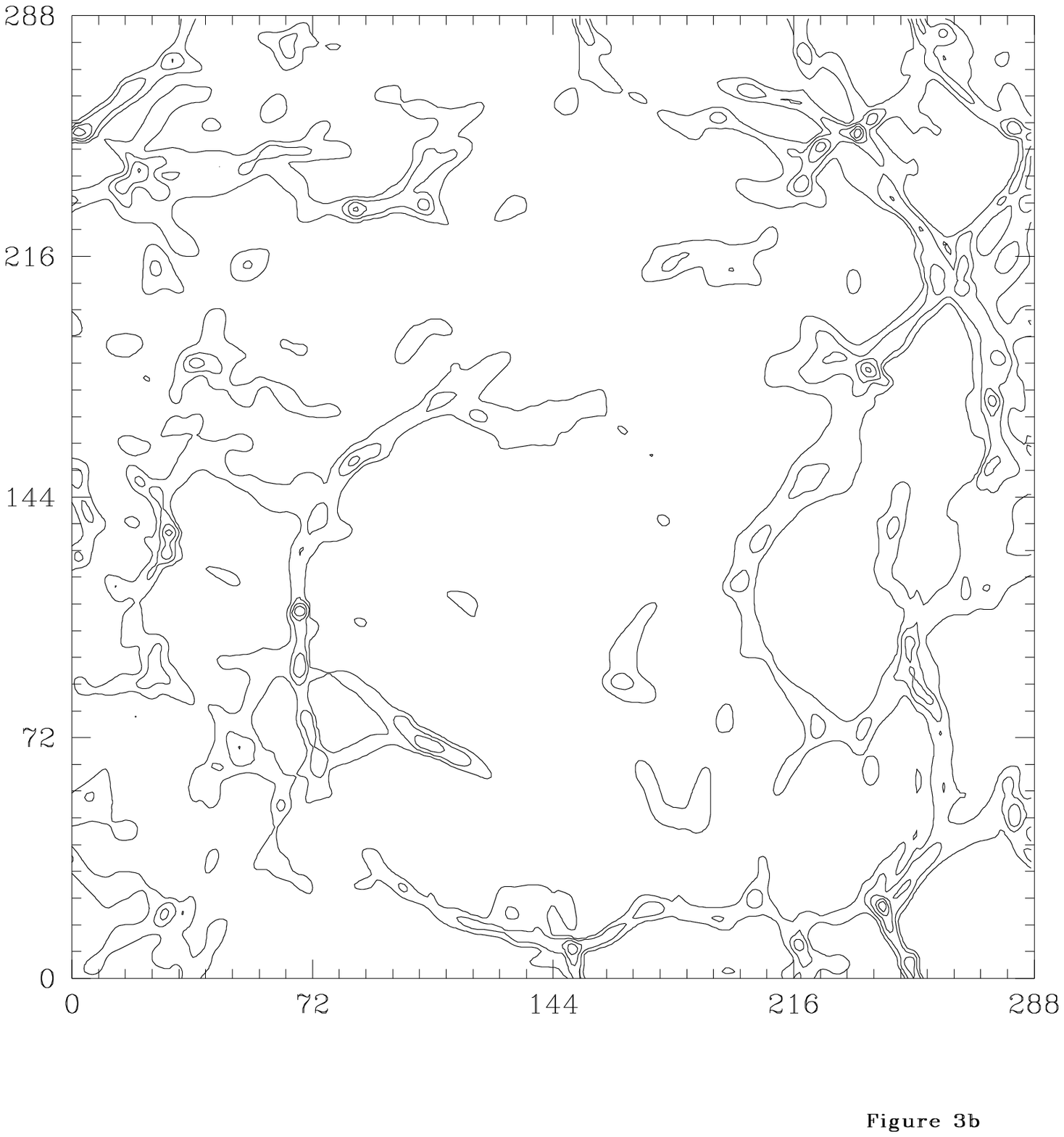}
}
\vskip -0.7truecm
\centerline{
\hskip -0.5truecm
\epsfxsize=2.9in \epsfbox{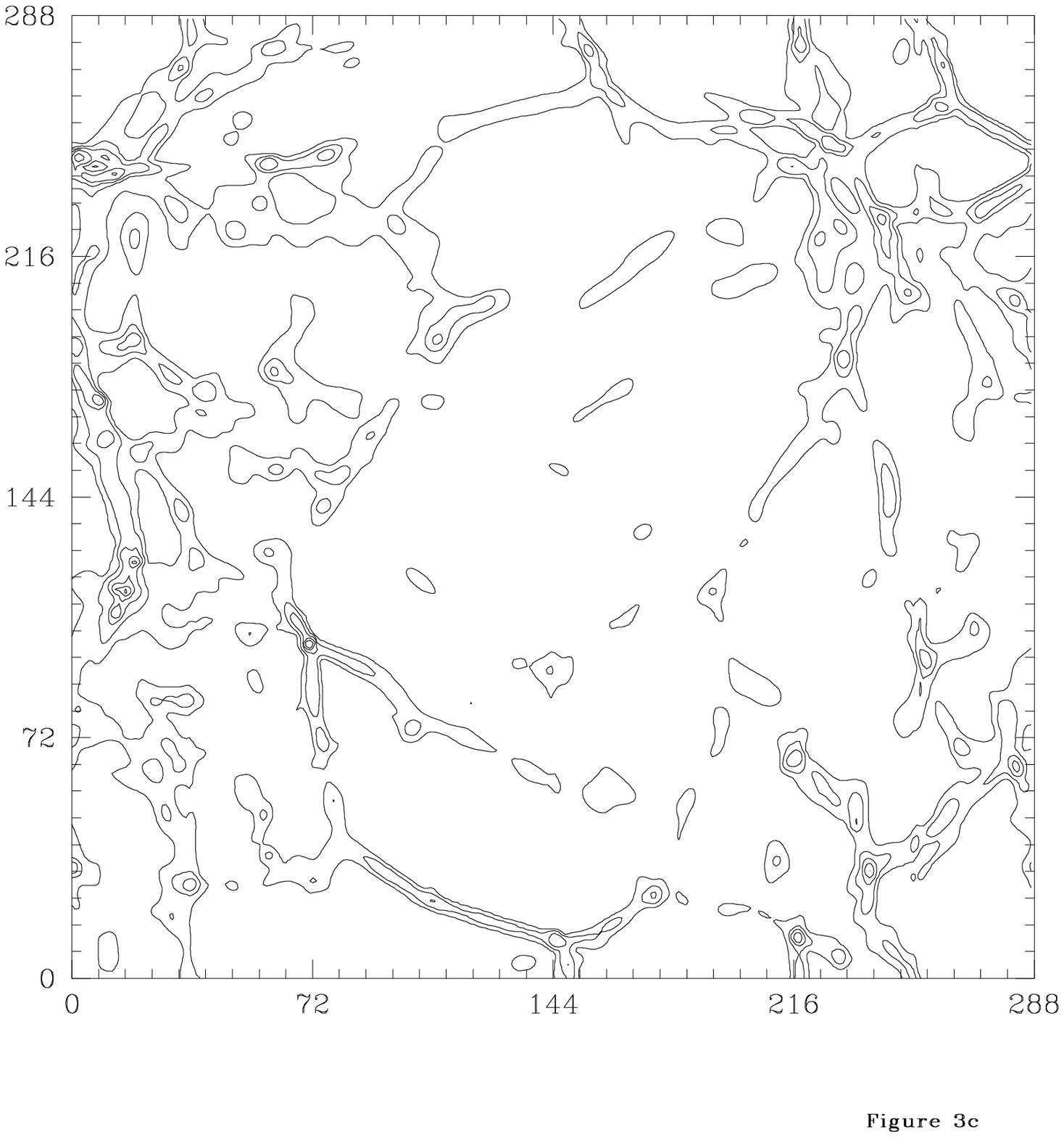}
\epsfxsize=2.9in \epsfbox{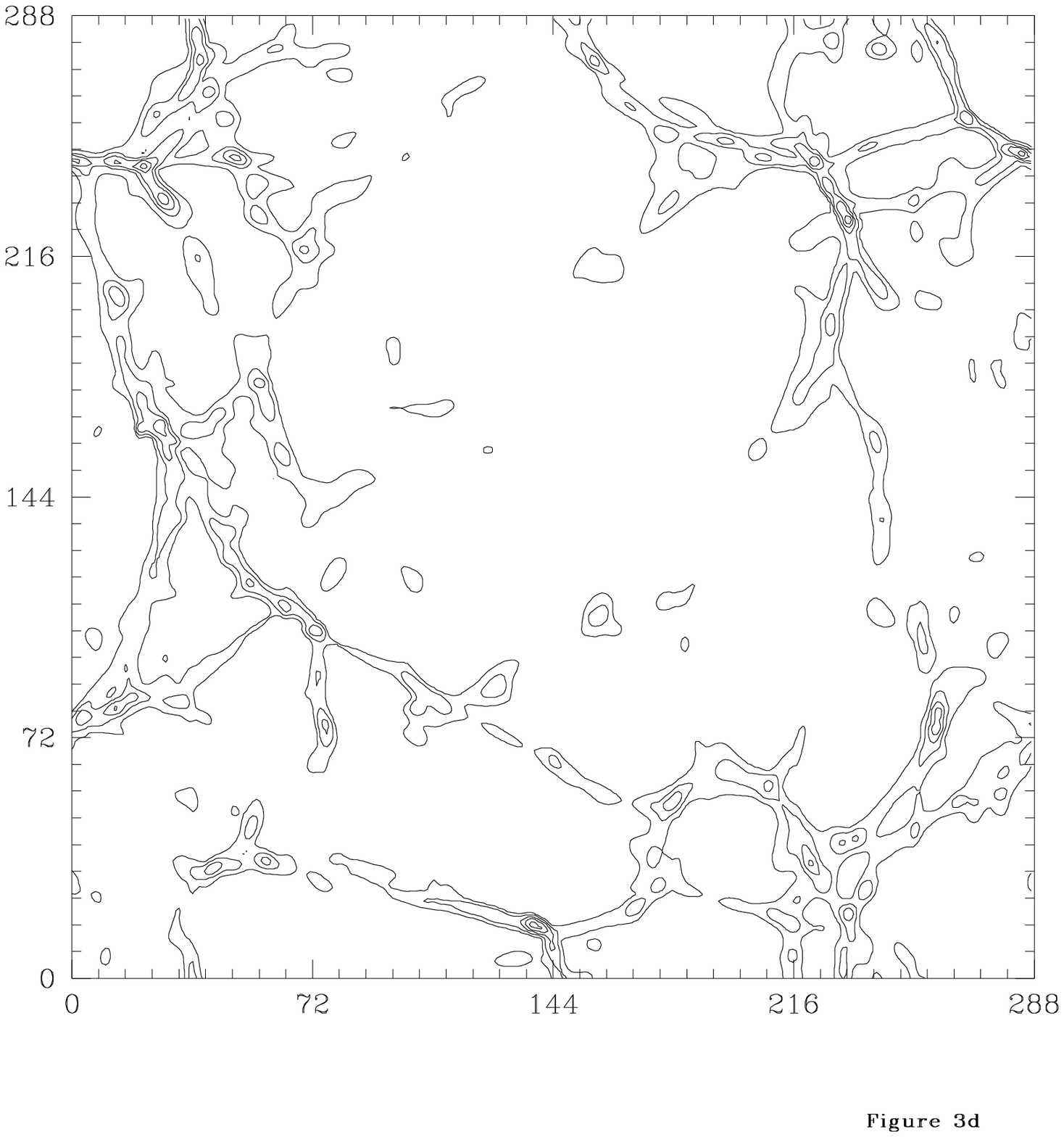}
}
\vskip -0.7truecm
\centerline{
\epsfxsize=2.9in \epsfbox{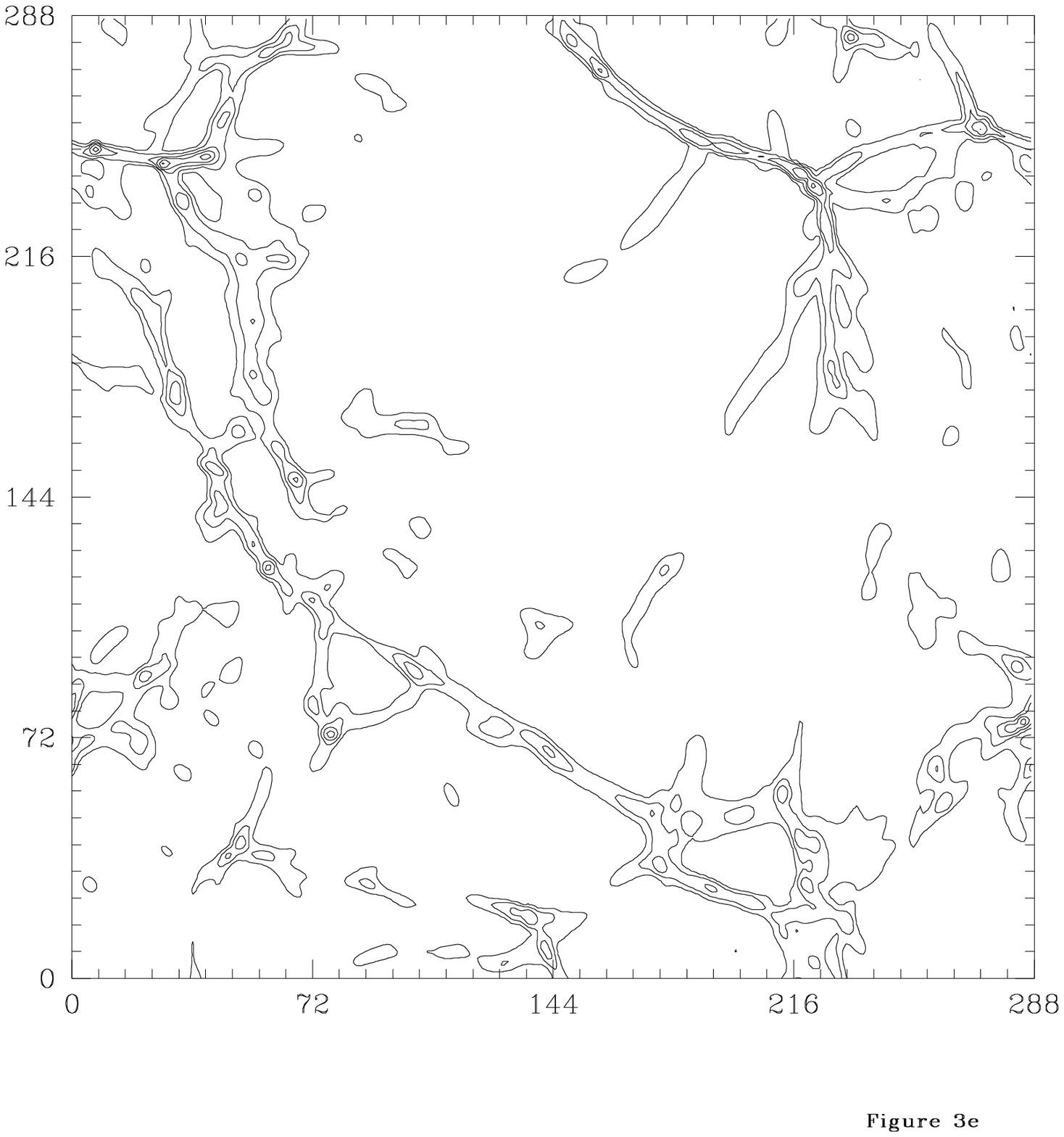}
}
\baselineskip=12truept \leftskip=3truepc \rightskip=3truepc
\noindent{\bf Fig. 3:} Gas density in five slices with thickness of only
one cell, (numbers 4,13,22,31, and 40 among the 48 of Fig. 2), with the
same contours as in Fig. 2a but all shown in solid curves. The
separation between two consecutive slices is $78^{-1}$kpc in proper
units.
\endinsert

  In Figures 3(a,b,c,d,e), we show the gas density in five slices of
one cell thick.
The contours are the same as
in Figures (2a,b) ($\rho/\bar\rho = 10^{0.5(i-1)}$, i=1,2,3,...).
The five slices shown are separated by 9 cells, or
$78h^{-1}$kpc; Figure 2 is averaged over 48 cells.
A filamentary structure located
near the bottom of Figure 3a connecting northwest-ward to the halo at
position (72,72) is clearly seen to extend through all five slices.
Another inter-connected structure located at the top right corner
is also seen in five slices.
The thin structures shown as
the contour level equal to the mean density
typically are sheet-like, which
gradually turn into filaments as density levels go
higher, and finally become spheroids in the vertices, the
likely sites for formation of galaxies.
The typical objects collapsing in the
simulation at this redshift form from regions of $\sim 1 h^{-1} \mpc$,
containing total masses of $10^{11} h^{-1} \msun$ (for $\Omega=0.4$),
and should therefore be the progenitors of relatively low-mass galaxies.

\topinsert
\centerline{
\epsfxsize=2.9in \epsfbox{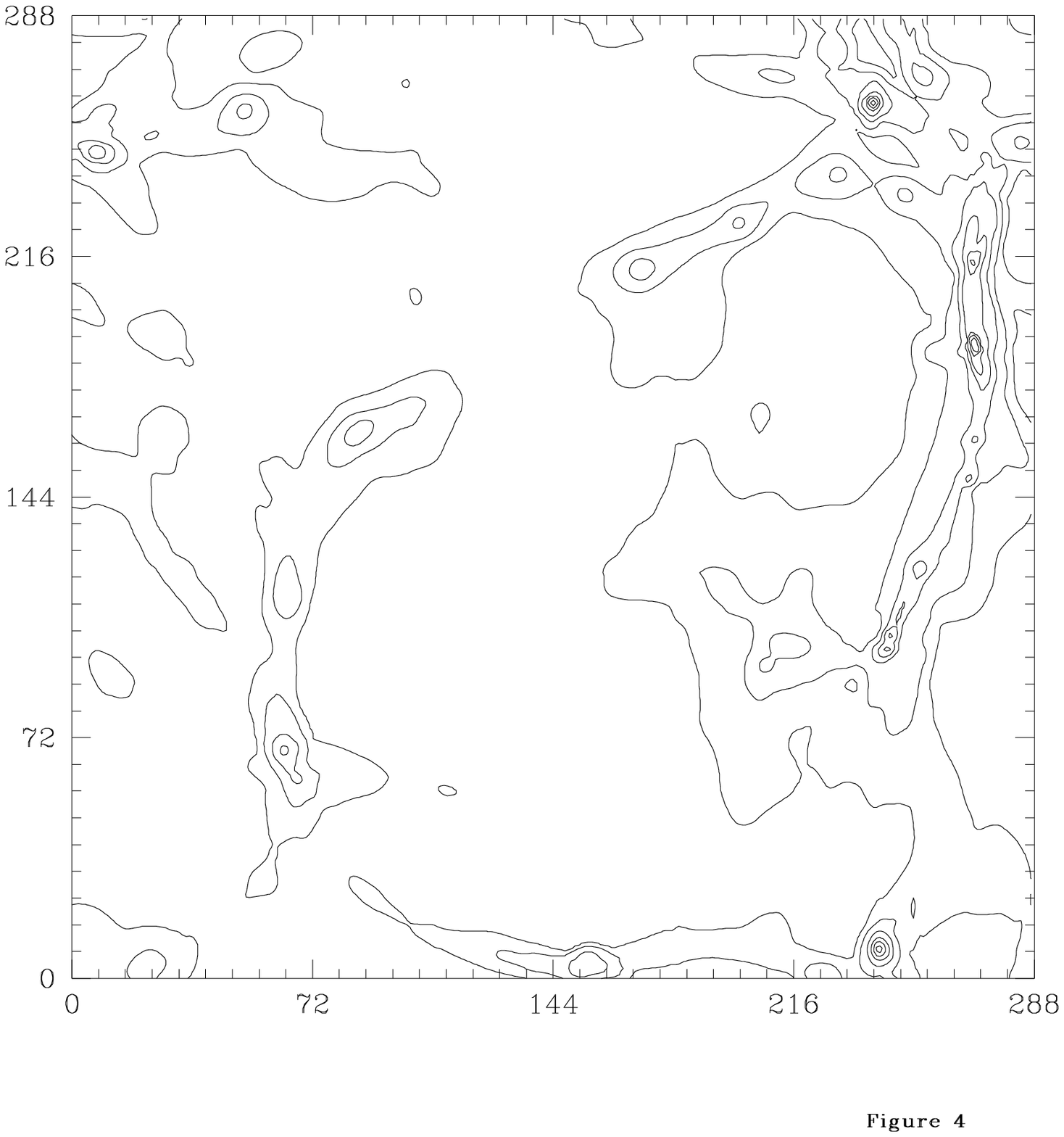}
}
\vskip -0.2truecm
\baselineskip=12truept \leftskip=3truepc \rightskip=3truepc
\noindent{\bf Fig. 4:} Slice of one cell showing gas density in the L3
simulation, with the same contours as in Fig. 2.
\endinsert

  The gas density in a slice of the L3 simulation, in the $3 h^{-1}
\mpc$ box, is shown in Figure 4. This slice is also one cell thick and
the density contours are the same as in Figure 3. The only regions where
the density is varying on a small scale are the halo centers, where the
density is highest. Filamentary structures are already well resolved
in the L10 simulation. The gas is intrinsically smooth in such regions
due to its own pressure. The characteristic thickness of these
structures is also an intrinsic physical property, independent of the
box size: $\sim 250 h^{-1} \kpc$ in comoving scale. The characteristic
thickness of the structures is related to the scales that are
collapsing at a given epoch, in a similar way as the thickness of the
filaments of the large-scale distribution of galaxies is related to
the scales from which clusters are collapsing at the present epoch.

\topinsert
\centerline{
\epsfxsize=6.7in 
}
\vskip 0.4truecm
\baselineskip=12truept \leftskip=3truepc \rightskip=3truepc
\noindent{\bf Fig. 5:} (a) Lower half of the slice in Figure 3b is shown with
contours at $\rho/\bar\rho = 10^{0.25(i-1)}$, $i=1,2,3,... $ for solid
contours and $i=0$ for dotted contour. (b) Temperature contours
at $10^{4.2+0.1i} \kelvin$, $i=1,2,3,...$ for solid contours and $i=0$
for dotted contour. (c) Peculiar velocity divergence contours at
${\bmsy\grad} \cdot {\bf v}_p = -3H$ (solid contour, corresponding to
constant proper density), and ${\bmsy\grad} \cdot {\bf v}_p = 0$ (dotted
contour, corresponding to constant comoving density).

\endinsert

  We now take the lower half of Figure 3b to examine it in more detail.
In Figure 5, we show contours of the density, temperature and velocity
divergence of the gas in this half slice. The gas density (upper panel)
is the same as that shown in the lower half of Figure 3b, but with
more contours.
The solid contours are shown for
$\rho/\bar\rho = 10^{0.25(i-1)}$ ($i=1,2,3,...$);
the dotted contour is for $\rho/\bar\rho = 10^{-0.25}$.
The middle panel shows the temperature with
solid contours for $T=10^{4.2+0.1i}\kelvin,~i=1,2,3,...$,
and the dotted contour for $T=10^{4.2}\kelvin$.
The bottom panel shows contours of
peculiar velocity divergence, with
${\bmsy\grad}\cdot {\bf v}_p = 3Hi,~(i=0, -1)$,
where $H$ is the Hubble constant of the model at $z=3$.
The first contour ($i=0$) is
shown as dotted, and it indicates the regions expanding at the Hubble
rate (i.e., where the comoving density stays constant). Regions outside
this contour are expanding faster than the Hubble rate, and are
typically in voids and low-density regions of sheets and filaments. The
solid contour ($i=1$)
is in stationary regions, with constant proper density.

  In the relatively dense regions along sheets or filaments, a double
shock is often formed. Initially, the pressure rises in the center, and
then two shock fronts emerge and separate with increasing velocity,
so the gas is heated to progressively higher temperatures as the infall
velocity increases, causing generally a temperature minimum between the
shocks (see Sunyaev \& Zel'dovich 1972).
The middle panel shows the shock heated gas as on mesa like plateaus
with sharp boundaries indicating the outward propagating shocks.
The double shock structure is
seen along several of the filaments, in the velocity divergence contour
map as two parallel peaks.
The fact that the gas is photoionized implies that it starts with
an initial entropy, and that when the infall velocities are small the
collapse of the gas can be stopped by its own pressure. The shocks are
formed later, when the velocities are high enough, but they are
generally weak (i.e., the shock velocities are of order or only a few
times the sound speed of the gas), and the temperature minimum between
the double shocks is often very small or non-existent (notice, though,
that very thin slabs of gas in thermal equilibrium with the radiation
might be present and not be reproduced with our resolution).
In low-density
regions along the filaments, the gas expands as it falls towards higher
density clumps, and it can also expand into the voids as the
ram-pressure of infalling material declines. At some point, the shocks
surrounding the filament may become pressure waves.
In fact, several
examples can be seen in Figure 5 where the expansion rate accross a thin
filament has two minima around the center, but the gas is not
contracting anywhere. Some of these low-density objects are isolated
within voids (as the small density enhancement towards the center of the
panels in Figure 5); they probably correspond to structures below
the Jeans scale where the gas is never shocked and it can bounce back
from the dark matter potential well due to its own pressure. In this
case, the gas should probably experience a damped oscillation around the
dark matter, although on average it expands as the surrounding gas
density in the void declines. The details of these processes and the
distinction between weak shocks and pressure waves is difficult to see
in the simulation owing to the limited resolution; this will be examined
in more detail in Gnedin \etal (1995) for the case of a uniform,
plane-parallel sheet or filament.

  We also see that the lowest temperature gas occurs in the voids, and
the gas is generally hotter in the high density regions owing to
adiabatic and shock heating, except for the
features caused by double shocks and cooling mentioned before.
The gas in filamentary structures is generally
expanding in proper coordinates,
except at the shocks, but the overdensity still rises (in other words,
the gas contracts in comoving coordinates). As expected, the rate of
contraction increases towards high density regions.

\topinsert
\centerline{
\epsfxsize=3.0in \epsfbox{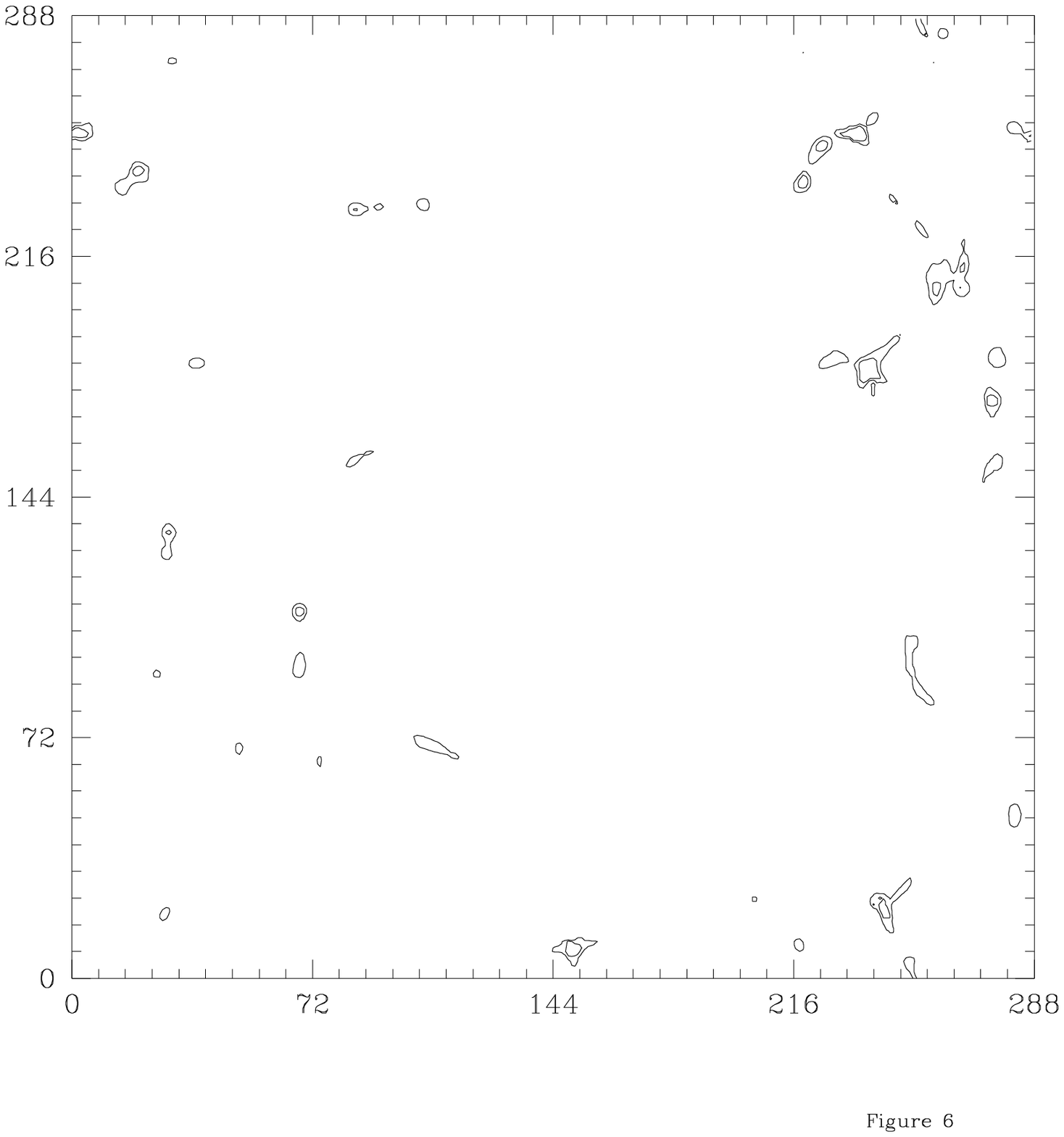}
}
\baselineskip=12truept \leftskip=3truepc \rightskip=3truepc
\noindent{\bf Fig. 6:} Contours of cooling time in the same slice
as Figure 3b, for $t_{cool} = 10^{-0.5i}\, H^{-1}$, i=0,1,2,...

\endinsert

  As the
gas continues to collapse along filaments, it is shocked again as it
merges into halos and will typically go through a complicated history of
radiative cooling and shock heating. In Figure 6
we show the following contours of the cooling time: $t_{cool} =
10^{-0.5(i-1)} \, H^{-1}, ~ i=1,2,3,...$, in the same slice as in
Figure 3b. The high density regions are cooling over timescales much
shorter than the Hubble time. As the gas cools and contracts in these
regions, small inhomogeneities will grow and lead to the formation of
clouds photoionized by the ionizing background and pressure-confined by
the surrounding hot gas (e.g., Fall \& Rees 1985). This is probably the
first physical process affecting the photoionized gas which our
simulation is not able to adequately resolve. In any case, the cooled
gas should continue to fall to the halo center and form galaxies. The
high column density absorption systems that can arise in this situation
have been studied by Katz \etal (1995) in an SPH simulation, under the
assumption that all the gas accreting to the center remains in the
form of neutral, atomic gas.

\subsec{3.2}{Choice of the Ionizing Radiation Intensity}

  The density of neutral gas shown in Figure 2c depends on the assumed
intensity of the ionizing radiation field used in the simulation. Most
of the gas in the simulation is at low enough density and temperature
that photoionization dominates over collisional ionization, and so if
the intensity of ionizing photons is increased by some factor, the
neutral density would simply decrease by the same factor eveywhere.
This will alter the number of absorption lines at a fixed column
density that will be observed along a line of sight, and the average
absorption decrement implied by all the neutral gas.
The intensity is determined by the models of production of ionizing
photons from the gas particles that form stars (see \Sec 2).
The value obtained for the intensity depends on an arbitrary parameter
of the models that were used giving an efficiency of production
of UV photons that can escape from stars and quasars formed out of the
radiatively cooled gas. The redshift evolution of the intensity can also
vary depending on the rate at which the stars and quasars form. Since
the detailed physics determining these processes are unknown to us, we
can change the photoionization rate to a value that will imply an
average flux transmission in the $\lya$ spectra close to what is
observed.

  For the purpose of determining the neutral fraction, the important
quantity is the photoionization rate. We define the quantity $J_{HI}$
as
$$ \jhi\equiv \left[ \shia \right]^{-1} ~ \int_{\nu_{HI}}^{\infty}
{d\nu\over \nu}\, J(\nu) \shi ~, \qquad\qquad
\shia\equiv \int_{\nu_{HI}}^{\infty} {d\nu\over \nu}\, \shi ~,
\eqnam{\javdef}\eqno(\new) $$
where $\shi$ is the ionization cross section of \hi , and $\nu_{\hi}$ is
the frequency of the ionization potential. This is proportional to the
photoionization rate, but is expressed in the usual (cumbersome) units
in the literature. In this paper, we shall fix this value to $\jhi =
10^{-22} \erg\cm^{-2}\sec^{-1}\hz^{-1}\sr^{-1}$ at all redshifts (this
corresponds to a photoionization rate of $4.34\times 10^{-13}
\sec^{-1}$). As we shall see, this value will give us a number of
absorption lines and an average flux decrement due to $\lya$ absorption
close to what is observed.

\topinsert
\vskip 0.2truecm
\centerline{\bf TABLE 2: RADIATION INTENSITY}
\bigskip
\hrule\vskip0.1truecm\hrule
$$\vbox{\tabskip 1em plus 2em minus 5em
\halign to\hsize{\hfil # \hfil & \hfil # \hfil & \hfil # \hfil \cr
Simulation & z & $\jhiu$ \cr
\noalign{\medskip\hrule\medskip}
L10 & 2 & 0.97 \cr
L10 & 3 & 1.56 \cr
L10 & 4 & 0.76 \cr
L3 & 3 & 1.38 \cr
l3 & 3 & 0.83 \cr
\noalign{\medskip\hrule\medskip}
}}$$
\endinsert

  The values of $\jhi$ that were obtained from the emission models in
various simulations at different redshifts are given in Table 2. We see
that these values of $\jhi$ are much higher than the value we need to
assume to have the right amount of absorption. This probably indicates
that the sources of ionizing photons should be fainter than were
assumed, and also that the absorption is higher. A lower $\jhi$ would
imply an increase in the neutral fraction of all the systems, and
therefore an increase of the absorption. In addition, optically thick
systems are not treated adequately here, and, as we shall see, their
numbers in our simulation are below the observed ones, so the absorption
should be strongly increased relative to our model. It is also likely
that the sources of ionizing photons appear in dense regions where the
absorbers are located, further reducing the background intensity. An
intensity $\jhiu = 0.1$ ($\jhiu \equiv \jhi / (10^{-21}
\erg\cm^{-2}\sec^{-1}\hz^{-1}\sr^{-1}) $ )
is also significantly lower than the
estimated contribution of the observed quasars, and the values given by
measurements of the proximity effect. However, the same neutral
densities can be obtained by raising $\Omega_b$ for a higher $\jhi$;
this will be discussed in detail in \Sec 5. The neutral fractions in the
simulation have been recalculated for the new value of the
photoionization rate, including also collisional ionization. These
neutral fractions are used for all the results shown in this paper
(except Fig. 2c), instead of the ones computed as the simulation was
evolved.

\subsec{3.3}{Examples of Simulated Spectra}

  We now select several rows along the half slice we have described, to
see the absorption lines that are produced. We take four horizontal rows
from the slice in Fig. 5 with cell number on the vertical axis equal to
20, 70, 96, and 133, indicated by the four horizontal dashed lines in
each panel. For each row we plot three panels in Figures 7(a,b,c,d). The
middle panels show the density along the row as the thick dotted line,
and the temperature as the solid line. The thin dotted line is the
pressure, plotted on the same scale as the density but with arbitrary
units. The coordinate along the row is shown as a velocity, where the
proper coordinate is $x \equiv v/H$ ($H=512 h \kms \mpc^{-1}$ at $z=3$
in our CDM$+\Lambda$ model). The total length in velocity of the L10
simulation at $z=3$ is $1280 \kms$, and a cell corresponds to
$4.44 \kms$. The peaks in density and temperature seen along these rows
can easily be identified in the contour plots of Figure 5.

\topinsert
\vskip -0.7truecm
\centerline{
\epsfxsize=5.5in \epsfbox{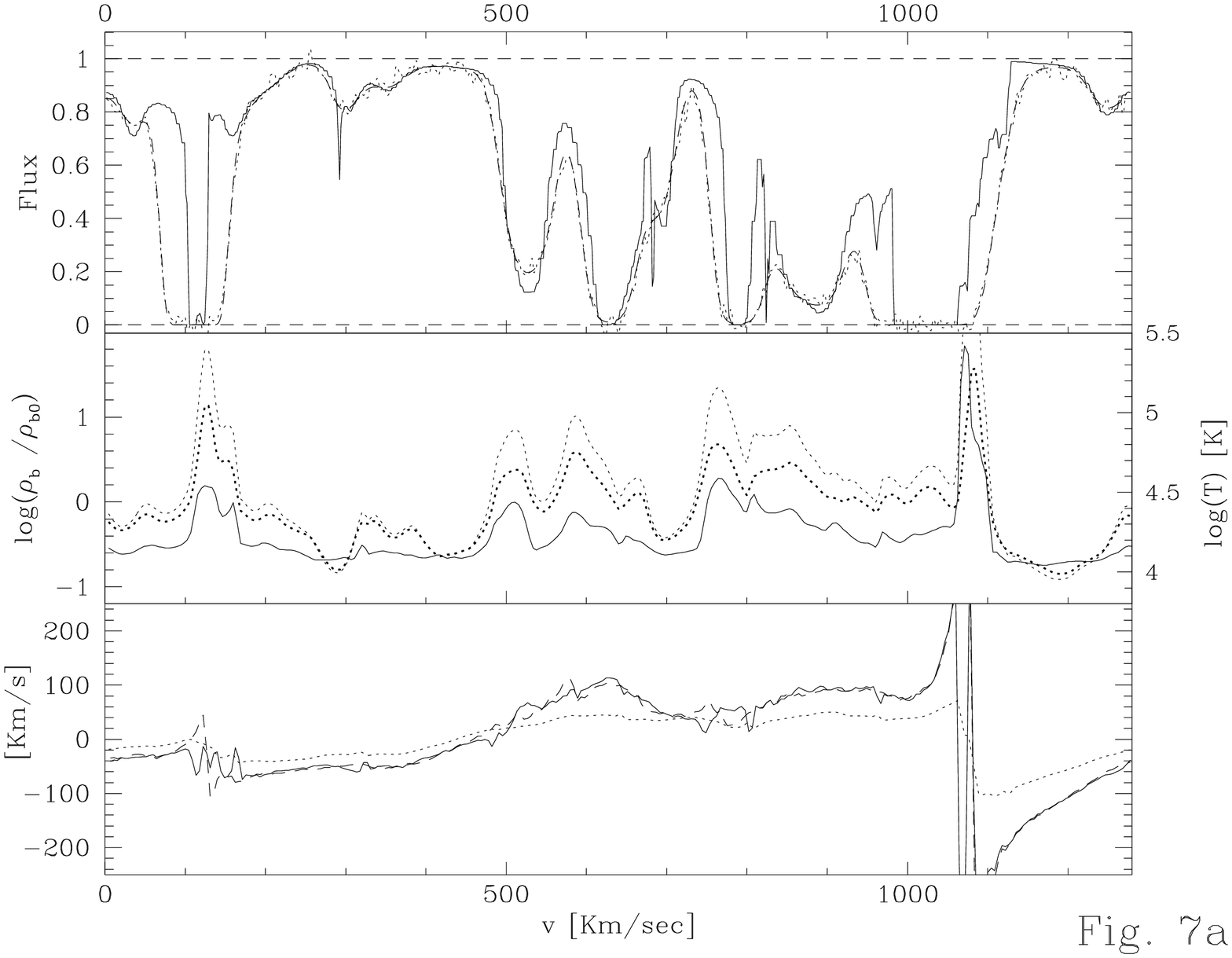}
}
\vskip -0.8truecm
\centerline{
\epsfxsize=5.5in \epsfbox{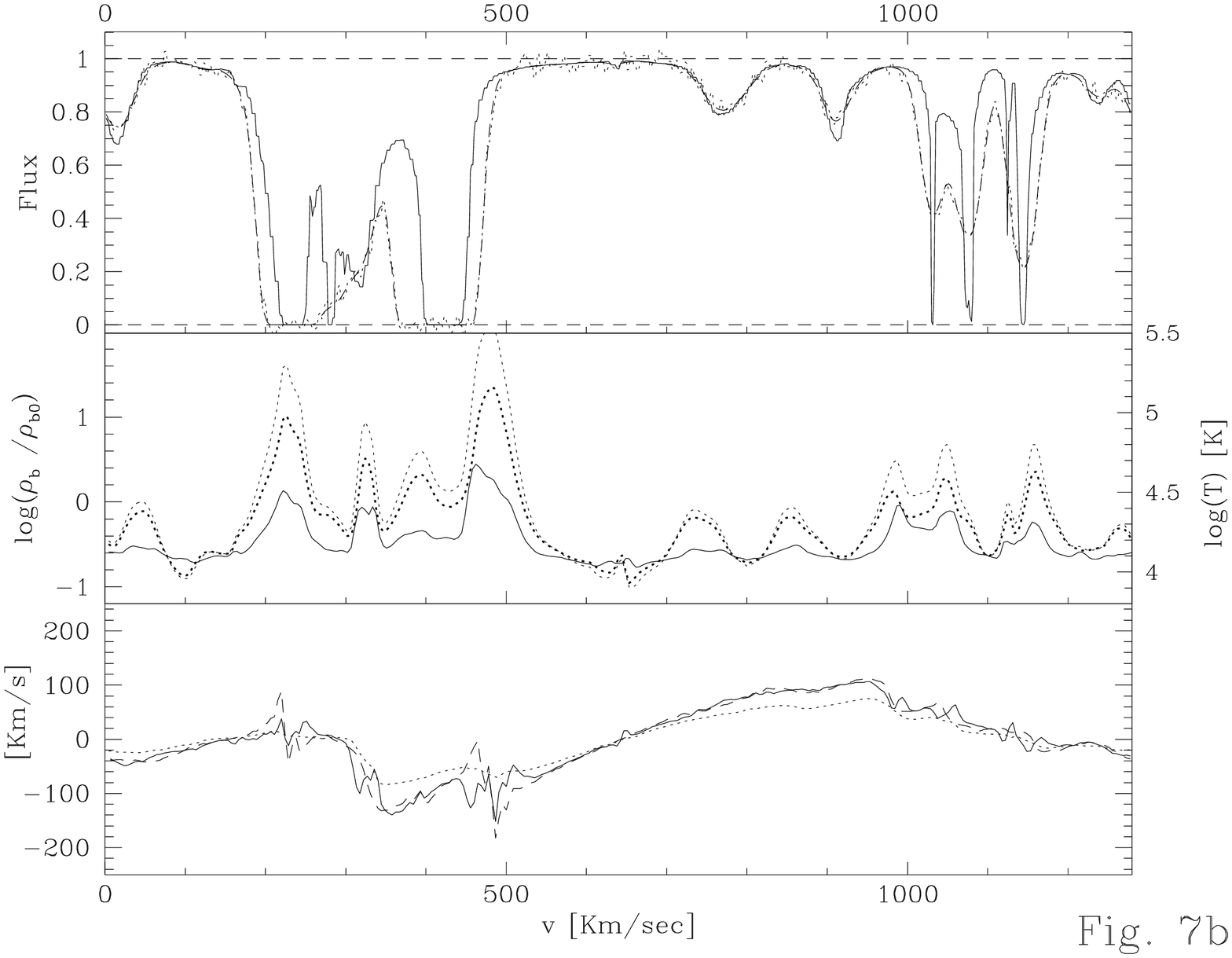}
}
\endinsert

\topinsert
\vskip -0.8truecm
\centerline{
\epsfxsize=4.9in \epsfbox{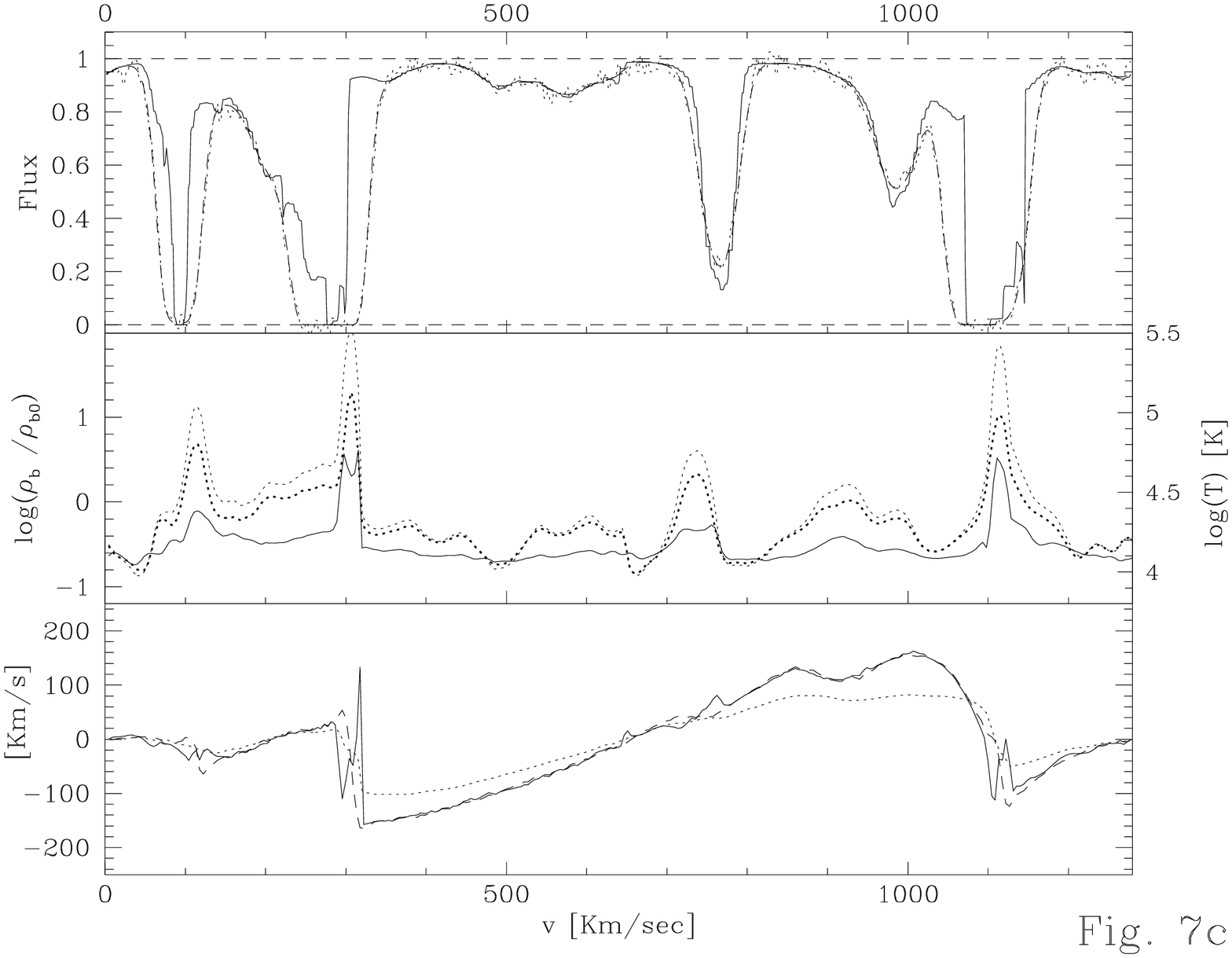}
}
\vskip -0.9truecm
\centerline{
\epsfxsize=4.9in \epsfbox{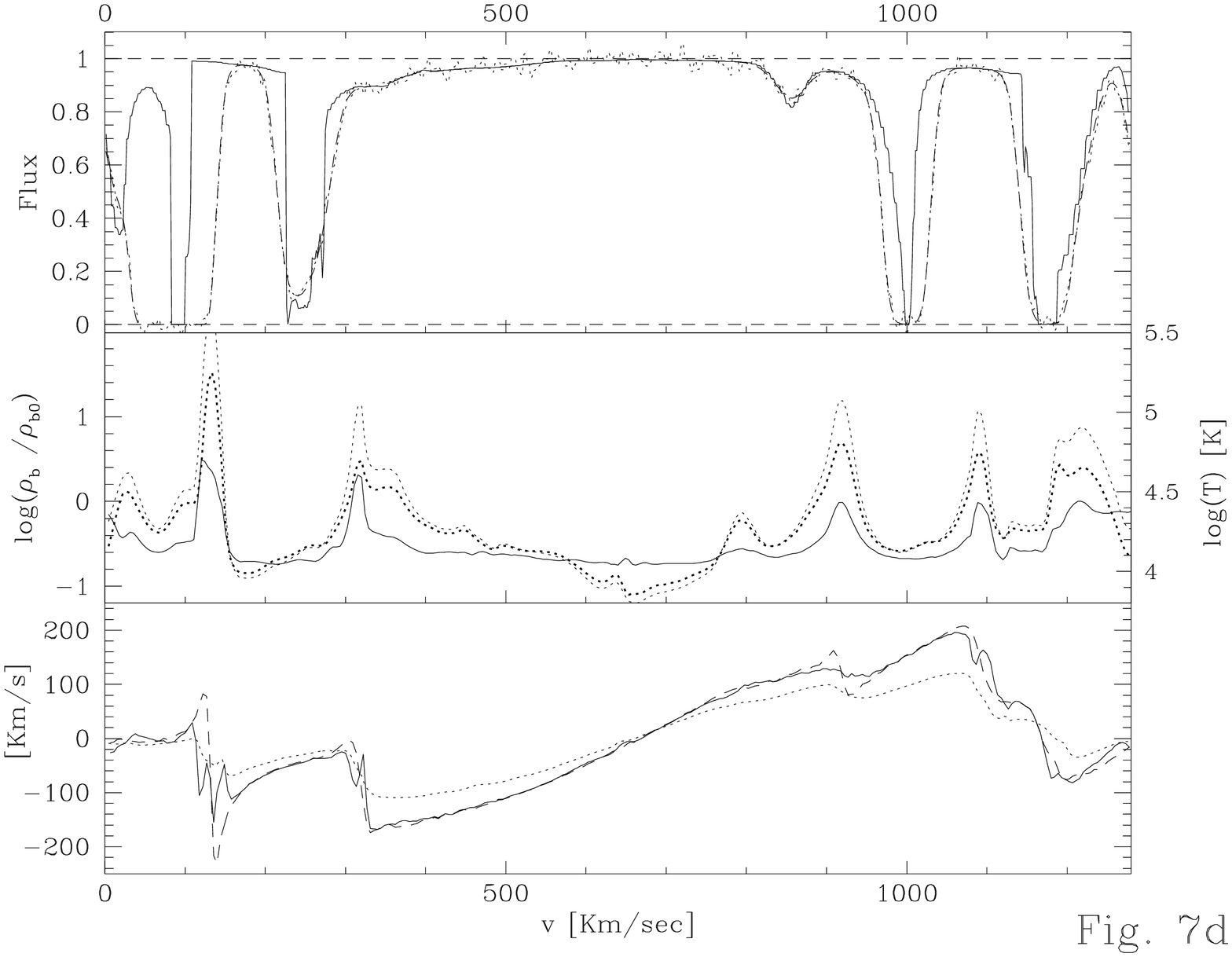}
}
\vskip -0.2truecm

\baselineskip=12truept \leftskip=3truepc \rightskip=3truepc
\noindent{\bf Fig. 7:} Middle panel in each figure shows the gas density
along a row, in units of the average gas density (thick dotted line; left
vertical axis), the gas temperature (solid line; right vertical axis),
and the gas pressure (thin dotted line; the same scale as density but arbitrary
units). Spatial coordinate in horizontal axis is $x = v/H$. Rows shown
in Figs. 7(a,b,c,d) are marked as dashed lines in slice in Fig. 5.
Calculated $\lya$ absorption spectrum is in top panel,
without thermal broadening (solid line), and including it (dashed line).
Peculiar velocity is shown as dotted line in bottom panel, together with
gravitational acceleration (dashed line) and total acceleration (solid
line) divided by the Hubble constant.

\endinsert

  The peculiar velocity of the gas along the row is shown as the dotted
line in the lower panel. We compute the resulting spectrum in the same
way as was done in Paper I; we use equations (1) and (2) of
Miralda-Escud\'e \& Rees (1993). The solid line in the upper panel
shows the spectrum without including thermal broadening, and the dashed
line includes thermal broadening. Finally, the dashed line in the lower
panel shows the gravitational acceleration divided by the Hubble
constant, $-(\grad \phi)/H$, and the solid line shows the total
acceleration $-(\grad \phi + \grad p)/H$ (notice that this acceleration
gives the time derivative of the {\it peculiar} velocity, since the
gravitational potential due to the average density is subtracted out).
Regions where the acceleration is increasing to the right are being
forced to expand in comoving coordinates, while they are being forced to
contract in comoving coordinates when the acceleration decreases to the
right.

\infsec{3.3.1}{Nature of the Absorption Features}

  We see in Figure 7 that, generally, the absorption features in the
spectra are caused by density peaks along a row. Velocity caustics
tend to produce narrow features when thermal broadening is not included,
but these are smoothed out by the thermal velocities. Usually, the
dynamical structure of the objects causing the absorption lines consists
of two shocks surrounding the central density peak, as mentioned above.
This is seen in the variation of the total acceleration (solid line in
the lower panel) as we move accross an object: on the left side,
there is usually a large negative peak, corresponding to the infalling
gas being stopped on a shock. The same thing occurs on the right side,
with a positive peak in the total acceleration. In the region between
these two shocks, the total acceleration can show a variety of
different behaviors, indicating sometimes the presence of other shocks.
In general, the variation of the total acceleration within the two
shocks is less than the gravitational acceleration, showing a tendency
for pressure gradients to balance gravity; however, gas motions are
always important. In any case, these regions are not very well resolved
by our code, since often there are only a few cells between the two
bounding shocks in any given structure.

  As seen in Figure 5, most of the objects giving rise to absorption
lines are sheets and filaments where the gas continues to fall to denser
regions in halos. The object on the lower left corner of Figure 5,
producing the first absorption line on the left edge in Figure 7a, seems
to have a more spherical nature, but it is actually an edge-on filament.
The weaker absorption lines tend to be produced by small density
enhancements in the voids (e.g., the small object towards the center
of Fig. 5, causing the absorption line at $v=750 \kms$ of the
spectrum in Fig. 7c). These correspond to potential wells where the
velocities are too low for the gas to be shocked. The smooth pressure
variation of the photoionized gas is sufficient to stop the collapse,
and the gas pressure oscillates in a wave. In this case, the gas density
can never reach very high values until there is a merger with a higher
velocity structure.

Voids are common features in the simulations.
They are typically underdense by factors of a few; density maxima along
them, which may not reach the mean density, are still visible in
absorption resembling absorption lines as they stand out from the even
lower background density.
In the spectra they tend to appear as low column density
$\nhi \sim 10^{13} \cm^{-2}$  relatively broad features.
Variations in the density of voids should also cause weak large scale
differential absorption, which could be difficult to detect in real data
as polynomial fits to apparently line free regions of the spectrum are
used to determine the continuum of the QSO.

\infsec{3.3.2}{Line Profile Analysis}

\topinsert
\vskip 0.2truecm
\tabskip 1em plus 2em minus 0.5em
\centerline{{\bf TABLE 3: ABSORPTION LINE PARAMETERS}}
\bigskip
\halign to \hsize{
\hfil#\hfil&\hfil#\ &\hfil\ \ # &\hfil\ #\ \ &\hfil#&\hfil#&\hfil#&
\hfil#&#\hfil&#\hfil&\hfil#\hfil&\hfil#\hfil&\hfil#&\hfil#&\hfil#
\cr
\omit{\hfil \# \hfil}
&\omit{\hfil\ \ \ \  \ v [km/s]}
&\omit{\hfil$b$\ \ \ \ }&\omit{\hfil$\pm$\ \ \ }&
\omit{\hfil$\log \nhi$\hfil}&\omit{\hfil$\pm$\hfil}
\cr
\noalign{\smallskip\hrule\smallskip}
& &Fig. 7a& (z = 3) &&& \cr
\noalign{\smallskip\hrule\medskip}
 1 &      34.1&    32.1   &    4.2&  13.06&   0.11\cr
 2 &     109.6&    21.8   &    0.5&  15.06&   0.04\cr
 3 &     139.3&    63.2   &   10.0&  13.41&   0.17\cr
 4 &     295.2&    20.4 &    2.0&  12.72&   0.09\cr
 5 &     343.4&    45.9 &    8.1&  12.84&   0.08\cr
 6 &     434.6&    23.7 &    8.3&  11.96&   0.14\cr
 7 &     526.1&    34.2 &    0.4&  13.89&   0.00\cr
 8 &     626.4&    27.6 &    1.2&  14.13&   0.04\cr
 9 &     650.1&    50.5 &    2.2&  13.72&   0.14\cr
10 &     693.2&    20.8 &    2.8&  13.10&   0.15\cr
11 &     788.1&    25.0 &    0.4&  14.33&   0.02\cr
12 &     856.9&    37.6 &    8.7&  13.98&   0.15\cr
13 &     899.2&    27.5 &    2.5&  13.86&   0.17\cr
14 &    1021.1&    53.3 &    0.5&  14.96&   0.01\cr
15 &    1150.0&    65.9 &   37.6&  12.48&   0.21\cr
16 &    1244.7&    28.7 &    2.8&  12.88&   0.06\cr
\noalign{\smallskip\hrule\medskip}
& &Fig. 7b& (z = 3) && \cr
\noalign{\smallskip\hrule\medskip}
 1 &    15.0&  24.8&    1.7&  12.94&   0.06\cr
 2 &  119.0&  11.7&    8.5&  11.52&   0.31\cr
 3 &  229.0&  22.6&    0.5&  14.84&   0.05\cr
 4 &  274.7&  59.3&    0.9&  14.27&   0.01\cr
 5 &  353.5& 202.1&   19.5&  13.34&   0.07\cr
 6 &  412.3&  25.6&    0.3&  15.66&   0.04\cr
 7 &  770.5&  37.3&    1.8&  12.99&   0.03\cr
 8 &  920.3&  23.8&    1.2&  12.86&   0.03\cr
 9 &  946.3& 153.7&   77.8&  12.91&   0.18\cr
10 &  1030.0&  18.2&    0.4&  13.32&   0.01\cr
11 &  1073.6&  21.3&    0.6&  13.49&   0.02\cr
12 &  1142.5&  19.7&    0.4&  13.61&   0.01\cr
13 &  1225.0& 104.2&   59.8&  12.87&   0.12\cr
14 &  1236.2&  18.6&    4.0&  12.38&   0.19\cr
\noalign{\smallskip\hrule\medskip}
}
\endinsert

\topinsert
\tabskip 1em plus 2em minus 0.5em
\centerline{{\bf TABLE 3: ABSORPTION LINE PARAMETERS - CONTINUED}}
\bigskip
\halign to \hsize{
\hfil#\hfil&\hfil#\ &\hfil\ \ # &\hfil\ #\ \ &\hfil#&\hfil#&\hfil#&
\hfil#&#\hfil&#\hfil&\hfil#\hfil&\hfil#\hfil&\hfil#&\hfil#&\hfil#
\cr
\omit{\hfil \# \hfil}
&\omit{\hfil\ \ \ \  v [km/s]}
&\omit{\hfil$b$\ \ \ \ }&\omit{\hfil$\pm$\ \ \ }&
\omit{\hfil$\log \nhi$\hfil}&\omit{\hfil$\pm$\hfil}
\cr
\noalign{\smallskip\hrule\medskip}
& &Fig. 7c& (z = 3) && \cr
\noalign{\smallskip\hrule\medskip}
1  &   91.4&   21.7&    0.2&  14.18&   0.01\cr
2  &  144.3&   14.6&    2.7&  12.28&   0.08\cr
3  &  244.0&   65.8&    3.3&  13.88&   0.05\cr
4  &  277.7&   24.8&    0.6&  15.06&   0.04\cr
5  &  379.6&   49.5&   37.6&  12.31&   0.46\cr
6  &  482.3&   30.5&    3.4&  12.59&   0.06\cr
7  &  575.9&   61.8&    3.1&  13.08&   0.02\cr
8  &  713.0&   24.3&    9.0&  12.15&   0.17\cr
9  &  763.6&   25.6&    0.3&  13.72&   0.01\cr
10 &  850.1&   31.6&   11.9&  12.02&   0.12\cr
11 &  913.5&   20.7&    2.9&  12.39&   0.06\cr
12 &  984.4&   35.6&    0.6&  13.51&   0.01\cr
13 & 1084.9&   28.2&    0.6&  14.54&   0.06\cr
14 & 1128.0&   20.7&    1.6&  13.75&   0.11\cr
15 & 1226.6&   78.7&   34.1&  12.75&   0.23\cr
16 & 1273.7&   24.2&   17.3&  12.04&   0.64\cr
\noalign{\smallskip\hrule\medskip}
& &Fig. 7d& (z = 3) && \cr
\noalign{\smallskip\hrule\medskip}
 1 &   17.0&   21.6&    1.2&  13.28&   0.04\cr
 2 &   82.7&   23.9&    0.3&  15.71&   0.04\cr
 3 &  229.8&   17.7&    2.1&  13.44&   0.22\cr
 4 &  252.0&   26.7&    2.8&  13.76&   0.11\cr
 5 &  326.2&   54.5&    9.8&  12.91&   0.07\cr
 6 &  419.1&   23.5&   10.1&  11.97&   0.27\cr
 7 &  491.6&   58.0&   12.4&  12.48&   0.08\cr
 8 &  853.1&   18.7&    4.5&  12.37&   0.21\cr
 9 &  861.4&   53.3&   15.9&  12.70&   0.09\cr
10 &  976.6&   37.7&    4.2&  13.02&   0.20\cr
11 &  998.2&   19.9&    0.5&  14.30&   0.01\cr
12 & 1174.3&   23.6&    0.3&  14.29&   0.01\cr
13 & 1216.2&  276.0&   62.3&  13.26&   0.05\cr
14 & 1216.6&   22.9&    2.8&  13.06&   0.09\cr
\noalign{\smallskip\hrule\medskip}
}
\endinsert

\topinsert
\tabskip 1em plus 2em minus 0.5em
\centerline{{\bf TABLE 3: ABSORPTION LINE PARAMETERS - CONTINUED}}
\bigskip
\halign to \hsize{
\hfil#\hfil&\hfil#\ &\hfil\ \ # &\hfil\ #\ \ &\hfil#&\hfil#&\hfil#&
\hfil#&#\hfil&#\hfil&\hfil#\hfil&\hfil#\hfil&\hfil#&\hfil#&\hfil#
\cr
\omit{\hfil \# \hfil}
&\omit{\hfil\ \ \ \  v [km/s]}
&\omit{\hfil$b$\ \ \ \ }&\omit{\hfil$\pm$\ \ \ }&
\omit{\hfil$\log \nhi$\hfil}&\omit{\hfil$\pm$\hfil}
\cr
\noalign{\smallskip\hrule\medskip}
& &Fig. 8a& (z = 2) && \cr
\noalign{\smallskip\hrule\medskip}
1  &   18.4&   33.0&    8.3&  12.47&   0.10\cr
2  &   96.5&   21.7&    0.4&  14.14&   0.02\cr
3  &  128.4&   35.0&   13.6&  12.77&   0.37\cr
4  &  232.4&   23.8&    4.7&  12.07&   0.08\cr
5  &  294.7&    9.2&    8.2&  11.28&   0.29\cr
6  &  421.4&   26.7&    0.3&  13.36&   0.01\cr
7  &  594.2&   41.5&    2.7&  12.74&   0.03\cr
8  &  696.3&  326.4&   60.5&  12.97&   0.06\cr
9  &  725.7&   23.6&    0.3&  13.45&   0.01\cr
10 &  881.6&   21.1&    0.8&  13.10&   0.02\cr
11 &  937.7&   28.9&    0.2&  14.21&   0.01\cr
12 & 1089.3&   23.2&    2.7&  12.49&   0.08\cr
\noalign{\smallskip\hrule\medskip}
& &Fig. 8b& (z = 4) && \cr
\noalign{\smallskip\hrule\medskip}
1  &    1.9&   60.0&   21.4&  13.65&   0.25\cr
2  &   44.4&   24.7&    1.6&  13.78&   0.08\cr
3  &  118.2&   20.3&    0.6&  16.00&   0.12\cr
4  &  157.2&   79.9&    4.0&  14.02&   0.04\cr
5  &  252.7&   17.0&    2.3&  12.47&   0.08\cr
6  &  355.4&   19.0&    0.3&  13.61&   0.01\cr
7  &  398.2&   24.8&    0.7&  13.51&   0.01\cr
8  &  473.2&  132.5&    8.6&  13.57&   0.03\cr
9  &  580.8&   28.7&    1.6&  14.00&   0.08\cr
10 &  609.7&   20.6&    1.2&  14.00&   0.09\cr
11 &  722.8&   19.2&    0.7&  15.42&   0.12\cr
12 &  756.3&  122.8&    2.8&  14.43&   0.01\cr
13 &  823.9&   20.0&    0.6&  14.68&   0.05\cr
14 &  978.9&   35.4&    0.6&  15.00&   0.03\cr
15 & 1086.4&   33.7&    2.6&  14.56&   0.03\cr
16 & 1187.5&   25.1&    0.4&  16.16&   0.07\cr
17 & 1314.6&   75.1&   57.6&  13.09&   0.50\cr
\noalign{\smallskip\hrule\medskip}
}
\endinsert

We now present the results of applying the same procedure used to
analyse observations of the absorption lines to the examples of
absorption spectra that we have shown. This analysis will be carried out
further over a large sample of the simulated spectra in Rauch \etal
(1995).
In Table 3 we give the result of Voigt profile fits to the four
spectra in Figure 7. The spectra were created by emulating
typical datasets obtained with the Keck telescope
high resolution spectrograph, by convolving them with
a velocity resolution of 8 km/s (FWHM)
and adding noise with a S/N per pixel of 50.
The fitting procedure was identical to the
one used previously on real data (Carswell et al.\ 1991), except that
here each simulated spectrum is short enough to be treated as a single
fitting region.
This has the advantage that we do not need to decompose in advance
the spectrum into several regions chosen on the basis of statistically
significant continuum depressions, a procedure which is generally not
done according to an automatic, well defined algorithm when lines
are severely blended.

  The absorption line width arises from both hydrodynamic and thermal
motions; the relative contributions vary substantially among the
absorption lines, as seen in Fig. 7.
The contributions are difficult to disentangle as the temperature
across a collapsing structure is not constant and the infall of gas
often is not symmetric about the density maximum.
The temperature in the densest
regions is generally higher than the thermal equilibrium temperature
for photoionized gas, due to shock heating. The combination of such
high temperatures and the hydrodynamic motions can explain the observed
large b-parameters, which would imply unrealistically low densities if
thermal equilibrium was assumed (e.g., Press \& Rybicki 1993).

  There are several lines with small bulk motion
(as apparent from the comparison
of the thermally broadened and unbroadened profiles),
and in these cases the Doppler parameter values
correspond to temperatures that are close to the peak temperature in
the system, usually occurring in the region of highest density.
See, for example, the first strong line in Fig. 7a (second line in
Table 3).
High temperatures like that of the strong line at $v\simeq 1000 \kms$
in Fig.7a are usually detected
as such, although with a larger error due to additional broadening by
infall of gas.
Often, however, the line profile of intermediate or lower column
density systems is dominated by bulk motion (e.g., the lines in the
center of the spectrum in Fig. 7c).
The weakest lines have widths dominated by bulk motion, and they
are produced by intergalactic gas in expansion.
Despite the low gas temperature, these lines are usually not narrow
due to the expansion velocities.

\infsec{3.3.3}{Departures from Voigt Profiles}

As a consequence of the dynamical character of the clouds, we find
numerous absorption lines which show departures from single Voigt
profiles. Typically, such a system may contain some cool, dense gas
producing a high column density, narrow component, and at the same time
hotter gas recently shock-heated at high velocity, which will produce
broad tails. Cooler and less dense gas falling into the system also
contributes to such tails, often making them asymmetric. For example,
the line at $v=250 \kms$ in Fig. 7d shows an extension to the right
which is due to accreting gas, as seen in Figure 5.  One can always
obtain a good fit to such a line by using superposed Voigt profiles,
although the added lines will not necessarily correspond to physically
distinct ``clouds''. These multi-component systems will give rise to
a correlation of line properties.
An individual system is often well represented by a pair of lines with a
small velocity splitting, consisting of a narrow central component
(cool gas) and a broader wing-like component caused by the hotter
and/or infalling gas. From an analysis of a large sample of simulated
spectra, we find that there is indeed a strong anticorrelation of
Doppler parameters on small (order 10-20 km/s) scales,
and there are indications that we are seeing it in real data as well;
we shall present this in detail in Rauch et al.\ (1995).

\infsec{3.3.4}{Temporal Evolution}

\topinsert
\vskip -0.5truecm
\centerline{
\epsfxsize=5.5in \epsfbox{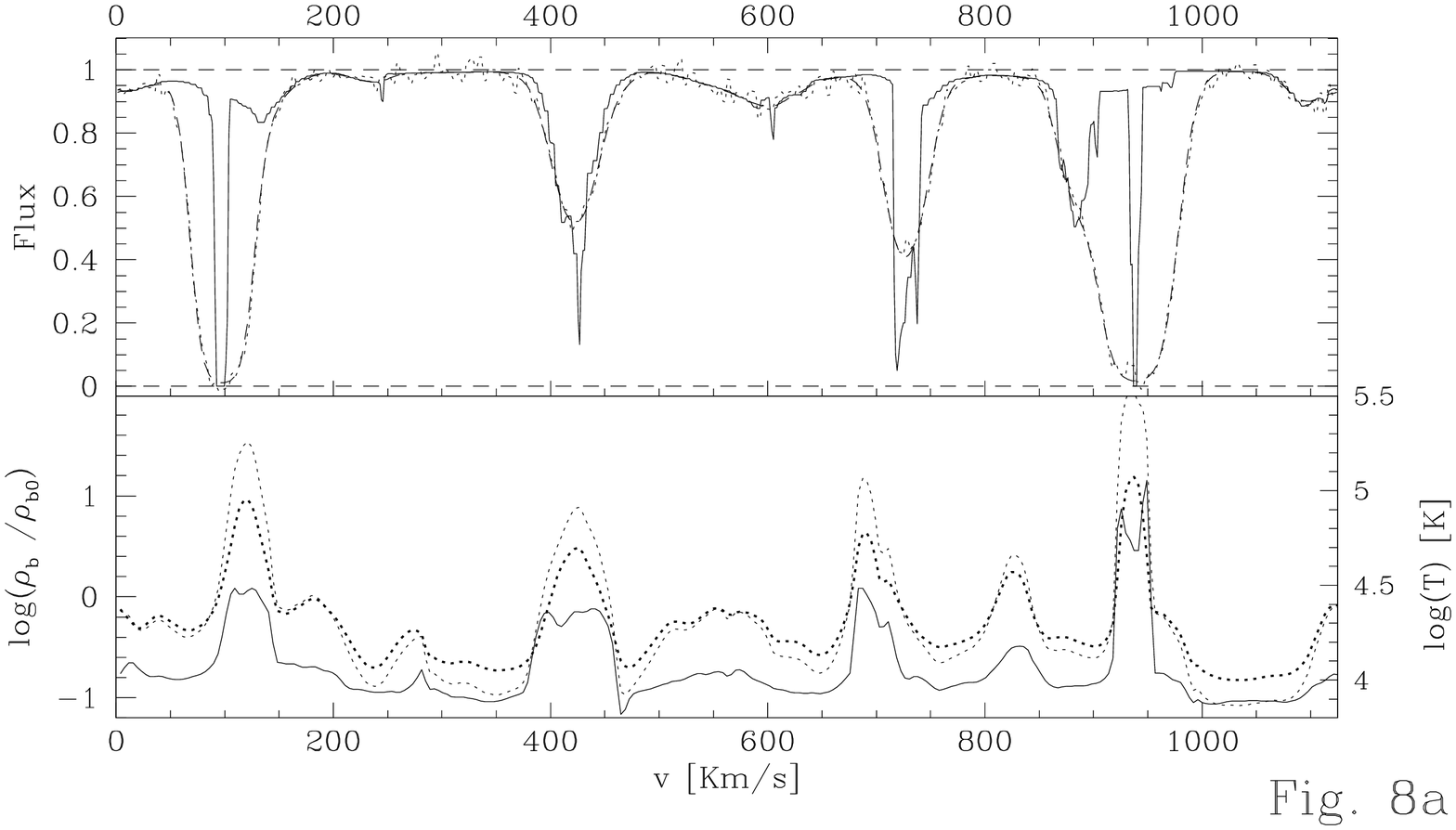}
}
\vskip -3.5truecm
\centerline{
\epsfxsize=5.5in \epsfbox{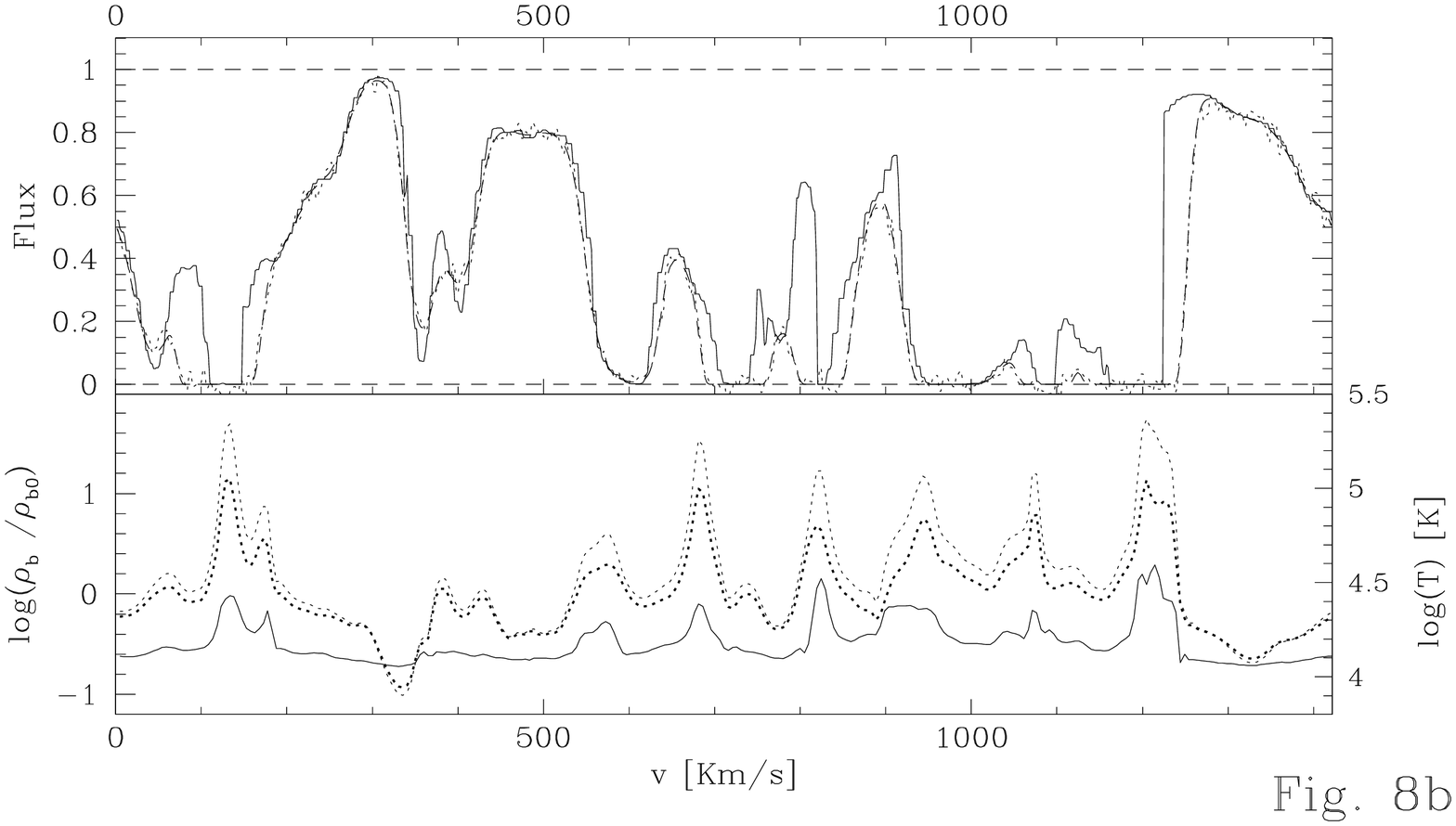}
}
\vskip -2.5truecm

\baselineskip=12truept \leftskip=3truepc \rightskip=3truepc
\noindent{\bf Fig. 8:} Same as top and middle panels of the same row as
Fig. 7a, at $z=2$ (Fig. 8a) and $z=4$ (Fig. 8b).
\endinsert

Finally, we show in Figures 8(a,b) the density, temperature and
pressure of the same row as in Fig. 7a at $z=2$ and $z=4$, as well as
the calculated spectrum. The most striking change with redshift is
the increase of the average absorption, as in the observations. This
is mostly due to the general increase of the optical depth corresponding
to a fixed overdensity as $(1+z)^{4.5}$ for a constant
proper intensity $\jhi$ (in the same way as the Gunn-Peterson effect).
As we shall see below, most of the gas in the absorbers tends to be
expanding at rates close to the Hubble rate, and the overdensities
are increasing only slowly. In addition, the gas is also becoming
hotter due to the increasing shock velocities as larger scales collapse,
which further reduces the neutral fraction.

In addition to the average increase of the optical depth, the systems
yielding absorption lines are evolving as they collapse gravitationally,
with the gas becoming more overdense as it flows to denser regions, and
the structures merging with each other. Transverse motions accross the
row are also responsible for changes (especially for the high column
density systems, which vary over small scales).
Thus, the evolution of an individual
absorption system is difficult to interpret, but the spectra give an
overall impression of the expected evolution with redshift of the
general population of absorbers which seems to agree with what is
observed.
Most of the absorption systems can be traced from $z=4$ to $z=2$, as
their column density decreases.

  The fitted lines in Table 3 show a general decrease of the Doppler
parameter with increasing redshift,
 in agreement with recent
observations of redshift 4 QSOs by Williger et al.\ (1994) and
Lu (1995, in preparation).
The gas temperatures are in fact
lower at high redshift, owing to the higher velocities and
temperatures of the gas on the larger scales of collapse at lower
redshifts (see \Sec 3.4). However, lower Doppler parameters might also
arise spuriously due to the increased blending at high redshift.

\subsec{3.4}{Physical Conditions of the Absorbers}

  The gas yielding $\lya$ absorption is spread throughout all the
volume in our simulation, and therefore there is no clear way to
identify the gas belonging to a given ``$\lya$ cloud''. However, the
simulated spectra show absorption lines similar to the observed ones,
so it seems useful to adopt a criterion to identify the objects
producing these lines and analyze their physical state. Here, we
shall identify clouds by choosing a contour in real space
at a fixed neutral hydrogen
density, and identifying each interval along a line of sight within
the contour as a cloud. Our ``clouds'' are therefore not
three-dimensional entities, but regions which are defined only for one
particular line of sight; a connected region in space above the neutral
density of the contour could be disconnected along a line of sight
crossing this region, and would therefore be split into several
clouds. We measure the total and neutral column density along such
clouds, and the neutral-weighted temperature,
neutral fraction, and velocity divergence.
We choose the contour having a neutral density corresponding
to the value in photoionization equilibrium when the gas density is
equal to the average baryon density and at a temperature $T=10^4
\kelvin$. The distribution of neutral column densities obtained
in this way will be shown later in \Sec 5; we shall also identify lines
in the simulated spectra using an analogous procedure in \Sec 4.

\topinsert
\centerline{
\hskip -0.5truecm
\epsfxsize=3.0in \epsfbox{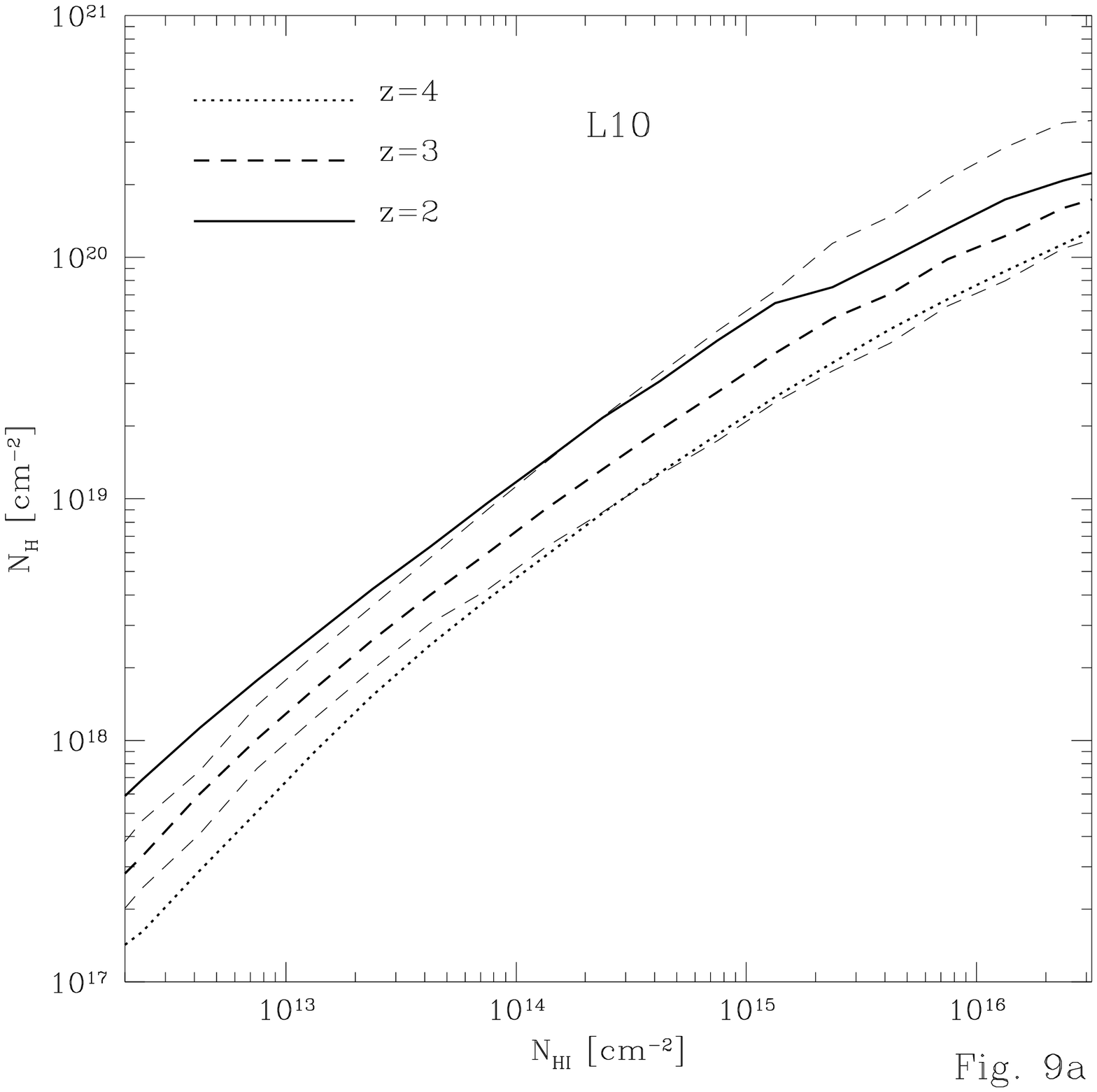} \hskip 0.3truecm
\epsfxsize=3.0in \epsfbox{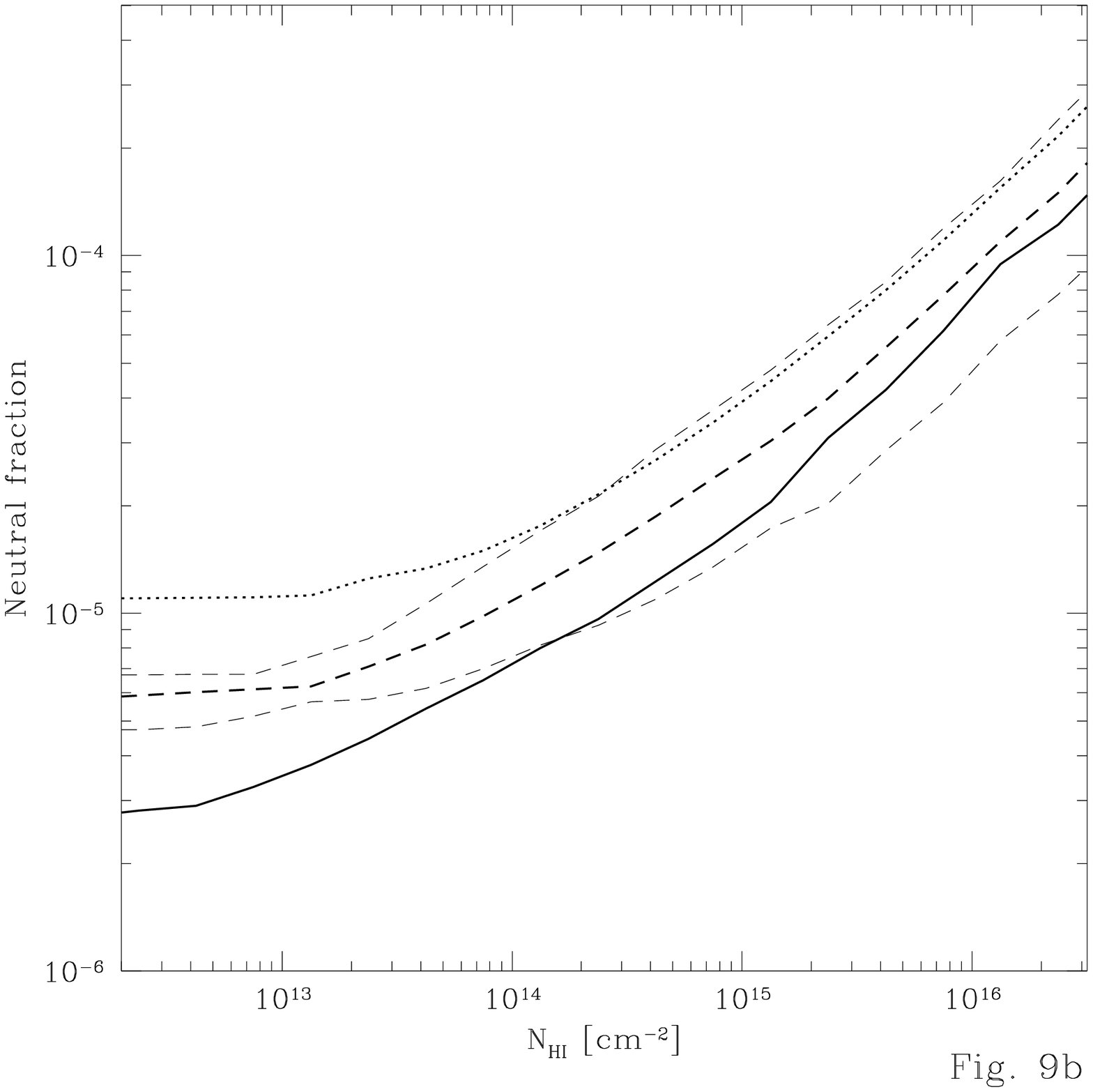}
}
\vskip 0.5truecm
\centerline{
\hskip -0.5truecm
\epsfxsize=3.0in \epsfbox{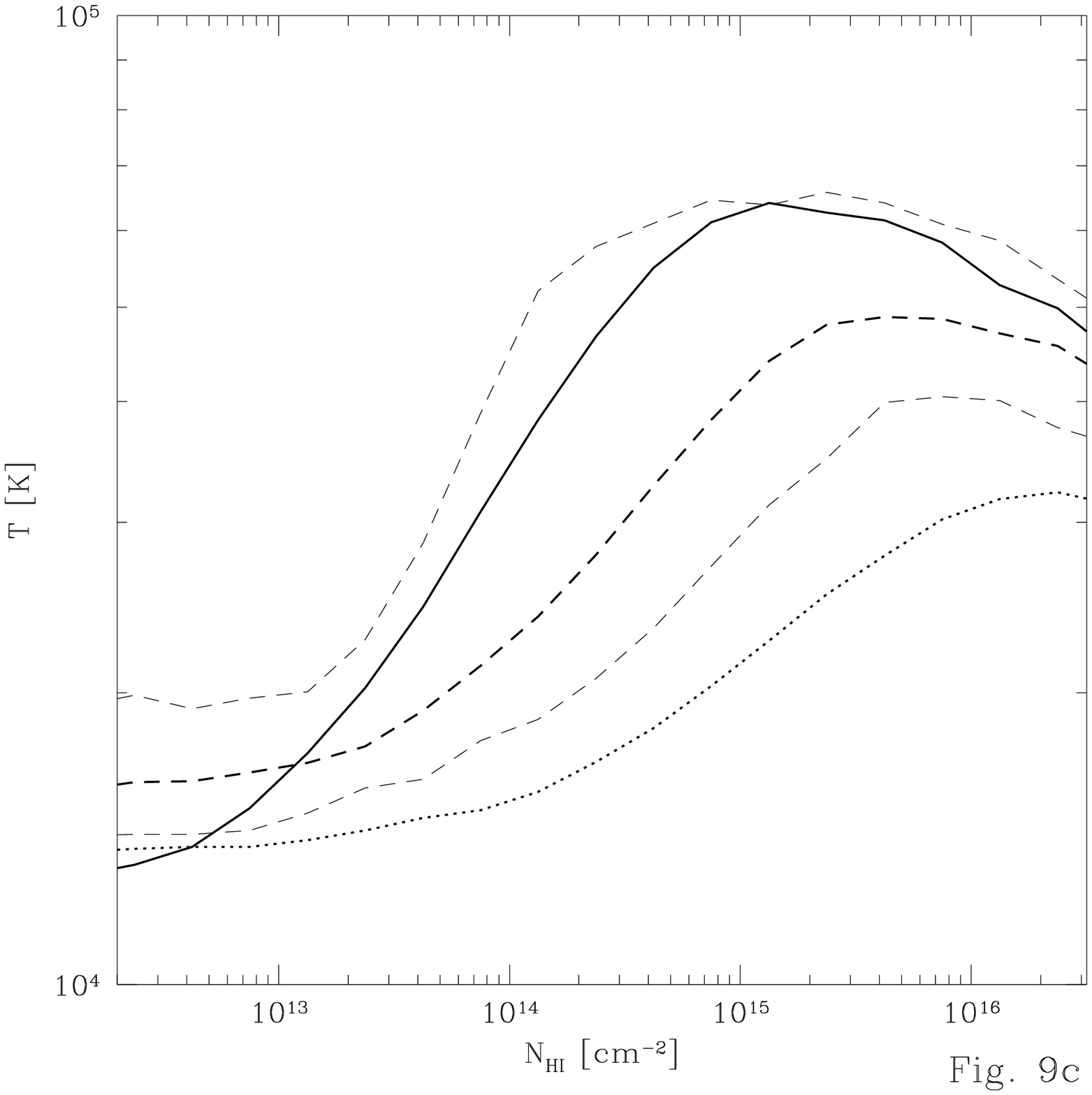} \hskip 0.3truecm
\epsfxsize=3.0in \epsfbox{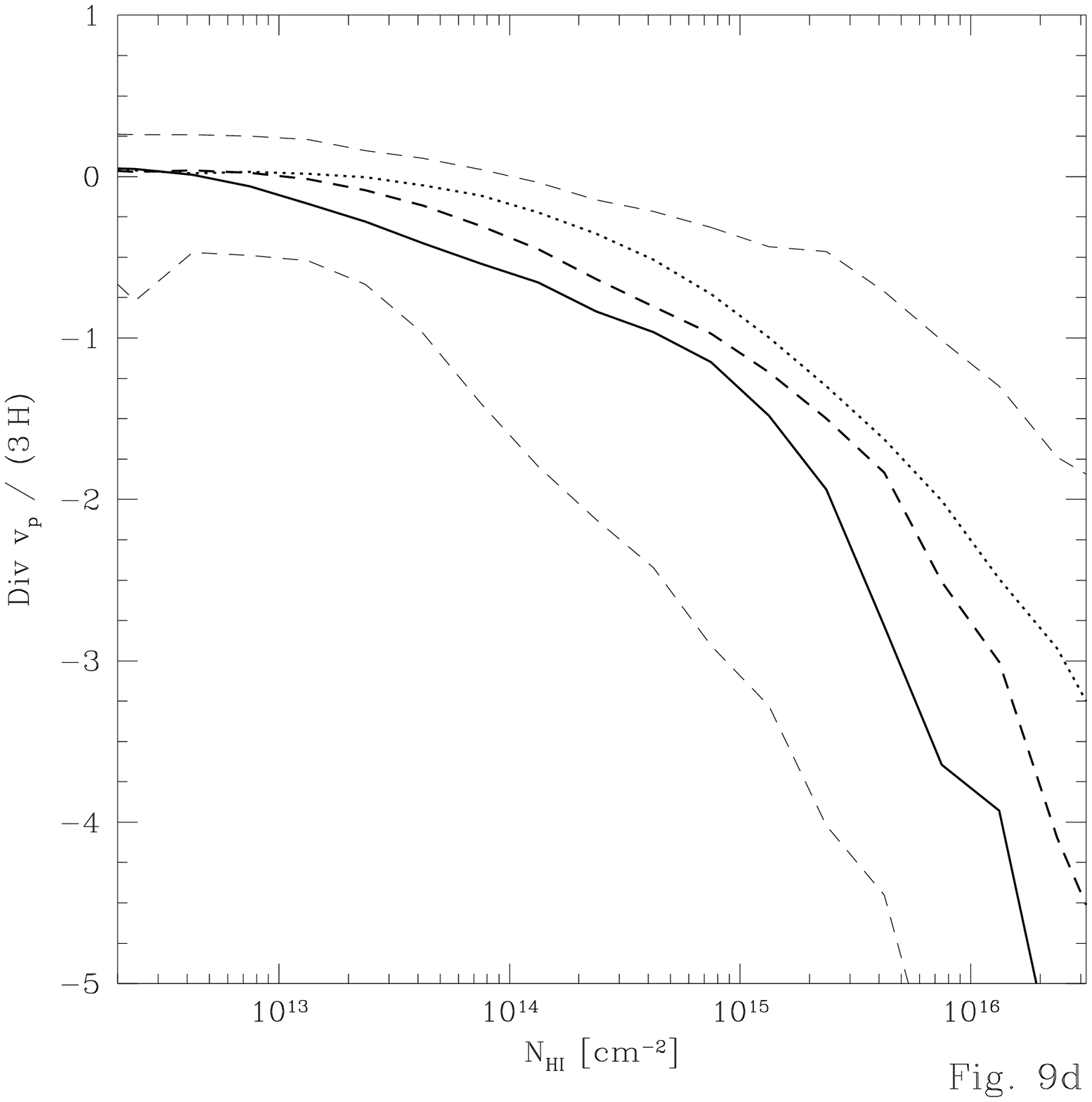}
}
\baselineskip=12truept \leftskip=3truepc \rightskip=3truepc
\noindent{\bf Fig. 9:} (a) Thick lines give the median total column
density at a fixed neutral column density, in the L10 simulation at
three redshifts. The two
thin dashed lines give the total column densities such that 5\% of
the clouds have a lower and higher column density. Both column densities
are measured within a fixed contour of neutral density in real space.
Other figures show the distribution of the neutral-weighted
(b) neutral fraction, (c) temperature, and (d) velocity divergence,
in the same way as Fig. 9a.

\endinsert

  Figure 9a shows the median value of the total column density, for
clouds of a fixed neutral column density. The results are shown for the
L10 simulation at three redshifts as thick lines, as indicated in the
figure. At $z=3$,
we also show two thin dashed lines such that 5\% of the systems have
total column densities below the lower line, and 5 \% above the upper
line. These two thin lines show that the dispersion of the total column
density at a fixed neutral column density is very small
(the dispersion is similar at other redshifts). The variation in column
density corresponds mostly to a variation of the gas density. The
dispersion is small because variations in pathlength for clouds of fixed
column density are not too large, and they cause both column densities to
change by the same factor, moving the location of a cloud in Fig. 9a in
a direction only slightly steeper than the dependence of the median
$N_H$ on $\nhi$. Variations of temperature at a fixed gas density
are also small.

  The distribution of the neutral-weighted neutral fraction for lines of
a given neutral column density is shown in Figure 9b (line types
have the same meaning as in Fig. 9a).
In Figure 9c, the distribution of the neutral-weighted
temperature is shown in the same way, and the distribution of the
neutral-weighted velocity divergence is shown in Figure 9d.
Regions expanding at the Hubble
rate have ${\rm div}\, v_p = 0$, while regions with constant proper
density have $({\rm div}\, v_p)/(3H) = -1$. These results have been
obtained from 18000 random lines of sight (selected as described below,
in \Sec 4.4) through the L10 box at
each redshift, and measuring all the quantities for every ``cloud''
identified as explained above.

  Higher column densities correspond to higher neutral fractions
of hydrogen. Thus both increasing the total column density and increasing
the neutral fraction combine together to produce the larger observed
column densities of neutral hydrogen, as the gas density through a line
of sight rises. If the median pathlength and temperature were constant
for clouds of different column density, the neutral fraction should rise
as $\nhi^{1/2}$, which is approximately the dependence seen in Fig. 9b.
In reality, the pathlength tends to slowly decrease with column density
(because high column densities correspond to lines of sight reaching
closer to the halo centers), but the effect is roughly cancelled by the
increasing temperatures (Fig. 9c). The temperature increase is due to
higher shock velocities in the denser regions.
The flattening of both median temperature
and neutral fraction at low column densities is an artifact of our
algorithm for cloud identification, since there is a minimum neutral
density, and therefore a minimum gas density and neutral fraction,
for the gas inside clouds. This also gives a minimum temperature, since
temperature correlates well with gas density.

  The velocity divergence
shows that most of the low column density systems are expanding near
the Hubble rate. The gas in
sheet-like and filamentary structures is typically expanding along them
as it falls
towards higher density halos; the infall rate of gas from surrounding
voids declines, and the dark matter density also decreases, so both the
ram-pressure and gravitational confining forces decrease with time.
The gas still contracts in comoving coordinates for ($\nhi \gta 10^{13}
\cm^{-2}$), and at higher column densities
($\nhi \gta 10^{15} \cm^{-2}$) the gas is contracting in real space.
These systems are associated with denser filaments and halos, where
shocks occur frequently and the collapse can continue as
gas cools radiatively. We have found that the curves in Figs. 9(b,c,d)
change only minimally when the total-hydrogen-weighted quantities are
calculated.

\sect{4. ANALYSIS OF THE SIMULATED SPECTRA}

  In this Section we present quantitative predictions from our
simulations for various statistical quantities that can be measured from
the $\lya$ absorption spectra produced by the neutral gas. We first
analyze the distribution and correlation function of the transmitted
flux, and then present properties of absorption lines defined according
to a new algorithm which is precisely defined and easy to implement.
The results
will be shown for the L10 simulation (on a $10 h^{-1} \mpc$ box with
$288^3$ cells) at the redshifts $z=2$, $z=3$ and $z=4$, which will
indicate the expected redshift evolution (at lower redshifts, the
effects of fluctuations on scales larger than the box size becomes
more important). The results at $z=3$ will be compared with the L3
and l3 simulations (on a $3 h^{-1} \mpc$ box with $288^3$ and $144^3$
cells, respectively) to show the effects of the large-scale fluctuations
and of the limited resolution. For each simulation at each redshift,
the statistical quantities presented are computed from 18000 spectra
corresponding to rows along one of the three axes in the simulation
(these are selected from 1500 random groups of 12 parallel spectra at
fixed distances, as described in \Sec 4.4).

\subsec{4.1}{Average Flux Decrement}

  The first statistical quantity that can be obtained from observed
quasar spectra is the distribution of the transmitted flux. We start
by considering the average transmission. This depends on our model
for the ionizing background through the resulting intensity $\jhi$, and
it therefore does not provide a strong test of the
models. Instead, we must use the observed average transmission to fix
$\jhi$, and then see if other predictions for the $\lya$ spectra
agree with observations. Since our simulations were evolved for only
one value of $\jhi$ (which, as mentioned in \Sec 3, is much too high
to reproduce the observed flux decrement), and it is impractical to
run many simulations with different $\jhi$, we must assume that varying
$\jhi$ would only alter the neutral fraction in the gas, but the effects
on the gas temperature and the consequent different dynamical evolution
of the gas can be neglected. While it is reasonable that the small
variation of the equilibrium temperature with $\jhi$ will not
drastically change the results, this will have to be tested with more
simulations in the future.

  More generally, if we take the density, temperature and velocity
fields of the gas in the simulations as constant and let the
normalization of the
gas density and the intensity of the ionizing background vary, then the
density of neutrals changes proportionally to $(\Omega_bh^2)^2/\jhi$
neglecting collisional ionization. The optical depth to $\lya$
scattering depends only on the column density of the gas, and therefore
varies as $\mu^2 \equiv (\Omega_bh^2)^2/(h \jhi)$.
Thus, we can approximate the variation in the predictions for the $\lya$ forest
given by our simulations when we vary $\Omega_b$ and $\jhi$ by
multiplying all the optical depths by the variation in the factor $\mu^2$,
if we neglect collisional ionization, the self-gravity of the baryons,
and the effects in the evolution of the gas due to the different
cooling rate when the neutral fraction changes (notice that changing the
Hubble constant would also imply a change in the power spectrum of our
CDM$+\Lambda$ model).

  As explained in \Sec 3, all our results on the simulated $\lya$
spectra will be presented assuming $\jhiu = 0.1$, and $\Omega_b=0.0355$
and $h=0.65$ as used
in the simulation. The neutral fractions have been recalculated at
every cell in the simulation for this value of $\jhi$, including both
photoionization and collisional ionization. However, it is of interest
to see how the predicted average flux decrement varies with the factor
$\mu^2$, because the observational determination of the flux decrement is
subject to some uncertainties due to the intrinsic noise in the
observations, the extrapolation of the quasar continuum spectrum, and
the contribution of metal lines and damped absorption lines (which our
simulation does not include). For this purpose, we can assume that the
optical depth varies proportionally to $\mu^2$, neglecting collisional
ionization, because the regions where collisional ionization is
important are always at high densities and they have a very large
optical depth to $\lya$ absorption in any case.

\topinsert
\centerline{
\epsfxsize=4.0in \epsfbox{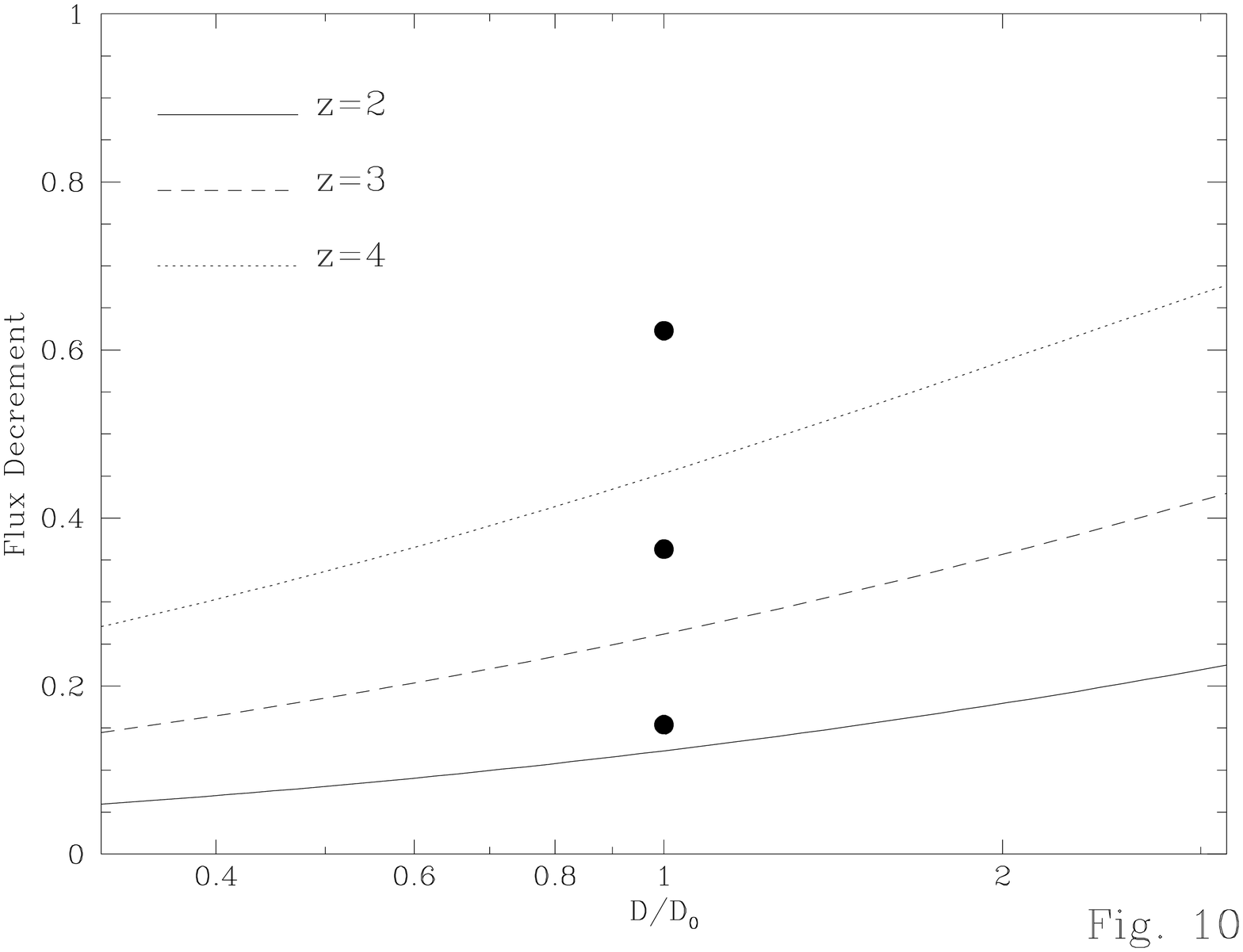}
}
\baselineskip=12truept \leftskip=3truepc \rightskip=3truepc
\noindent{\bf Fig. 10:} Average flux decrement in the $\lya$ spectra from the
L10 simulation at $z=2$, 3 and 4, as a function of the factor $D/D_0$ by
which the neutral densities are
uniformly multiplied. We use $D/D_0=1$ in all subsequent figures,
corresponding to the neutral fractions calculated assuming ionization
equilibrium with a background intensity $\jhiu = 0.1$, for
$\Omega_b = 0.0355$ and $h=0.65$, as in the simulation. The change with
the factor $D/D_0$ is approximately the same as the change with
$\Omega_b^2/\jhi$, except for the effects of collisional ionization
(which is negligible for the regions making most of the contribution to
the average flux decrement), self-gravity of the baryons, and the change
in the equilibrium temperature.

\endinsert

  The predicted flux decrement when the optical depths are multiplied
by a factor $\mu^2/\mu_0^2$ is shown in Figure 10
($\mu_0^2$ is the value of $\mu^2$
when $\jhi$ and $\Omega_b$ have our standard values). The filled circles
show the observed flux decrement determined in Press, Rybicki, \&
Schneider (1993). These observed values are still slightly higher than
our results, but they include the contribution from metal lines and the
damped wings of high column density systems. It is therefore not
completely clear if the difference from the observations is real.
A constant value of $\jhi$ comes close to reproducing the observed
evolution with redshift of the flux decrement; this is also similar
to the dependence of $\jhi(z)$ obtained from our emission models (a
precise fit to the values of Press \etal requires $\jhi$ to increase
by a factor $\sim 1.5$ from $z=4$ to $z=2$).
Figure 10 can be used to see the amount by which $\jhi$ or $\Omega_b$
need to be changed to fit any observed value for the average decrement.
The value $\mu^2=\mu_0^2$ is assumed throughout the rest of this paper.

\topinsert
\vskip 0.2truecm
\centerline{\bf TABLE 4: FLUX DECREMENTS}
\bigskip
\hrule\vskip0.1truecm\hrule
$$\vbox{\tabskip 1em plus 2em minus 5em
\halign to\hsize{\hfil # \hfil & \hfil # \hfil & \hfil # \hfil & \hfil #
\hfil & \hfil # \hfil \cr
& L10, z=2 & L10, z=3 & L10, z=4 & PRS \cr
\noalign{\medskip\hrule\medskip}
$D_\beta/D_\alpha$ & 0.30 & 0.37 & 0.43 & $0.48\pm 0.05$ \cr
$D_\gamma/D_\alpha$ & 0.13 & 0.19 & 0.24 & $0.35 \pm 0.07$ \cr
$D_\delta/D_\alpha$ & 0.073 & 0.11 & 0.15 & $0.16 \pm 0.08$ \cr
$D_\epsilon/D_\alpha$ & 0.045 & 0.073 & 0.10 & $0.12 \pm 0.07$ \cr
\noalign{\medskip\hrule\medskip}
}}$$
\endinsert

  We also give in Table 4 the ratios of the average flux decrements
of all the Lyman series transition lines, up to ${\rm Ly}\epsilon$,
to the $\lya$ average flux decrement, for the L10 simulation
at three different redshifts \footnote{$^1$}{We thank David
Weinberg for suggesting to calculate these ratios}.
These are calculated by simply
multiplying the optical depths by the ratio of the oscillation strengths
times the ratio of wavelengths of the Ly$(\beta,\gamma,\delta,\epsilon)$
lines to the $\lya$ line (equal to $0.16$, $0.056$, $0.027$, and $0.015$,
respectively), and computing the average flux decrement. Determinations
of these flux decrements were given by Press \etal (1993) (the average
redshifts of their sample of $\lya$ absorption spectra is $\sim 3.4$).
These are sensitive to the column density distribution:
the more abundant the lines of high column
density are relative to low column density, the higher a fraction of the
$\lya$ decrement is contributed by heavily saturated lines, so the
larger the decrements of the high order lines should be. As we can
see, our predicted ratios of the decrements of high order Lyman lines
to the $\lya$ decrement are smaller than observed, suggesting that the
column density distribution we obtain is too steep. We shall return to
that in \Sec 4.6.

\subsec{4.2}{Flux Distribution}

\topinsert
\centerline{
\hskip -0.5truecm
\epsfxsize=3.0in \epsfbox{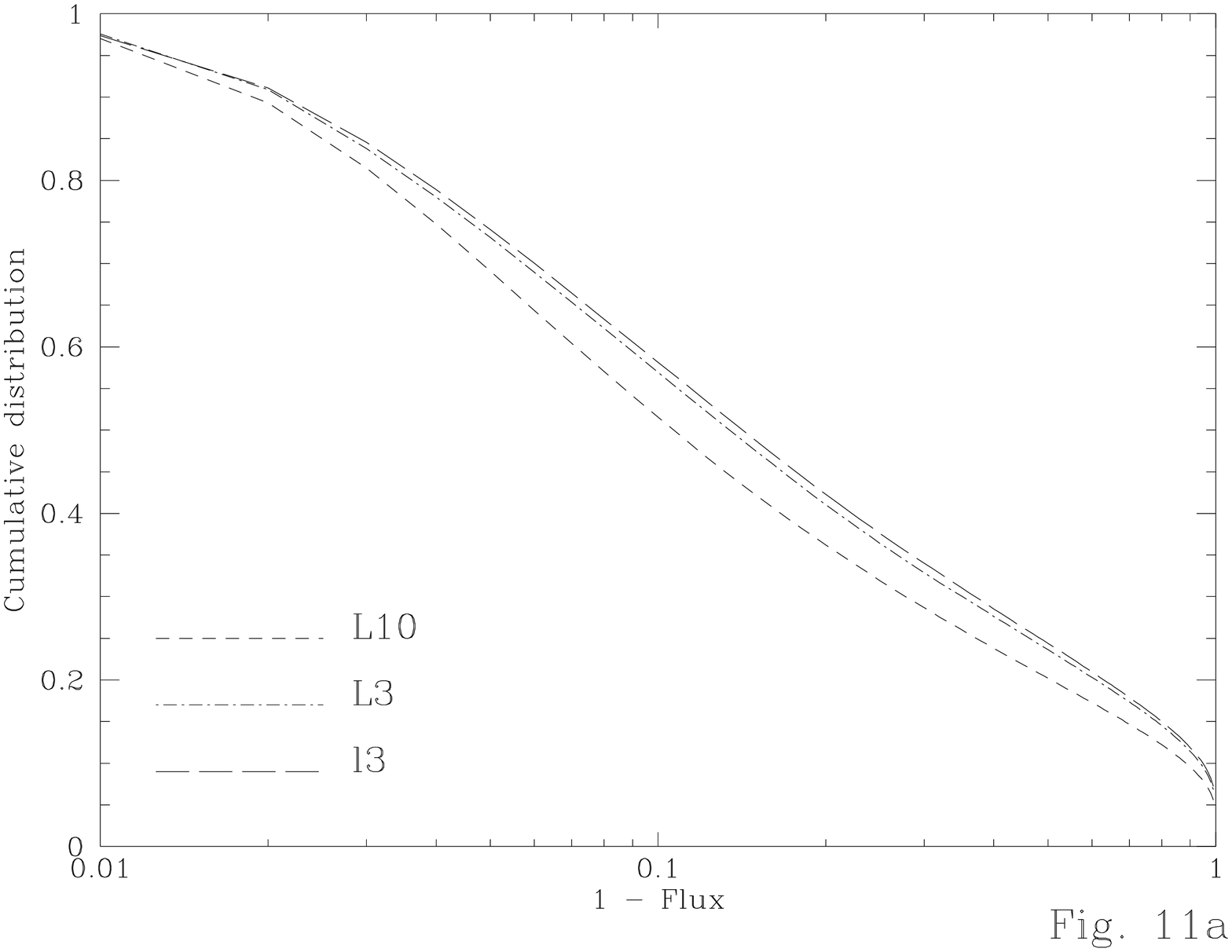} \hskip 0.3truecm
\epsfxsize=3.0in \epsfbox{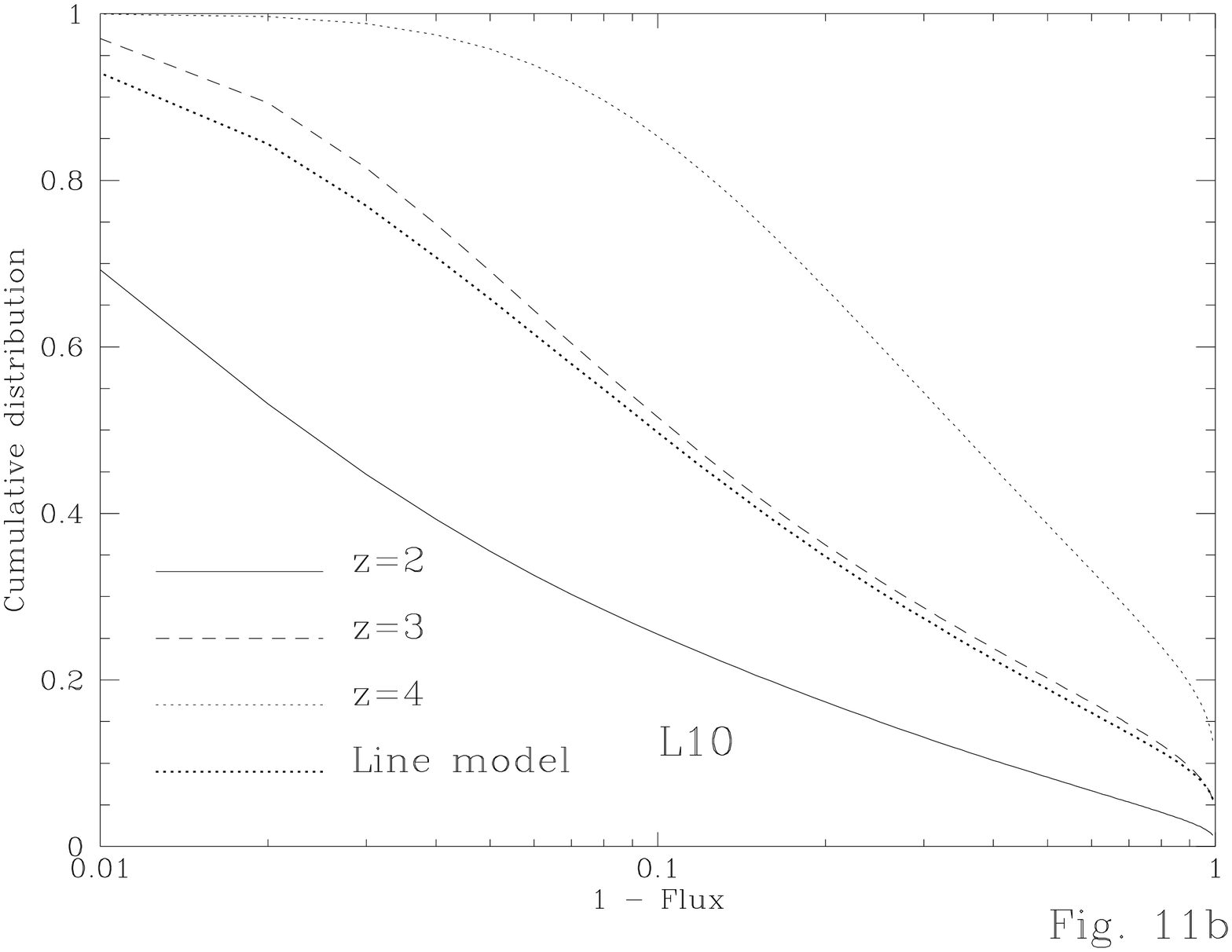}
}
\baselineskip=12truept \leftskip=3truepc \rightskip=3truepc
\noindent{\bf Fig. 11:} Cumulative distribution of the absorbed flux (equal to
one minus the transmitted flux), for the simulations and redshifts as
indicated in the figure (line types for each simulation and redshift are
the same in all figures). The thick dotted line in figure 10b is
calculated from randomly generated spectra containing superposed lines
with Voigt profiles with a distribution of column densities and
b-parameters as described in \Sec 4.2, with the line density normalized
to the observed one at $z=3$, and it has a similar shape as the flux
distribution from the simulations.

\endinsert

  The cumulative distribution of the transmitted flux in the spectra is
shown in Figures 11(a,b).
As seen in Figure 11a, the simulations with the smaller box yield more
absorption. This is a consequence of the missing large-scale power in
the small box simulation. One cause for the difference is the higher
temperature, and the correspondingly lower recombination coefficient in
the L10 simulation; the other is the fact that the distance separating
successive lines is limited by the size of the box.
The typical transmission in the $\lya$ spectra is mostly sensitive to
the gas distribution in the voids, rather than the denser regions, since
the voids occupy most of the volume in the universe. In the small box
simulation, the size of the voids is limited and the separation between
lines is smaller.
The effect of the finite resolution is also to
increase the average absorption, because the inclusion of collapse on
small scales results in a lower gas density over most of the volume
(more of the gas is concentrated in high-density, small-scale objects).
The difference between the L3 and l3 simulations is, however, very
small, because the scales that are unresolved are already below the
Jeans mass of the photoionized gas.
These results imply that {\it the effect of the finite resolution and
size of the simulated box can only be to increase the amount of
absorption that we obtain, so the true predicted absorption
of the model should
be lower than the results of the L10 simulation}. Judging from the
differences between the three simulations, the correct prediction is
probably not lower by more than $\sim 10\%$ relative to L10. However, a
possible effect that could increase the absorption found in our
simulations is if the gas temperatures in the absorption lines were
significantly lower due to additional cooling for the low value of
$\jhi$ we need to assume, which could also cause additional compression
of the gas. This could be particularly significant if thin, cool slabs
of gas between shocks were common and could not be resolved in our
simulation.

  The observations of the $\lya$ forest have customarily been reduced by
decomposing the observed spectrum into a discrete number of absorption
lines, assuming they are described by Voigt profiles (or gaussians for
the optical depth, since the damped wings are not important for the low
column density lines we are interested in). One then obtains
distribution functions of the column density and b-parameters. In order
to compare with observational results, we shall use a model for the
distribution of $\nhi$ and the b-parameters of the lines and then
compute the distribution of the flux from artificially generated
spectra, assuming that the lines are uncorrelated. Our model (which
shall be called ``line model'' in our figures) has a column density
distribution with $ f(\nhi)\, d\nhi \propto \nhi^{-\beta}$, where $\beta
= 1.5$ for $\nhi < 10^{14.8} \cm^{-2}$, and $\beta = 1.9$ for $\nhi >
10^{14.8}\cm^{-2}$. The total number of lines per unit redshift above
$\nhi = 10^{14}\cm^{-2}$ is fixed to 100 at $z=3$ (we shall only show
results for this model at $z=3$). The b-parameters are assumed to be
uncorrelated to the column densities, and their distribution is chosen
as a gaussian with the peak at $b=27\kms$, a dispersion of $9\kms$, and
a cutoff at $b=17\kms$, so that no lines have b-parameters lower than
the cutoff. This distribution is close to what is favored by the
observations (e.g., Rauch \etal 1992; Petitjean \etal 1993, Hu \etal
1995), and as we shall see it provides a relatively good fit to the
spectra obtained from our simulation.

  The thick dotted line in Figure 11b shows the cumulative distribution
implied by our line model, calculated from 1800 randomly generated
spectra with the same velocity length as the L10 simulation at $z=3$.
Its shape is very similar to what is obtained
from the L10 simulation, and the average decrement implied is $0.249$,
only slightly below the result of our simulation for $\mu^2/\mu_0^2=1$
($0.263$). The shape of this distribution is a prediction of the
CDM$\Lambda$ model which should be directly testable with observations
(see Jenkins \& Ostriker 1991; Webb \etal 1992).

\subsec{4.3}{The Fluctuating Gunn-Peterson Effect}

  We see in Figure 11b that there is a fast evolution in the
distribution of transmitted flux: the absorption increases rapidly with
redshift. This is mostly due to the rapid increase of the optical depth
that would be observed if the gas was uniformly distributed,
$\tau_u = 4.65\times 10^5\, \Omega_{HI}h\, [H_0/H(z)]$ (Gunn \& Peterson
1965; here, $\Omega_{HI}$ is the fraction of the critical density in
neutral hydrogen in a uniform medium, $H_0$ is the
present Hubble constant and $H(z)$ is the Hubble constant at redshift
$z$). For the parameters of our simulation and $\jhiu = 0.1$, this
optical depth would be $\tau_u = 0.212$, $0.785$ and $2.15$
at $z=2$, 3 and 4, respectively. Because the gas moves towards collapsed
structures, leaving most of the volume with a density below the average,
we expect that the optical depth over most of the spectrum is
considerably below that
derived for a uniform medium with the average density.

\topinsert
\centerline{
\hskip -0.5truecm
\epsfxsize=3.0in \epsfbox{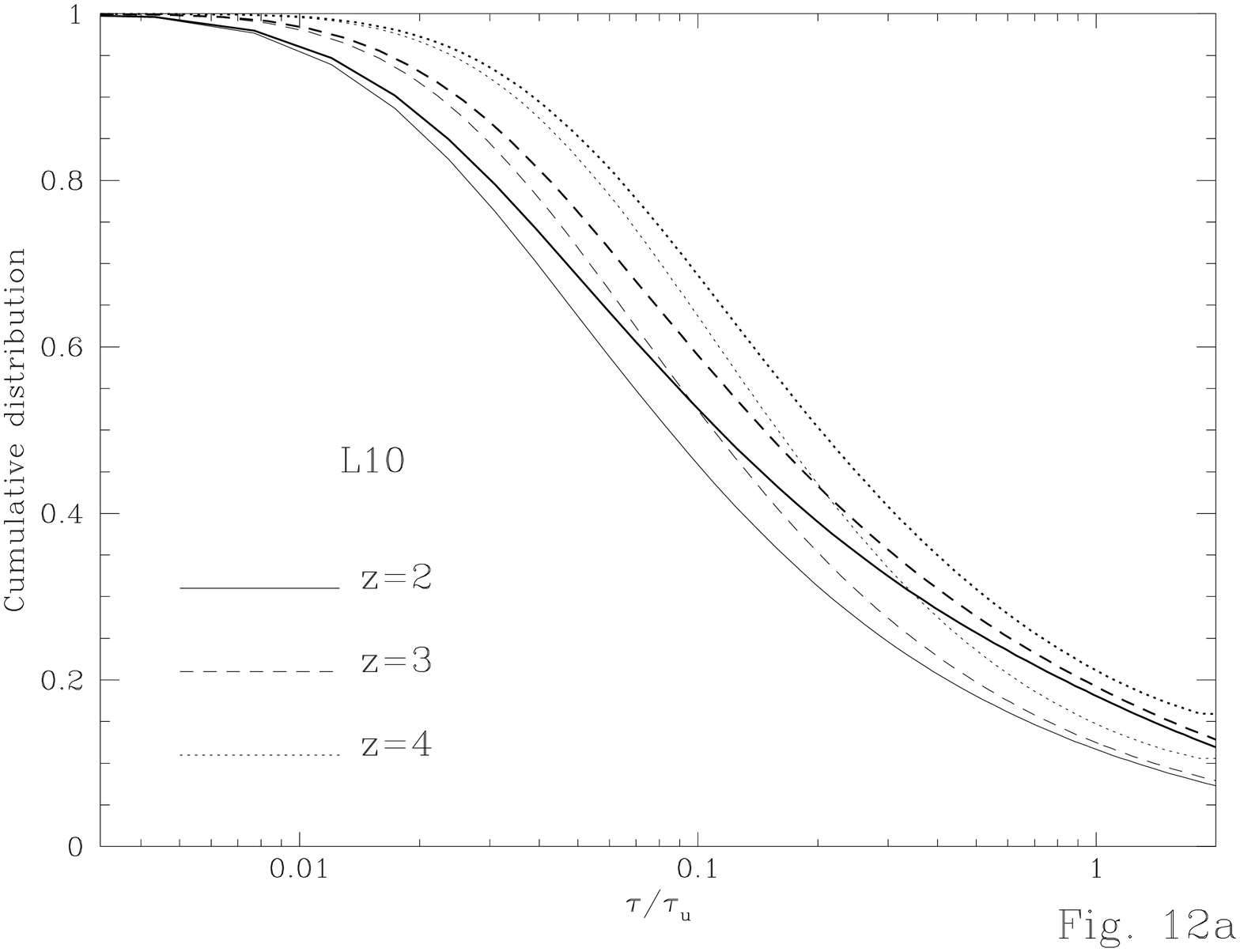} \hskip 0.3truecm
\epsfxsize=3.0in \epsfbox{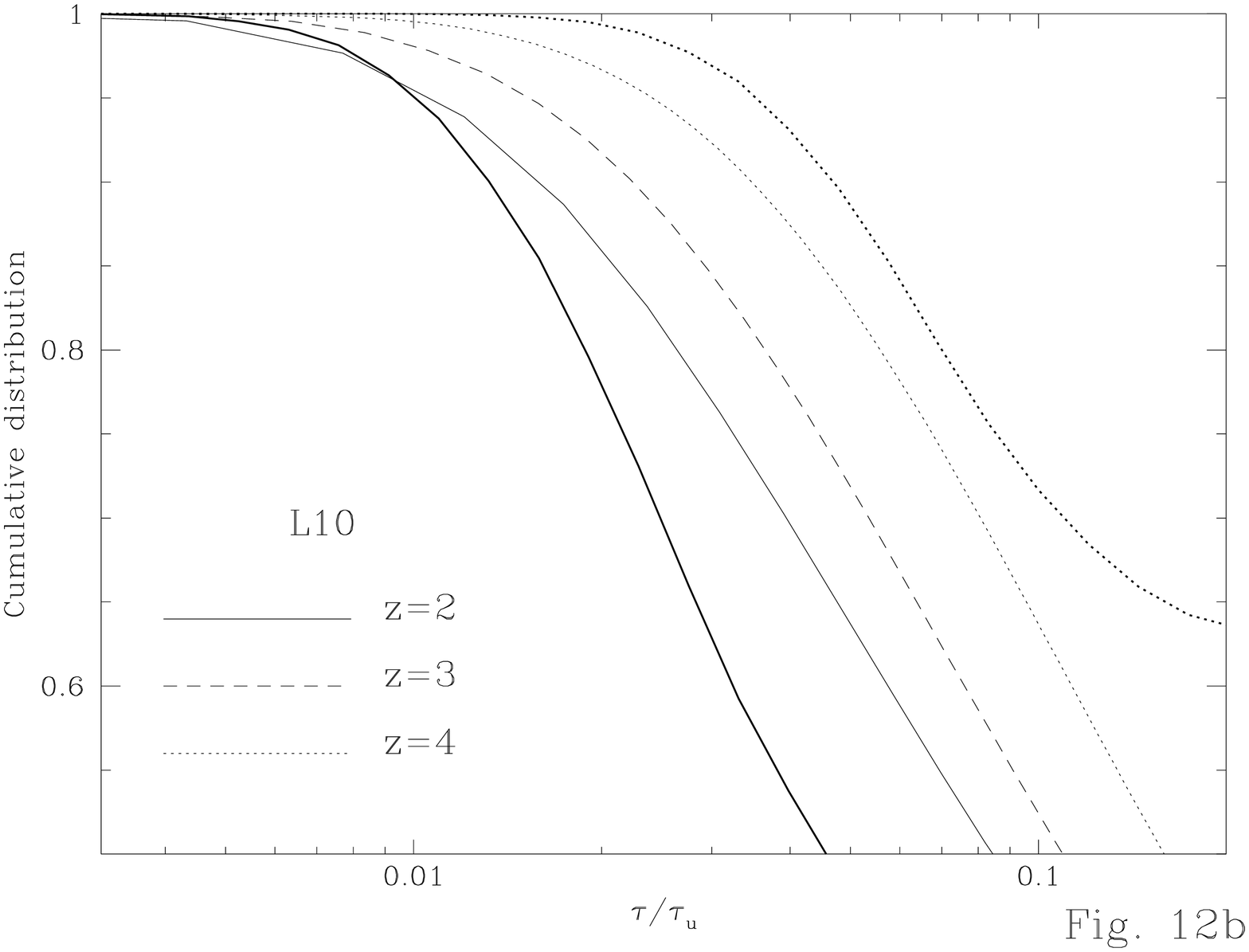}
}
\baselineskip=12truept \leftskip=3truepc \rightskip=3truepc
\noindent{\bf Fig. 12:} (a) Thick lines are the same cumulative distribution
as in Fig. 10, but as a function of $\tau/tau_u$, where $\tau$ is the
optical depth and $\tau_u$ is the uniform optical depth that would be
observed for a uniform intergalactic medium of density $\Omega_b$, and
for the same intensity $\jhi$. The thin lines are the same cumulative
distribution when thermal broadening is not included. (b) Thin lines are
as in Fig. 11a, and thick lines are the analytical prediction from the
``modified Zel'dovich approximation'' in Reisenegger \& Miralda-Escud\'e
(1995).

\endinsert

  The cumulative distribution of the ratio $\tau/\tau_u$ is shown in
Figure 12a as thick curves for the simulation L10.
This shows the true evolution that results from the gravitational
evolution of the shape of the density field: as the redshift decreases,
the gas continues to move away from the voids into collapsed objects,
and the fraction of the spectrum with $\tau/\tau_u$ below a fixed value
increases. However, the evolution in $\tau/\tau_u$ is quite small
compared to the evolution seen in Figure 11b. Most of the rapid
increase of the typical absorption with redshift is due to the change in
$\tau_u$, which evolves as $(1+z)^{4.5}$ when $\Omega(z)=1$ and
$\jhi$ is constant.

  We also show as thin lines the distribution of $\tau/\tau_u$ which is
obtained if the thermal broadening due to the gas temperature is not
included when the $\lya$ absorption spectra are computed. This removes
the thermal wings of the absorption lines, and therefore increases the
fraction of the spectrum with low optical depths. The difference from
the thin to the thick lines is relatively small. In particular, we see
in Figure 12a that most of the absorption in the regions with the
lowest optical depths comes from a true fluctuating ``Gunn-Peterson
effect'' arising from the most underdense regions of the intergalactic
medium, rather than from thermal wings of the lines.

  The distribution of $\tau/\tau_u$ was calculated analytically by
Reisenegger \& Miralda-Escud\'e (1995), using an approximation they
proposed, the ``Modified Zel'dovich Approximation'' (MZA). The
CDM$\Lambda$ model we have adopted in the simulations was also used by
them, among various other models. In Figure 12b, we show again the thin curves
of Figure 12a together with the MZA prediction given in Figure 4 of
Reisenegger \& Miralda-Escud\'e for $z=2$ and 4 (plotted as thick
curves; this is only shown in the region of validity of the MZA, for low
optical depths). The agreement between the curves is probably better
than could be expected given the approximations involved in the MZA
(neglecting the gas pressure and assuming that the deformation tensor of
any gas element conserves its shape), and given also the possible
effects of the missing large-scale power in the simulation.

  The distribution of the fluctuating Gunn-Peterson optical depth should
be a sensitive probe to the amplitude of the primordial density
fluctuations that are collapsing at the redshifts of observation. For
high amplitudes, the voids should be much more empty and much lower
optical depths should cover a large fraction of the spectra. The
relatively good agreement of the numerical simulation and the analytical
results of Reisenegger \& Miralda-Escud\'e (1995) is also encouraging,
since the predictions of different models can be anticipated
approximately without the need for expensive computer simulations.

\subsec{4.4}{Correlation of Flux Along and Across the Line of Sight}

  We now take parallel rows along the simulation separated by a fixed
number of cells. We define the correlation function of the flux as:
$$\xi_f (\Delta v, \Delta r) \equiv
< { (F_{\bf r}(v_0) - < F > ) -
    (F_{{\bf r}+\Delta {\bf r}}(v_0 + \Delta v) - < F > ) \over
  < F^2 > - < F >^2 } > ~, \eqno(\new) $$
where $F_{\bf r}(v)$ is the fraction of transmitted flux along a line of
sight defined by a transverse vector ${\bf r}$, and $v$ is the velocity
along the spectrum. The average is done over the ensemble of all
possible pairs of lines of sight.

\topinsert
\centerline{
\hskip -0.5truecm
\epsfxsize=3.0in \epsfbox{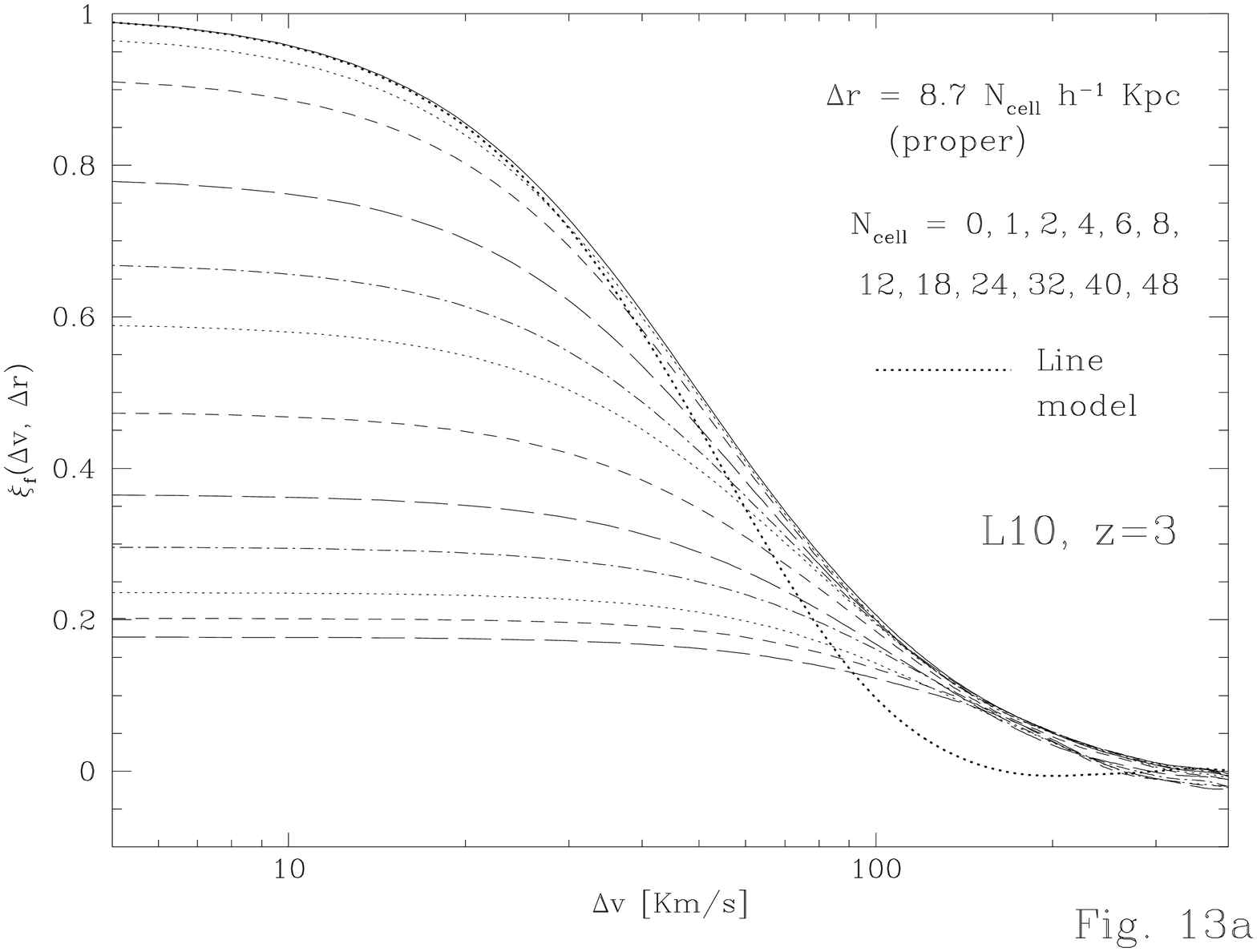} \hskip 0.3truecm
\epsfxsize=3.0in \epsfbox{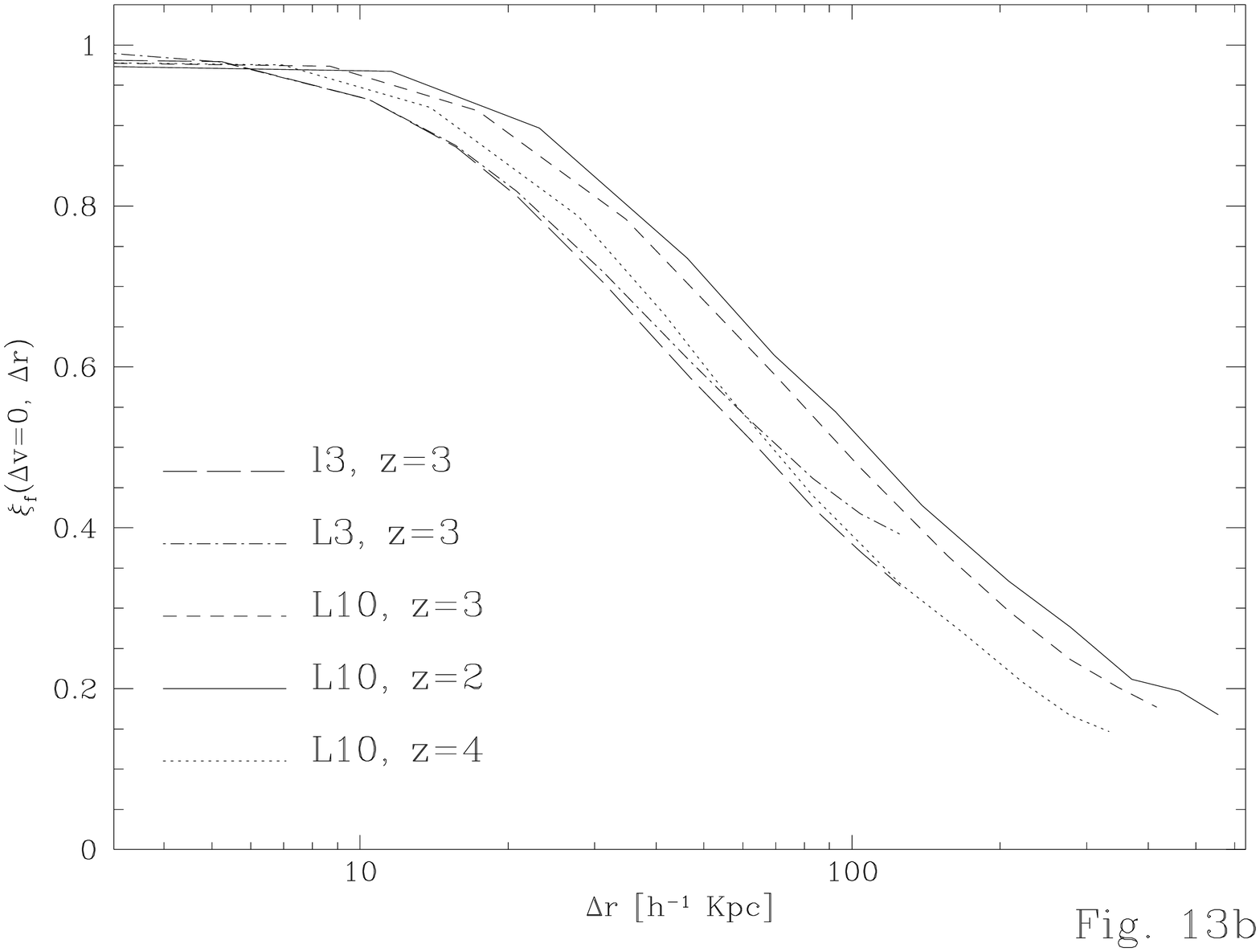}
}
\baselineskip=12truept \leftskip=3truepc \rightskip=3truepc
\noindent{\bf Fig. 13:} (a) Solid line shows autocorrelation function of the
transmitted flux as a function of velocity separation $\Delta v$ in the
spectrum, and other lines are the correlation function for spectra
separated by a transverse distance $\Delta r$, as given in the Figure.
Thick dotted line is the autocorrelation of the line model. (b) Flux
correlation on parallel spectra as a function of transverse separation
$\Delta r$, when $\Delta v = 0$.
\endinsert

  The correlation function $\xi_f$ is shown in Figure 13a for the L10
simulation at $z=3$, as a function of $\Delta v$, at a number of
different transverse separations $\Delta r$. To calculate the
correlation, we have randomly selected 1500 groups of 12 rows from the
simulation. We select one of the possible 3 axis of projection, and then
one among $288^2$ possible rows. Thus, the total number of choices is
248832 (since we did not avoid repeated choices, there is an average of
4.5 groups of 12 rows that appear twice in the sample).
Within each group, the last 11 rows are separated from the
first by the number of cells (and the proper separation $\Delta r$)
indicated in Figure 13a. The correlation for $\Delta v=0$ declines
monotonically with $\Delta r$, so the curve corresponding to each
separation $\Delta r$ is easily identified in the figure. The curve for
$\Delta r = 0$ gives the autocorrelation function of the flux along a
spectrum, and is plotted as a solid line. Also shown as a thick dotted
line is the autocorrelation function of the flux for our line model
described in \Sec 4.2 (we compute it from the same 1800 randomly
generated spectra).

  The autocorrelation for our line model agrees well with the simulation
for $\Delta v < 30 \kms$. For these low velocities, the flux
autocorrelation depends mostly on the distribution function of
b-parameters and column densities, and the agreement suggests that the
spectra in our simulation should yield b-parameters similar to the
observed ones (see also \Sec 4.8) if analyzed in the same way as the
observations, since the b-parameters of our model are similar to the
observed distribution (e.g., Rauch \etal 1992, Cristiani \etal 1995, Hu
\etal 1995). An analysis of the simulated spectra using the same
deblending techniques to fit lines as have been used in the observations
will be done in a subsequent paper.

  The flux correlation in our line model drops very fast for velocities
larger than the maximum width of the lines. Our simulation results are
clearly above the line model at $\Delta v > 50 \kms$. This could
indicate the presence of a correlation of the lines, or a tail in the
distribution of b-parameters for high values above the gaussian in our
model. The latter possibility, however, would also raise the correlation
at low velocities.
In any case, the flux autocorrelation is a
directly observable quantity which our model predicts. In general, the
distribution of b-parameters and the correlation function of the lines
obtained from deblending techniques are quite sensitive to the
detailed algorithm used to identify the lines.

  The cross-correlation function can also be plotted for $\Delta v=0$
as a function of $\Delta r$, the spacing of spectra perpendicular to
the line of sight as measured by comparing absorption lines
in quasars that appear close together on the sky.
This is shown in Figure 13b. The
correlation of the transmitted flux is strong up to large transverse
separations: it declines to 0.5 only at $80 h^{-1} \kpc$, and there is
still a significant
correlation at $\sim 500 h^{-1} \kpc$. The correlation simply
arises because the absorption is produced by the large-scale structure
of the universe; in particular, the
filaments or sheets that cause the absorption (\cf Figure 2c)
show extended linear structures.
A simple ``cloud'' model as adopted by most previous
investigators (usually considering spherical structures in dynamical
equilibrium) would clearly
predict a much smaller correlation on the large scales shown in Figure
13b. The absorption lines, as we can see directly in Figures 2 and 3,
arise from intersection of the lines of sight with filaments and sheets,
as well as clumps which might be approximated as ``clouds''.
We notice that the correlation tends to increase at
lower redshift, but only by a modest amount. The smaller box simulation
shows less correlation, owing to the missing large-scale power. There is
practically no difference between the L3 and l3 simulations, since the
structures are already well-resolved in the lower resolution simulation.
The small difference shows that as the resolution is increased, the
correlation on transverse scales also increases. This may at first
sight seem paradoxical, since one might have intuitively expected that
higher resolution should imply more small-scale variations of the
density, and therefore a loss of the correlation on large scales.
However, the structures causing the absorption are caustics where the
gas is shocked, which are intrinsically smooth due to the erasure of
small-scale power that the gas pressure causes on the gas density
field. The absorbing structures in the high-resolution simulation are
equally smooth, and the better correlation is due to a decrease in
the numerical noise arising from the finite number of cells accross the
absorbing clouds.

Our model is probably in agreement with recent
observations of line correlations (see Smette \etal 1992, 1995
Bechtold \etal 1994, Dinshaw \etal 1994, 1995),
although that will have to be tested
by line identification using the same procedure as in the observations.
We must also caution about the fact that the observed line correlation
has been found for strong lines, whereas our flux correlation is
sensitive to weaker lines.
If the spectra of a quasar pair can be observed with high
signal-to-noise, then the analysis of the flux correlation can also be
compared with our prediction. We notice that the correlation on
parallel lines of sight remains appreciable over a substantial range of
$\Delta v$ (see Figure 13a), and that one should not just search for
``line coincidences'', but use also the information which is in the flux
fluctuations in the spectra which do not correspond to significantly
detected lines.

\subsec{4.5}{Number of Lines}

    The statistical information contained in the spectra of the $\lya$
forest is entirely contained in all the N-point correlation functions of
the flux. In fact, if the spectra were one-dimensional gaussian fields,
as could be the case if the collapsing structures producing the
absorption were still in the linear regime, then the one-point
correlation function would already provide all the information
and all higher order correlations could be derived from this.
In practice, the absorbers are already at an advanced nonlinear stage of
the collapse and the spectra are highly non-gaussian, so that
statistical measures other than the higher order correlation functions
may be more useful in comparing observations with theoretical
predictions. One of the characteristics of the $\lya$ forest is the
presence of discrete absorbers, which correspond to individual
collapsing objects in our simulations. However, the shape of the
absorption profiles is determined by the distribution of hydrodynamic
velocities within the system in addition to the gas temperature, and
therefore the total velocity distribution is generally non-gaussian. If
the shape of an absorption profile is ``deblended'' into several Voigt
profiles to obtain a fit, the separate profiles do not necessarily
correspond to individual physical components, and the fitted
b-parameters and column densities are not
necessarily related to physical quantities. On the other hand, when two
such lines are only partially ``blended'', but they still show two clear
minima in the transmitted flux, they often correspond to separate
physical systems.

  Thus, a measure of the number of lines should give another useful
statistical quantity. The detection procedure to identify lines should
be performed using a well defined algorithm which is simple to
implement, but its
relation to fitted Voigt profiles is less important. Such a detection
is straightforward in low redshift, low signal-to-noise spectra, where
the flux fluctuations are dominated by noise over most of the spectrum
and the detected lines are well separated (e.g., Schneider \etal 1993):
one simply imposes a minimum equivalent width to any detected lines.
But here we are interested in the regime where the true fluctuating
continuum of absorption is observed and any identified ``lines'' are
blended among them and with this continuum (notice that in the theory
of gravitational collapse we are simulating, a true fluctuating
continuum is present also at low redshift, although a much higher
signal-to-noise is required to observe it). Here, we propose an
algorithm based on the same concept of a cloud introduced in \Sec 3 in
real space, which was also presented and used in HKWM.
For any given flux threshold $F_t$, we identify all the regions where
the flux is
below this value and call them ``lines'' (the flux threshold plays an
equivalent role to the contour of neutral density in real space).
All the lines are
therefore bounded by two points in the spectra where $F=F_t$.
As an example, Table 5 (in \Sec 4.7) gives the lines identified
in the spectra plotted in Figures 7(a,b,c,d) and 8(a,b), when $F_t=0.7$.
The first column shows the central velocity of the lines, and the other
two columns give the column densities and b-parameters measured as
explained below (\Sec 4.6 and 4.7). Many fewer lines are detected
compared to the deblending method, since the lines in Table 3 are
blended and the weakest ones do not reach below the threshold $F_t$.
To apply this algorithm to observational data, one
should probably impose the additional requirement of a minimum
equivalent width.

\topinsert
\centerline{
\hskip -0.5truecm
\epsfxsize=3.0in \epsfbox{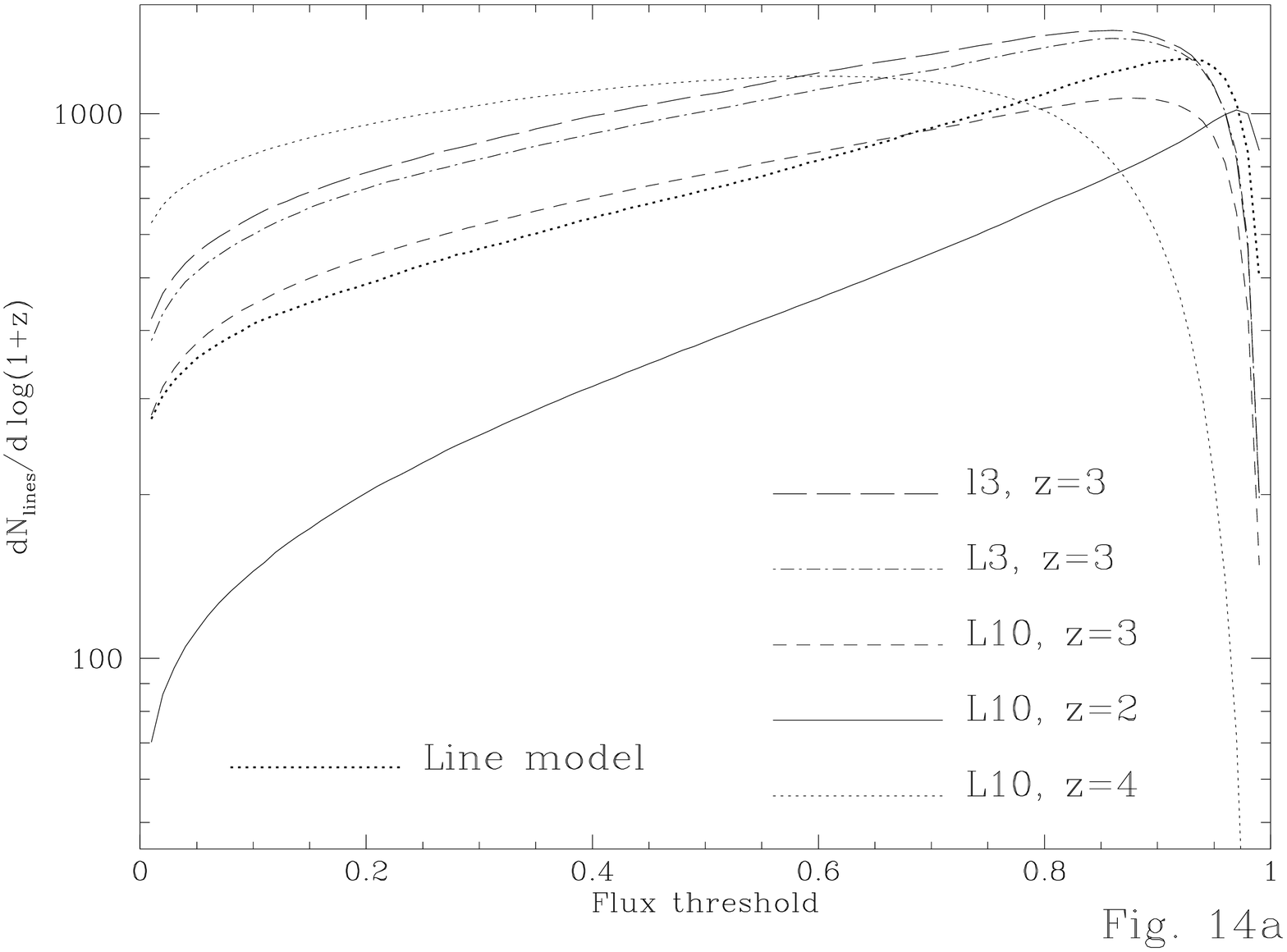} \hskip 0.3truecm
\epsfxsize=3.0in \epsfbox{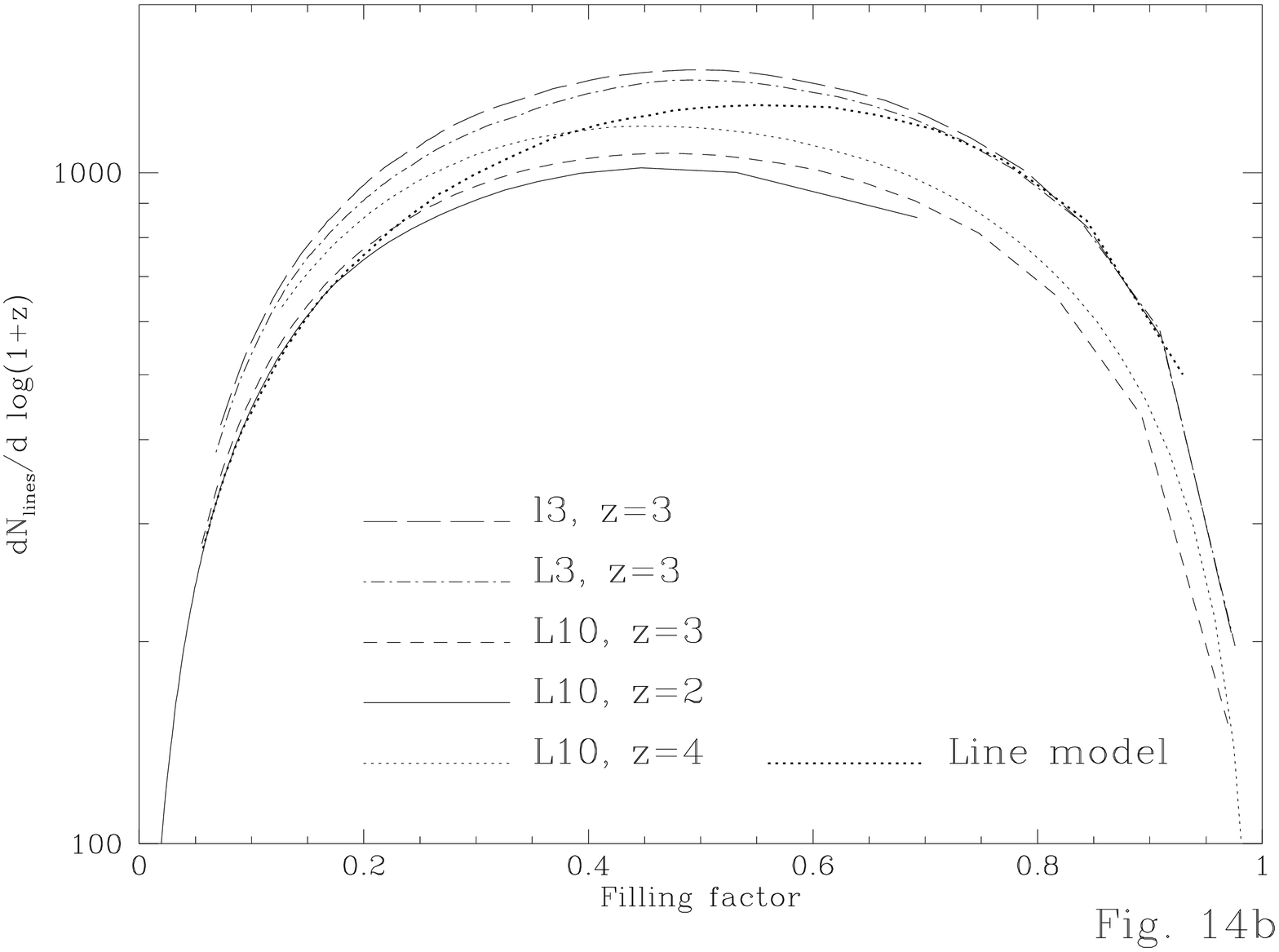}
}
\baselineskip=12truept \leftskip=3truepc \rightskip=3truepc
\noindent{\bf Fig. 14:} (a) Number of lines per unit $\log(1+z)$, as a
function of flux threshold $F_t$, where a line is any region of a
spectrum where the flux is $F < F_t$ continuously. (b) Same number of
lines is shown as a function of the filling factor in the spectra of
the regions with $F < F_t$.
\endinsert

  We start discussing the number of such lines in the simulated spectra.
Figure 14a shows the number of lines per unit $\log(1+z)$ detected
according to this algorithm. Each curve was
computed from the same 18000 spectra described in \Sec 4.4, and the
curve for our line model was computed from 1800 random spectra. The
number of lines is shown as a function of the flux threshold. Strong
lines are counted for low flux threshold, since the optical depth has
to reach a lower minimum in the line profile. The number of lines
increases as the flux threshold is raised, and it starts decreasing at
some point due to blending when the lines start covering most of
the spectrum. The same number of lines is shown in Figure 14b as a
function of the filling factor, defined as the fraction of the spectrum
occupied by the regions with $F<F_t$, identified as lines. The filling
factor was shown as a function of the flux threshold in Figures
11(a,b). Again, low filling factor corresponds to strong lines.

  We notice that the curves shown in Figures 14b are totally
independent of the factor $\mu^2$ by which we may multiply the optical
depth in order to vary the average flux decrement. Thus, the number of
lines as a function of the filling factor is a completely new piece of
information once we have fixed the average flux decrement, in contrast
to the number of lines above a fixed column density. These curves are
an independent testable prediction of our model, related to the scale
of the fluctuations in the absorption.

  As before, we see that the difference between the l3 and L3
simulations is extremely small, and that the L10 simulation has $\sim
25\%$ fewer lines. Again, this is due to the large-scale power which is
absent in the small box simulations. The L10 simulation has larger
voids, and these result in long voids in the spectrum. The Jeans length
of the gas is larger in a large-scale void, so the dark matter
fluctuations within voids will often not affect the gas, and no lines
will be produced.

  We also see in Figure 14b that there is very little change in the
number of lines with redshift. Our simulations predict that {\it when
the number of absorption lines is measured at a fixed filling factor,
there is practically no evolution with redshift}.
Neither merging of structure, nor fragmentation, nor appearance of new
structures is a major effect over the redshift interval considered.
Most of the observed
evolution is due to the increase in the mean opacity of the universe
$\tau_u$ as expressed in classical Gunn-Peterson calculations.
The small evolution
seen in Figure 14b is probably due to the larger scales collapsing at
$z=2$, compared to $z=4$: there is more spacing between successive
structures, and the voids grow bigger and more underdense at low
redshift (see also Fig. 12a). We emphasize that since the curves in
Figure 14b are
independent of the parameter $\mu^2$, the predicted evolution of the $\lya$
forest in this plot is independent of the value of $\jhi$ and
$\Omega_b$, and is therefore a genuine prediction of the simulation.
Of course the evolution at a fixed filling factor is more difficult
to determine observationally, because very high signal-to-noise is
required at low redshift for high filling factor, and at high redshift
for low filling factor.

  The evolution of the $\lya$ forest has been measured for lines above
a fixed column density, or equivalent width, and it has been found that
the number of lines per unit $\log(1+z)$ increases rapidly with redshift
in this case, proportionally to $(1+z)^{\gamma+1}$, where $\gamma\simeq
2.5$ (e.g., Lu, Wolfe, \& Turnshek 1991). Our results say that this
evolution is mostly due to a general increase of the optical depth,
evolving similarly to the Gunn-Peterson optical depth: $\tau
\propto (1+z)^6\, H^{-1}(z)\, \Omega_b^2/J_{-21}$ (see also HKWM).
The evolution in the
``shape'' of the $\lya$ forest is what is shown in Figure 14b and is
very weak; the observed evolution is of the ``amplitude'' of the $\lya$
forest. For constant $\jhi$ and temperature, the optical depth will
increase as $(1+z)^{4.5}$ (notice that in our model,
$\Omega(z)$ is practically equal to one at $z > 2$), and if the gas in
the $\lya$ forest is expanding at the Hubble rate
this will cause an evolution of the number of lines above a fixed
column density proportional to $(1+z)^{4.5(\beta-1)}$, similar to the
observed evolution for $\beta = 1.75$. In reality, the overdensities
of the absorbing gas tend to increase (especially for high column
density systems; Fig. 9d), but the increasing temperatures at low
redshift compensate for that.

  Finally, we compare our result for the L10 simulation at $z=3$ with
our line model. Figure 14b shows that the line model has an excess of
lines at high covering factor, by about $25\%$. We might want to solve
this by making the column density distribution shallower at low column
densities; however, this would also change the distribution of flux
at very low optical depths. We see from Figure 11b that the fraction
of spectrum having very low optical depths is already substantially
higher for our line model than in the simulations. If the number of
weak lines is reduced, this fraction will be further increased. Thus,
a model with superposed Voigt profiles with uncorrelated positions
and similar b-parameters for weak and strong lines cannot reproduce
both the flux distribution and the number of lines as a function of
the covering factor. This is a sign of the presence of the
Gunn-Peterson effect, produced in our simulation by gas in the voids
that fills the volume between the structures yielding the absorption
lines. This Gunn-Peterson optical depth fluctuates owing to the
density variations accross the voids, but it is a smoother variation
than that caused by superposed weak lines; in particular, it has many
less minima in the transmitted flux, hence the lower number of lines
at high covering factor.

\subsec{4.6}{The Column Density Distribution}

  We now take the individual lines detected according to the
prescription specified above, and integrate the optical depth over
the interval in the spectrum between the two points where the flux is
below our threshold $F_t$. The column density within this interval is
then given by $\nhi = (m_e c)/(\pi e^2\, f\, \lambda) \int dv\, \tau =
[7.445\times 10^{11}\cm^{-2}(\kms)^{-1}]\, \int dv\, \tau$ (Gunn \&
Peterson 1965; here, $f=0.416$ is the oscillator strength, and $\lambda$
the wavelength of the $\lya$ line). Notice that the optical depth
contributed by any element of gas in a physical system will be only
partially included in the total column density of a line defined in this
way, since the optical depth
from the gaussian tails of the thermal velocity distribution will fall
outside the interval corresponding to the absorption line. As the
flux threshold is raised, the column density of each line will
increase as the tails in the profile are included, and as it merges
with its neighbors.

\topinsert
\vskip -0.2truecm
\centerline{
\hskip -0.5truecm
\epsfxsize=3.0in \epsfbox{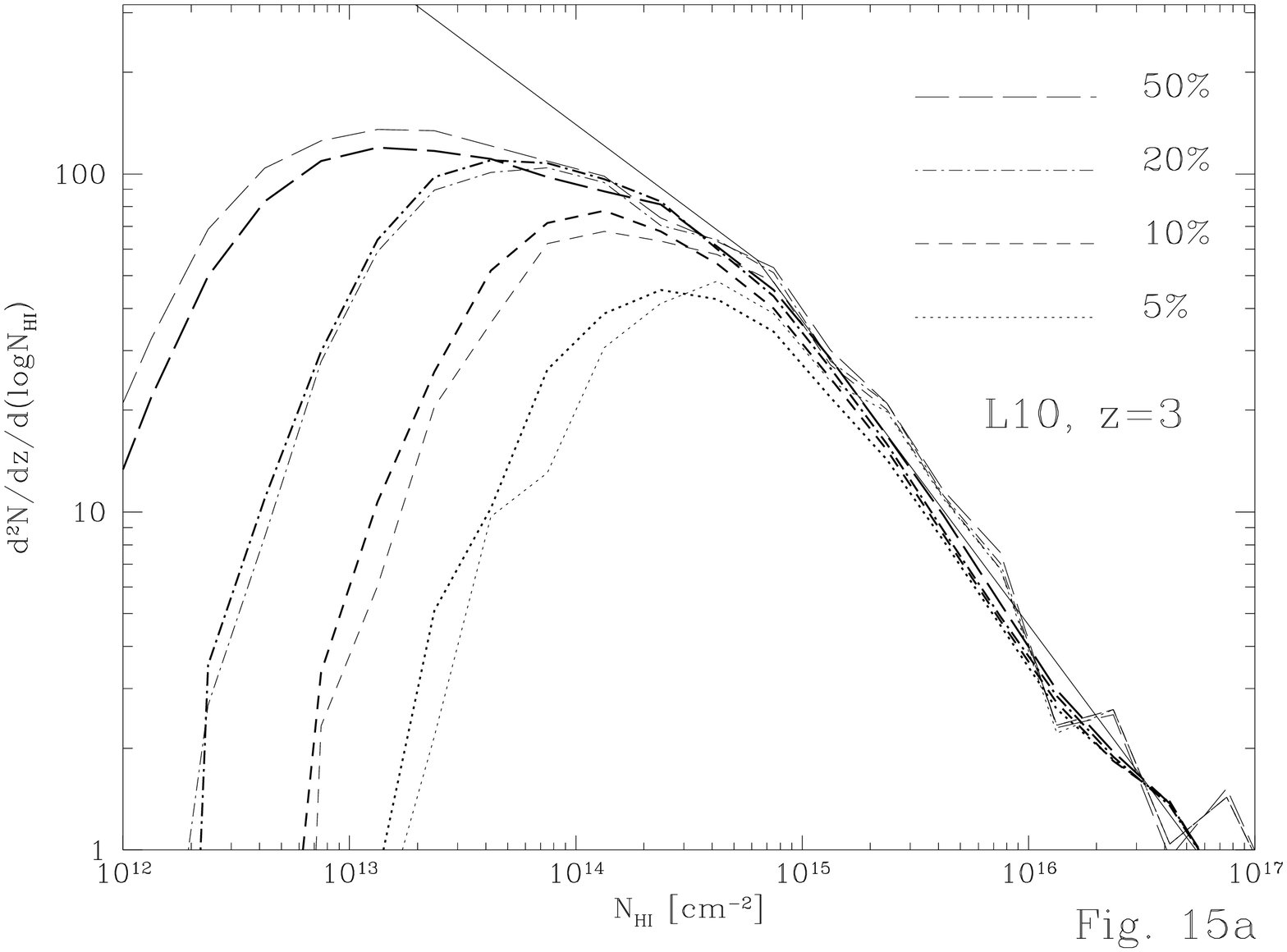} \hskip 0.3truecm
\epsfxsize=3.0in \epsfbox{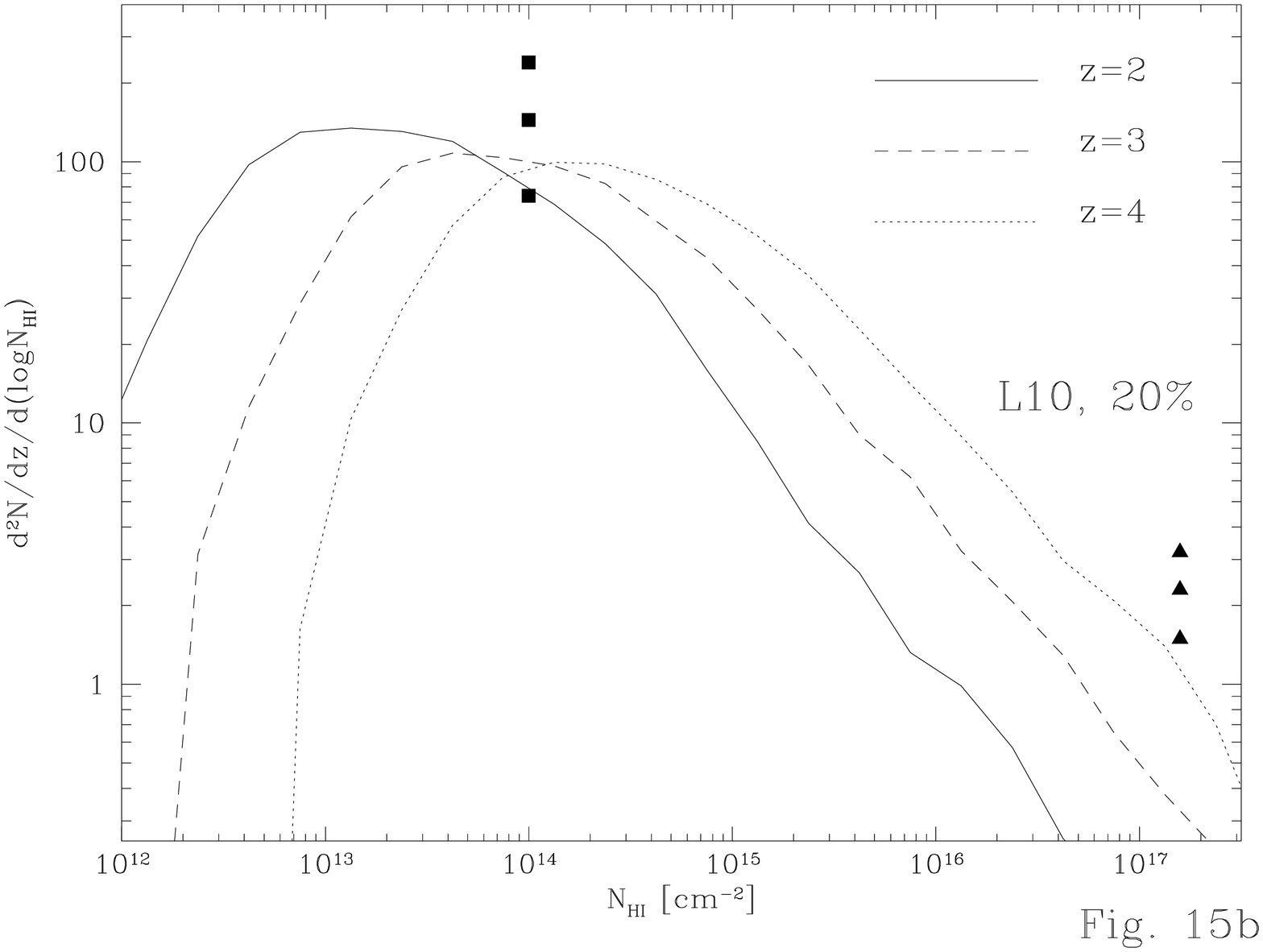}
}
\vskip -0.2truecm
\centerline{
\epsfxsize=3.0in \epsfbox{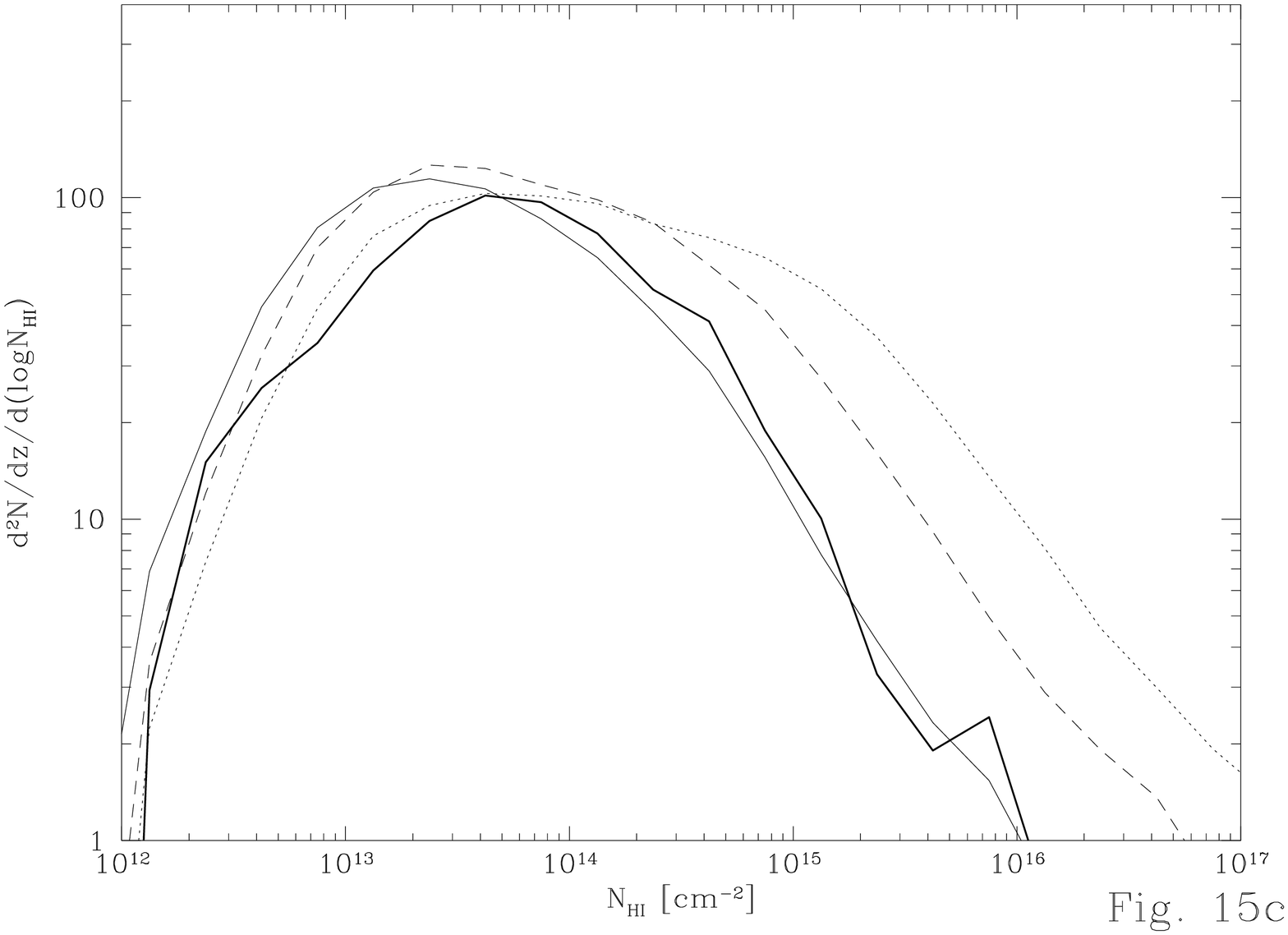}
}
\baselineskip=12truept \leftskip=3truepc \rightskip=3truepc
\noindent{\bf Fig. 15:} (a) Thick lines are column density distribution
at $z=3$ in the L10 simulation for filling factors 10\%, 20\% and 50\%
(corresponding to flux thresholds $F_t=0.107$, 0.486 and 0.891,
respectively). Results for line model are shown as thin lines, thin
solid line is the true column density distribution of line model.
(b) Column density distribution for filling factor of 20\%, at $z=2$, 3
and 4. Black squares are observed number of lines at $z=3$ $\nhi =
10^{14}\cm^{-2}$ from Petitjean \etal (1993), and assuming evolution as
$(1+z)^2.3$ for $z=2$ and $z=4$. Black triangles are number of Lyman
limit systems at $z=2$, 3 and 4 from Stengler-Larrea \etal (1995; we
assume a slope $\beta=1.5$ of the column density distribution to transform
their cumulative number of Lyman limit systems to a differential number).
(c) Column density distribution at a fixed flux threshold, $F_t=0.7$.
The thick solid line is the distribution obtained by HKWM (using an
SPH simulation and for a CDM model different from the one we adopt).
\endinsert

  The distribution of column densities found in this way is shown in
Figure 15a as the thick lines, for the L10 simulation at z=3, for four
different flux thresholds corresponding to the filling factors
indicated in the figure. Also shown as thin lines are the column density
distribution obtained in the same way for our line model. Finally the
thin solid line is the true distribution of column densities assumed in
our line model. There is generally good agreement among the curves
obtained from the simulations and those of the line model.  For $\nhi >
10^{15} \cm^{-2}$, the agreement of the solid line with the other thin
lines show that these column densities are, as expected, appropriately
recovered from our algorithm. At the same time, the fact that for low
column densities the column density distribution of the line model
obtained from our algorithm agrees well with the results from the
simulation shows that the turnover in the column density distribution at
low column densities is partly due to our algorithm. This is also as
expected: some lines are entirely missed when their flux minimum is
above the threshold, and the detected lines also lose the column density
in the tails of their profiles that are above the threshold. Since the
line model agrees well with the determinations of the column density
obtained using the deblending technique (Petitjean \etal 1993; Hu \etal
1995), it seems probable that our simulations should also yield results
in agreement with observations when analyzed in the same way.

  We also show in Figure 15b the column density distribution from the
L10 simulation at $z=2$, 3 and 4, fixing the filling factor to 20\%.
The black dots indicate the observed number of lines at $\nhi =
10^{14}\cm^{-2}$, taken from Petitjean \etal (1993) at $z=3$,
and assuming evolution with $\gamma = 2.3$ for the other two redshifts.
The evolution at a fixed column density probably agrees with
observations, once the effect of our algorithm shown in Figure 15a is
taken into account. Since we have assumed a constant intensity of the
ionizing radiation, the result suggests that if our simulation provides
a correct description of the $\lya$ forest $\jhi$ does not vary by a large
factor within this redshift range, which is corroborated
by the sources of ionizing photons that can be expected and the effect
of the absorption by the $\lya$ clouds (see Table 2, and Cen \& Ostriker
1992). The predictions for Lyman limit systems ($\nhi
\sim 10^{17} \cm^{-2}$) are, on the other hand, clearly discrepant with
observations.
The observed number of Lyman limit systems
at $z=2,3,4$ is shown as solid triangles
(Sargent, Steidel, \& Boksenberg 1989;
Storrie-Lombardi \etal 1995; Stengler-Larrea \etal 1995).
Not only we have too few of them in the simulation (by a factor 10 at
$z=3$), but also they evolve much faster than is observed.
A deficit of Lyman limit systems was also found in Katz \etal
(1995); it may indicate the presence of a second population of absorbers
at high column densities which are not reproduced in the numerical
simulations, and the self-shielding effects which we have not included.

  Finally, in Figure 15c we show the column density distribution at a
fixed flux threshold $F_t=0.7$, and we compare with the results at
$z=2$ for the CDM model in HKWM ({\it thick solid line}).
Interestingly, the number of lines at high column densities is very
similar (with a slight excess in the HKWM simulation) even though the
two simulations are of two different theories. This suggests that
the column density distribution is not a very sensitive test of the
model, once they are all normalized to the same average flux decrement
(the HKWM results were actually normalized to a slightly larger flux
decrement). The smaller number of weak lines in the HKWM results may
arise from their lower mass resolution, since weak lines tend to be
produced by the small-scale systems that survive in large-scale voids,
but it could also be a real difference of the models.

\subsec{4.7}{Other Properties of the Absorption Lines}

  We now measure equivalent widths and b-parameters of the absorption
lines, using also the same procedure as in HKWM.
In this paper, we shall only highlight the main results; a complete
analysis of properties of individual absorption lines will be presented
in a future paper, where various methods for measuring such
properties will be tested and applied to observations and to the
simulated spectra in the same manner. The results for this Section are
obtained from an independent set of 3600 random spectra of the L10
simulation at $z=3$.
Equivalent widths of absorption lines are easily measured when
lines are well separated in the spectrum. However, when lines are
blended the equivalent widths are no longer well defined and depend on
the ``deblending'' algorithm. In particular, if lines are fitted to
Voigt profiles, the number of components required will increase with
the signal-to-noise of the observation, and therefore the equivalent
widths of individual components will on average decrease with the
signal-to-noise.

\topinsert
\centerline{
\hskip -0.5truecm
\epsfxsize=3.0in \epsfbox{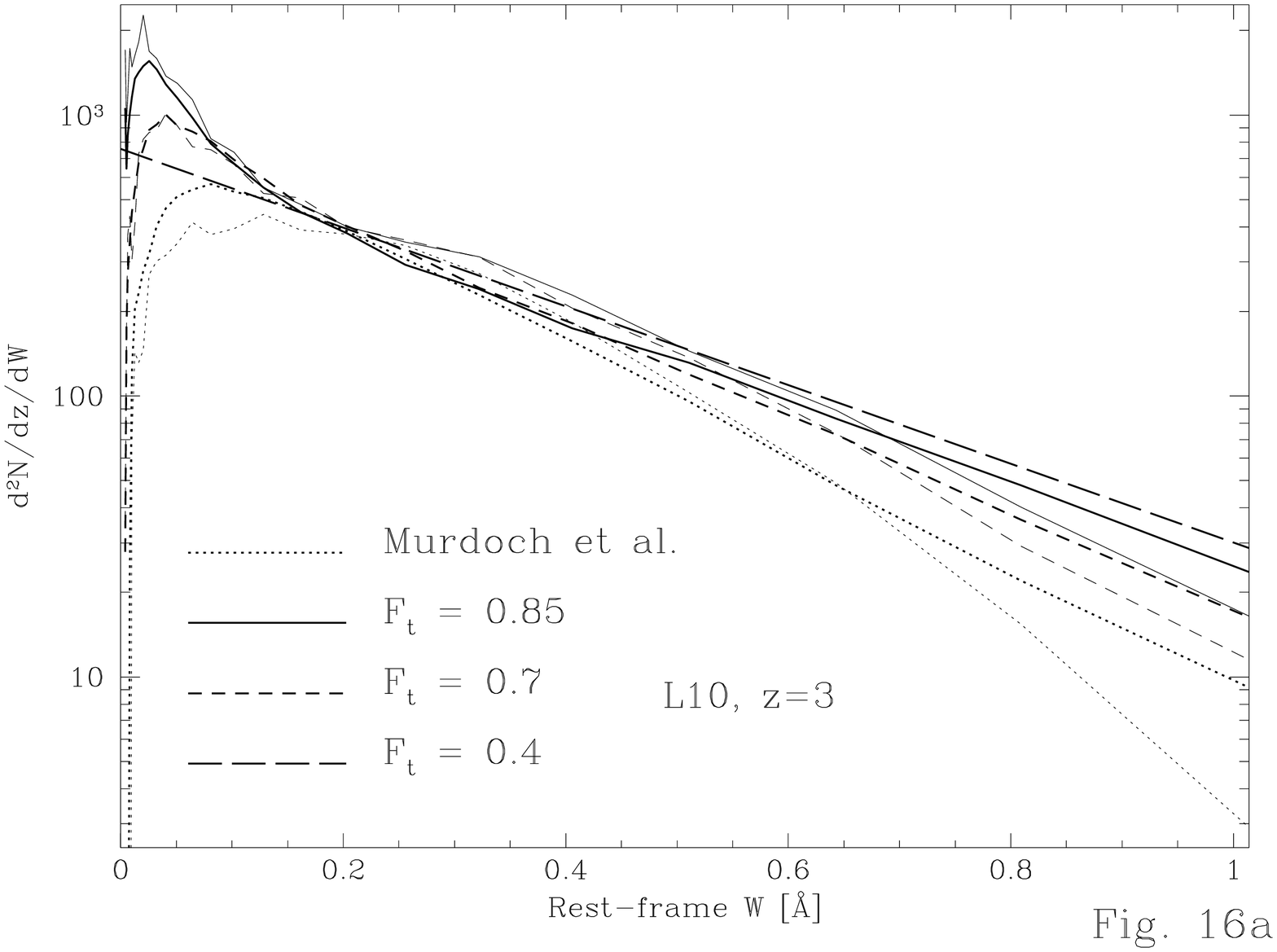} \hskip 0.3truecm
\epsfxsize=3.0in \epsfbox{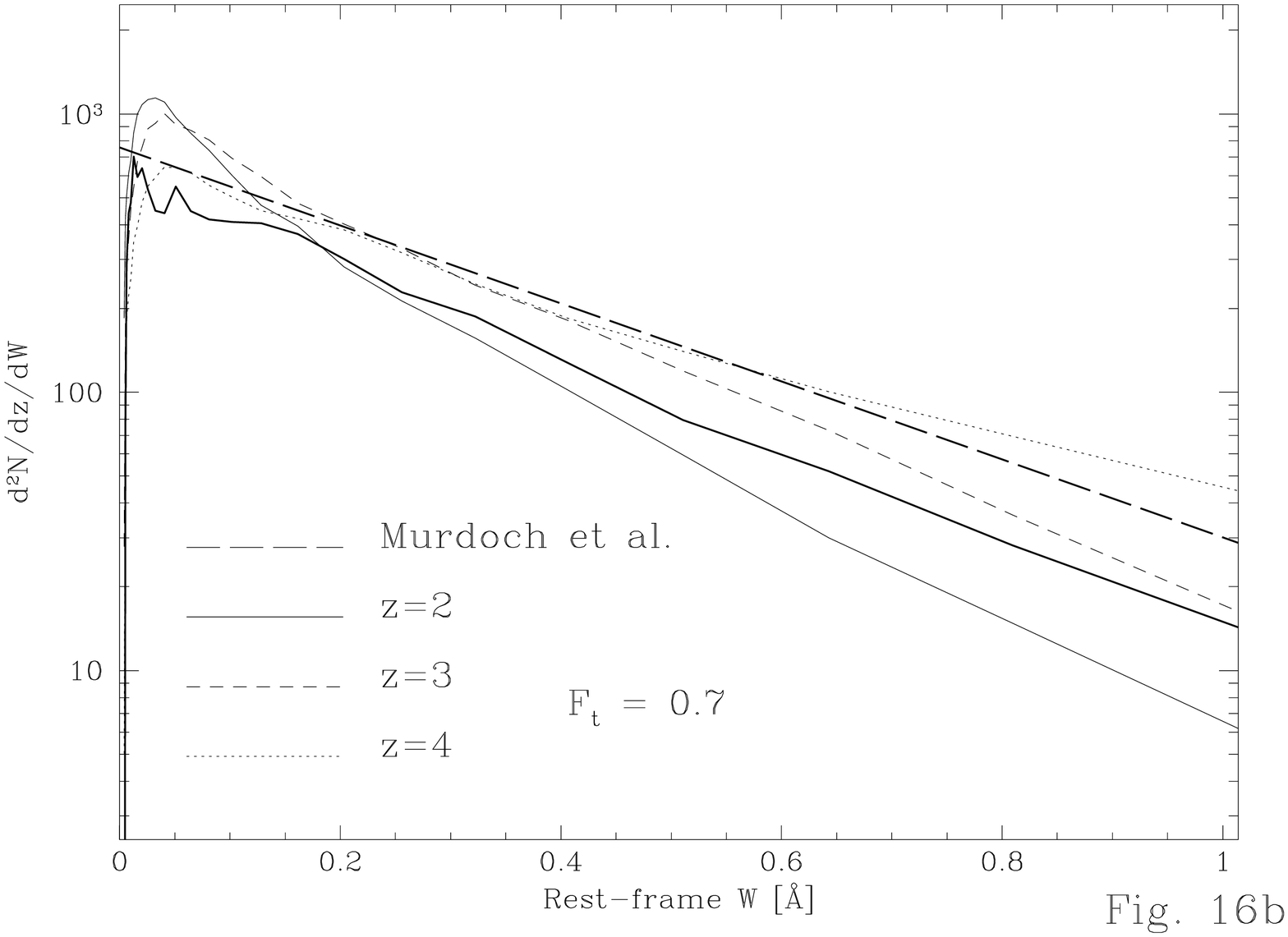}
}
\baselineskip=12truept \leftskip=3truepc \rightskip=3truepc
\noindent{\bf Fig. 16:} (a) Thick lines show the equivalent width
distribution at $z=3$ from the L10 simulation, at three different flux
thresholds. Thin lines are the same distribution obtained from the
line model. The long dashed line is the exponential distribution in
Murdoch \etal (1986). (b) Same distribution for the L10 simulation
at $z=2$, 3, 4, for $F_t = 0.7$. The thick solid line is the equivalent
width distribution obtained by HKWM at $z=2$.
\endinsert

  We therefore use a different definition of the equivalent width,
depending on the flux threshold used to select lines as described in
\Sec 4.5. The equivalent width is measured only within the region where
the flux is below the threshold, without attempting to add any tails of
the absorption profile outside this region and without deblending.
It is defined as the integral of $1-F$ over the region between the two
crossing points where $F=F_t$ ($F$ is the transmitted flux).
We show as thick lines in Figure 16a the distribution of equivalent
widths, for three different flux thresholds; also shown, as thin lines,
is the distribution obtained from our line model defined in \Sec 4.2.
The thick dotted line is the exponential fit to the observed
distribution given by Murdoch \etal (1986). The resulting distribution
depends sensitively on
the adopted flux threshold: it steepens as the flux threshold is
lowered, because the large equivalent widths lines are split.
This dependence on the flux threshold is well reproduced by our line
model.  An exponential profile provides a good fit to the equivalent
width distribution we obtain,
and the slope is similar to the observed one for $F_t =0.85$. The excess of
weak lines at low equivalent widths above the exponential fit is also
present in the observed $\lya$ spectra (Murdoch \etal 1986; Hu \etal
1995), and agrees well with what is obtained for the line model, which
incorporates weak lines with a power-law distribution of column
densities with $\beta = 1.5$.

  Shown in Figure 16b are the same equivalent width distributions for
$F_t=0.7$ at three different redshifts, and the equivalent width
distribution found in HKWM at $z=2$. We see again that our
simulation contains more weak lines than HKWM. At high
equivalent widths, the slope is slightly different. This results partly
from the larger b-parameters (as we shall see below), and partly from
the higher average flux decrement of the model of HKWM, since
for a higher flux decrement a fixed flux threshold will correspond to a
larger covering factor, which increases line blending and the number of
high equivalent width systems.

\topinsert
\centerline{
\hskip -0.5truecm
\epsfxsize=3.0in \epsfbox{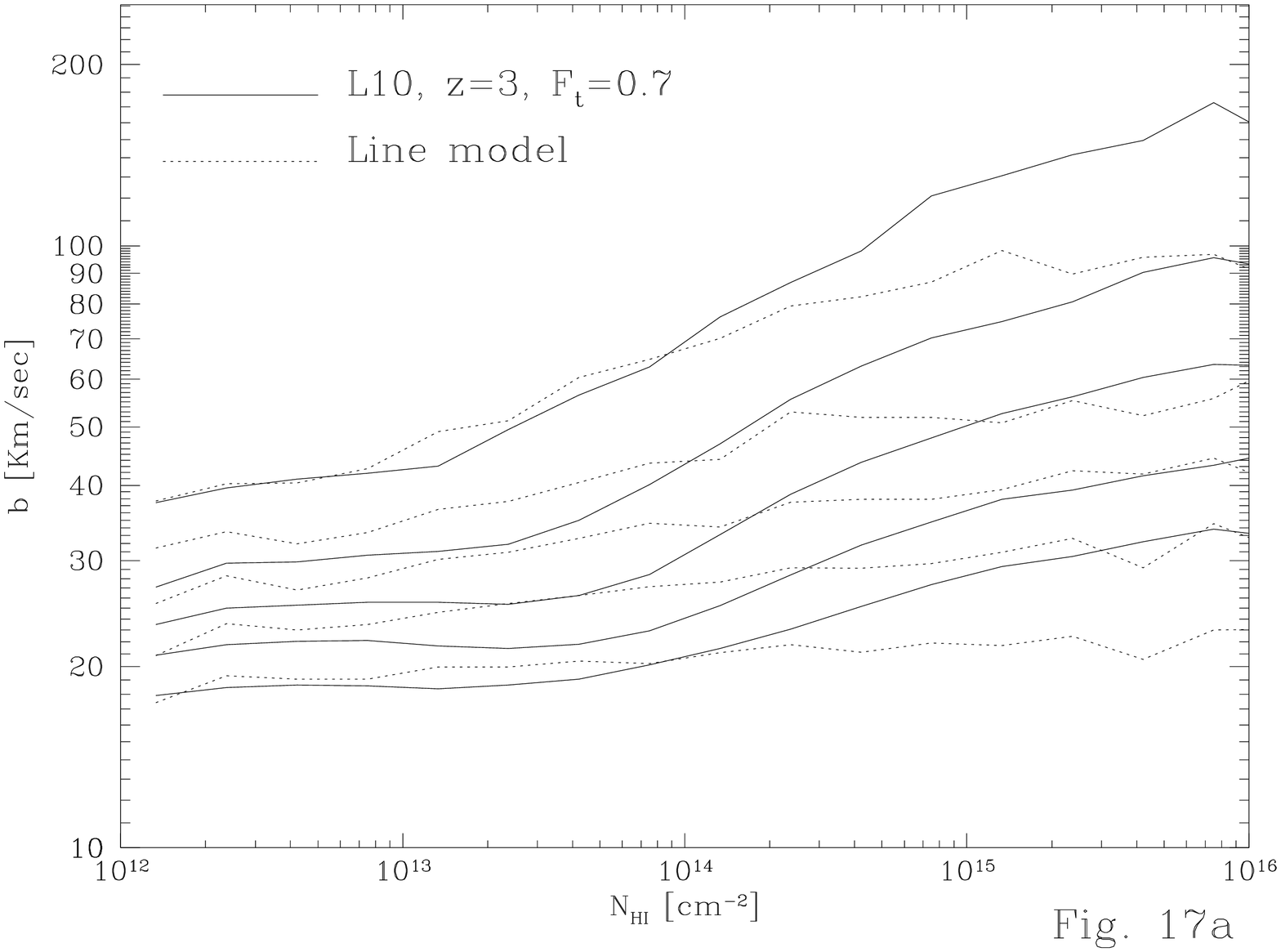} \hskip 0.3truecm
\epsfxsize=3.0in \epsfbox{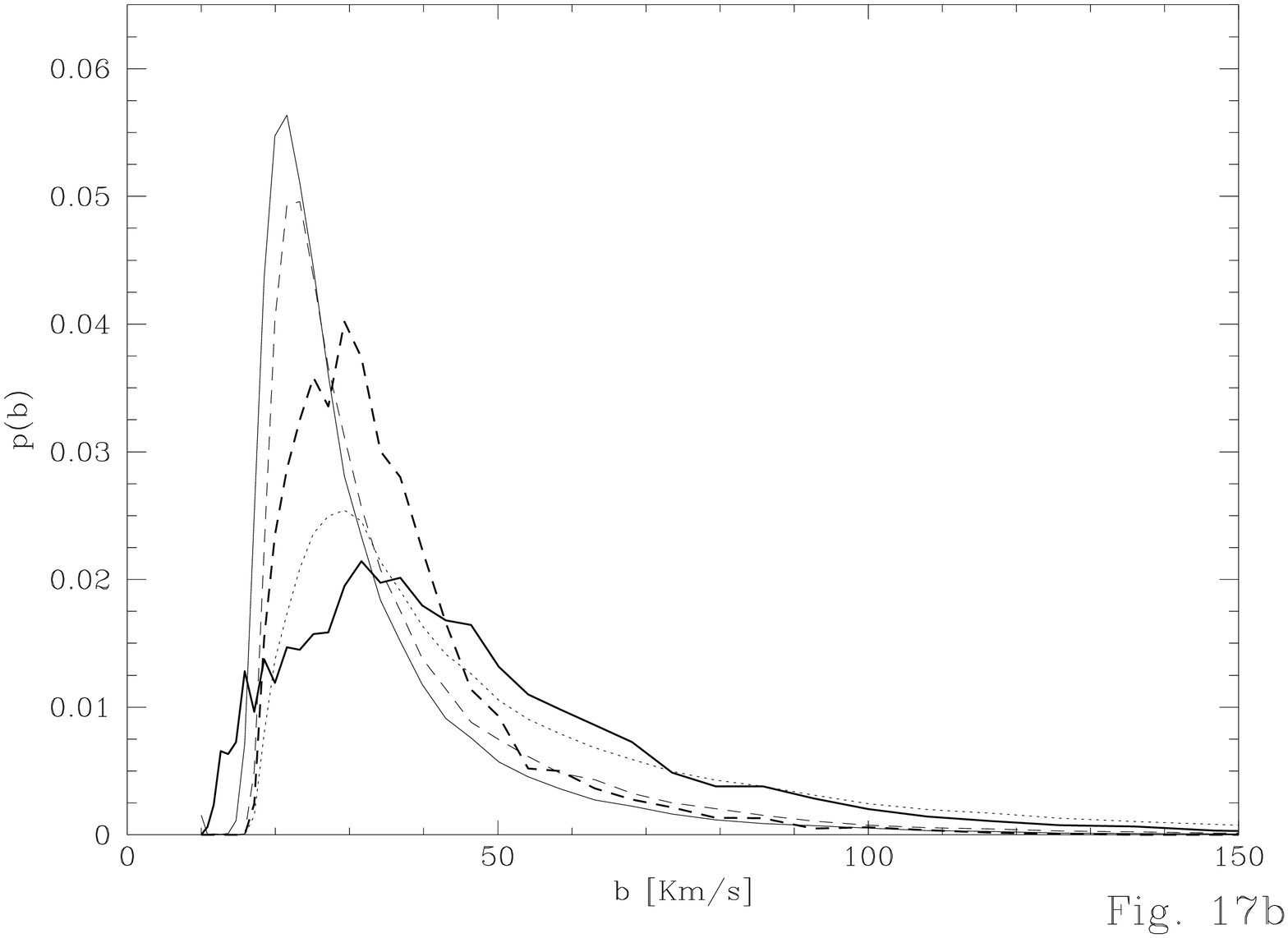}
}
\baselineskip=12truept \leftskip=3truepc \rightskip=3truepc
\noindent{\bf Fig. 17:} (a) Distribution of b-parameters of lines found
with a flux threshold $F_t = 0.7$. At each column density, the five
solid curves show where (5\%, 25\%, 50\%, 75\%, and 95\%) of the
absorption lines have lower b-parameters. Dotted lines are computed for
the line model. (b) Distribution of b-parameters for all lines, detected
with $F_t=0.7$. Thin lines show results from L10 simulation, thick solid
line is the result from HKWM, and thick dashed line is the result for our
line model.
\endinsert

  The b-parameters of the lines are defined as being the same as that of
a Voigt profile that has the same width of the interval where $F < F_t$,
and the same equivalent width measured as above (i.e., not including the
tails of the line profile where $F>F_t$). Figure 17a gives the
distribution of b-parameters as a function of column density, measured
as in \Sec 4.6. The five solid lines plotted indicate the value of $b$
such that (5\%, 25\%, 50\%, 75\%, 95\%) of the lines have a lower
b-parameter at each column density. The dotted lines are the same
results predicted by our line model. Most of the $b-N_{HI}$ correlation
seen in the figure is caused by our algorithm to measure $b$, as seen
from the curves for the line model (there is no intrinsic $b - \nhi$
correlation in the assumed line model).
However, for $\nhi > 10^{14}\cm^{-2}$,
some of the increase of $b$ with $\nhi$ seems to be real. This is
probably due to the presence of hot gas in the dense regions of the
high column density systems, caused by high velocity shocks, and also by
cool gas falling at high velocity towards the system. Deblending
techniques may not be able to recover such a correlation, since the
dense cool gas in the system will cause a narrow minimum of the line
profile, and the outer tails may be fitted with other weak lines; this
would cause a correlation of the weak lines with the strong lines. The
typical b-parameters obtained from the simulation and the line model
also agree well, suggesting therefore that they also agree with the
observations.

\topinsert
\vskip 0.2truecm
$$\vbox{\tabskip 1em plus 2em minus 5em
\halign to \hsize{
\hfil # \hfil & \hfil # \ & \hfil # \hfill \cr
\multispan{3}{\hfil {\bf TABLE 5: ABSORPTION LINE PARAMETERS} \hfil }
\cr
\noalign{\bigskip\hrule\vskip0.1truecm\hrule\medskip}
v [km/s] & $b$ & $\log \nhi$
\cr
\noalign{\medskip\hrule\medskip}
&Fig. 7a& (z = 3) \cr
\noalign{\smallskip\hrule\medskip}
 112.5&  30.9&  14.86 \cr
 600.0&  84.8&  14.43 \cr
 936.0&  96.4&  15.16 \cr
\noalign{\medskip\hrule\medskip}
&Fig. 7b& (z = 3) \cr
\noalign{\smallskip\hrule\medskip}
 328.7&  74.8&  15.52 \cr
1056.3&  40.4&  13.70 \cr
1143.9&  21.3&  13.59 \cr
\noalign{\medskip\hrule\medskip}
&Fig. 7c& (z = 3) \cr
\noalign{\smallskip\hrule\medskip}
  93.8&  23.4&  14.17 \cr
 259.0&  49.6&  14.94 \cr
 764.5&  26.4&  13.68 \cr
 987.2&  37.9&  13.39 \cr
1094.5&  36.6&  14.58 \cr
\noalign{\medskip\hrule\medskip}
&Fig. 7d& (z = 3) \cr
\noalign{\smallskip\hrule\medskip}
  71.7&  40.9&  15.35 \cr
 249.1&  29.8&  13.92 \cr
 998.1&  23.3&  14.24 \cr
1182.4&  31.4&  14.29 \cr
\noalign{\medskip\hrule\medskip}
&Fig. 8a& (z = 2) \cr
\noalign{\smallskip\hrule\medskip}
  99.2&  23.9&  14.12 \cr
 422.8&  27.4&  13.24 \cr
 727.3&  24.7&  13.38 \cr
 928.0&  40.1&  14.22 \cr
\noalign{\medskip\hrule\medskip}
&Fig. 8b& (z = 4) \cr
\noalign{\smallskip\hrule\medskip}
 105.4& 109.2&  15.46 \cr
 382.8&  41.1&  13.90 \cr
 895.4& 176.1&  15.85 \cr
\noalign{\medskip\hrule\medskip}
}}$$
\endinsert

  We also show in Figure 17b the probability density of the b-parameter
for all lines detected at a threshold $F_t = 0.7$, at three redshifts,
and compare the result at $z=2$ with the same distribution obtained by
HKWM. The b-parameters are remarkably different in the
two simulations. Our b-parameters are smaller and have a smaller
dispersion than in HKWM, and they show a very pronounced
lower cutoff at $b=17\kms$. The mean value we find at $z=2$ is
$\bar b = 33 \kms$, compared to $\bar b = 50 \kms$ in HKWM. The average
flux decrement in HKWM was higher, and this can contribute to the larger
Doppler parameters they find. However, raising the flux threshold to
$F_t=0.85$ (roughly equivalent to increasing the optical depths by a
factor of 2, which is what is needed to bring our flux decrement to the
value adopted by HKWM; see Fig. 10), our mean b increases only to
$\bar b = 36 \kms$. Thus, there is a genuine difference between the
b-parameters, which could be due to a difference in the models or in the
numerical methods of the simulations (in particular, the lower mass
resolution in HKWM may cause some narrow lines from low-mass systems
to be missed).

  In Table 5, we give the b-parameters and column densities measured
with the threshold method for the lines identified in the six spectra
in Figs. 7 and 8. Many fewer lines are identified compared to the
deblending method (see Table 3). The number of lines in Table 3 is
particularly large because the signal-to-noise assumed was very high,
so many lines are required to fit in detail the absorption profiles.
The values of the parameters agree only for isolated lines. If the
column densities of the blended lines are added together, they are
generally larger than the column densities in Table 5. The reason is
that high column density systems tend to have their profiles fitted
with superposed lines of lower column density, and a central highly
saturated line with a low b-parameter, and an overestimated column
density. This is highly sensitive to the assumed signal-to-noise,
but it shows that the parameters derived from deblending of complex
profiles are not necessarily related to the physical properties.

\subsec{4.8}{Correlation of Lines Along and Across the Line of Sight}

  The observations of correlations in the $\lya$ forest have most often
been done from line identifications. In general, the correlation that
is found will depend on the algorithm used to identify lines. In the
usual deblending algorithm, the correlation of the lines will depend on
the signal-to-noise of the observation. At low signal-to-noise, many
absorption features will be fitted adequately with only one Voigt
profile, and a
negative autocorrelation will be found for separations of the order
of the thickness of the lines. But at high signal-to-noise, several
superposed Voigt profiles may be required to fit the shapes of the
lines, and the correlation on small scales will be increased. In fact,
the line detection algorithm affects the measured correlation on
arbitrarily large velocity separations, because it can change the
density of lines in a way which is not necessarily linear.

\topinsert
\centerline{
\epsfxsize=4.0in \epsfbox{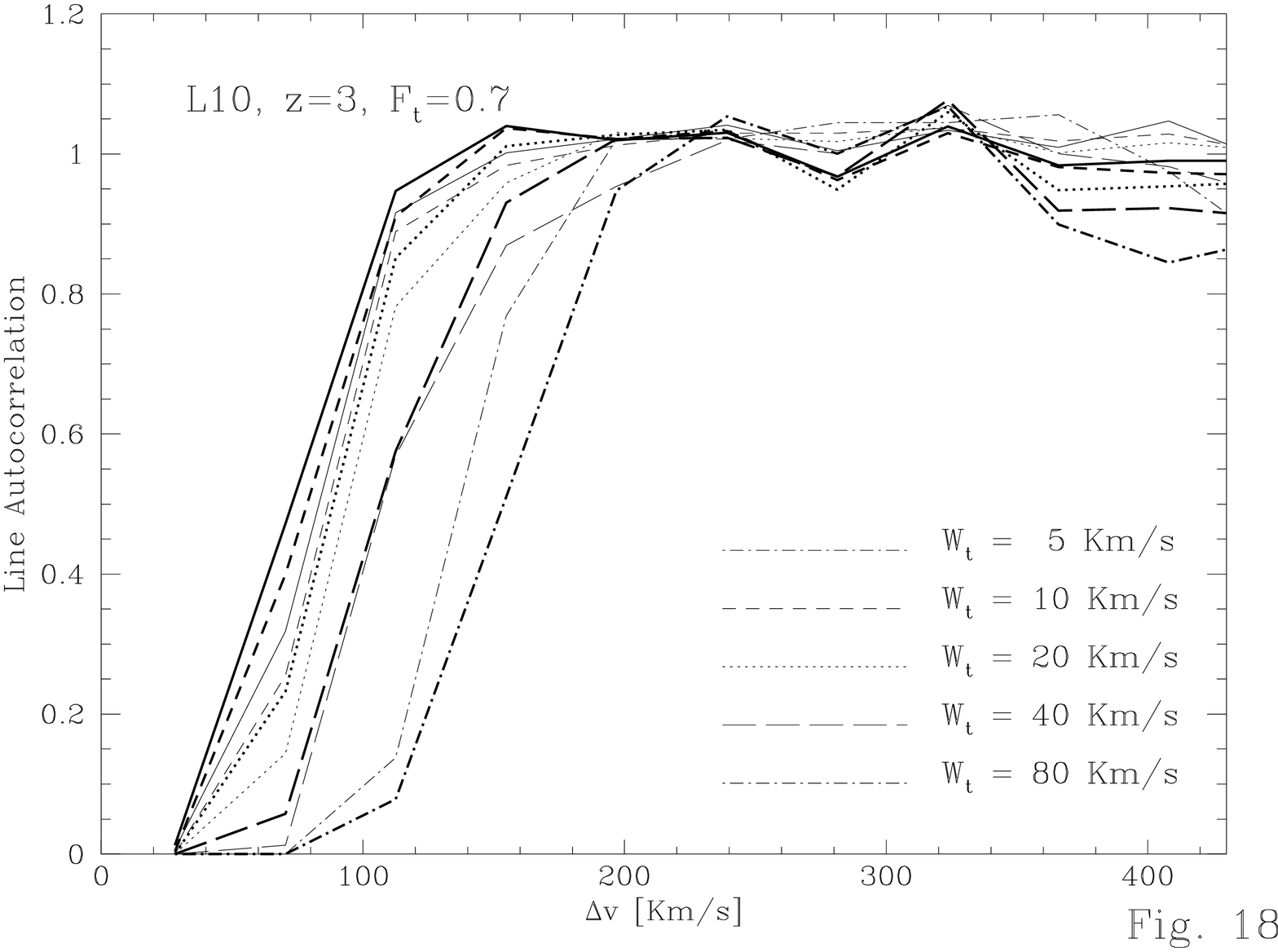}
}
\baselineskip=12truept \leftskip=3truepc \rightskip=3truepc
\noindent{\bf Fig. 18:} The autocorrelation function of lines along a
spectrum, for all lines detected with $F_t=0.7$ with equivalent width
higher than the specified threshold $W_t$. Thick lines are for the
L10 simulation, thin lines are for our line model. The anticorrelation
at small separations is due to the width of the lines.
\endinsert

   Here, we shall present the velocity correlation function of lines
along the line of sight selected with the algorithm we have proposed,
but we emphasize that this cannot be directly
compared with the correlation of
lines in the observations identified with other methods. We show the
line autocorrelation in Figure 18 (the number of lines at a separation
$\Delta v$ from another line, divided by the expected number given the
line density), where the lines are identified using
the fixed flux threshold $F_t=0.7$, and in addition we require the
equivalent width (defined as in \Sec 4.7) to be above a given
threshold $W_t$. The five curves are for five different thresholds in
equivalent width, as indicated in the figure. The thick lines are the
results from the L10 simulation at $z=3$, and the thin lines are
obtained from the line model described in \Sec 4.2 (in which
there is no intrinsic spatial correlation).
The thin and thick
lines are quite similar, indicating that the anticorrelation at small
velocities is entirely due to our algorithm for defining lines.
The only difference with
the line model appears to be that the simulation shows a larger spread
of the $\Delta v$ where the autocorrelation has a fixed value, as $W_t$
is varied. This may be caused by a correlation of weak lines with the
strong lines (which is expected from the infalling gas and shock-heated
gas in the $\lya$ absorbing systems, as described in \Sec 3.3), which
would increase the ``zone of avoidance'' around every strong line where
other strong lines cannot be found. However, the effect is small, and
the similarity of the curves with the line model emphasizes the
importance of working directly with the flux correlation function.
Unfortunately, we cannot investigate the correlations on large scales
due to the small size of the box.

\topinsert
\centerline{
\hskip -0.5truecm
\epsfxsize=3.0in \epsfbox{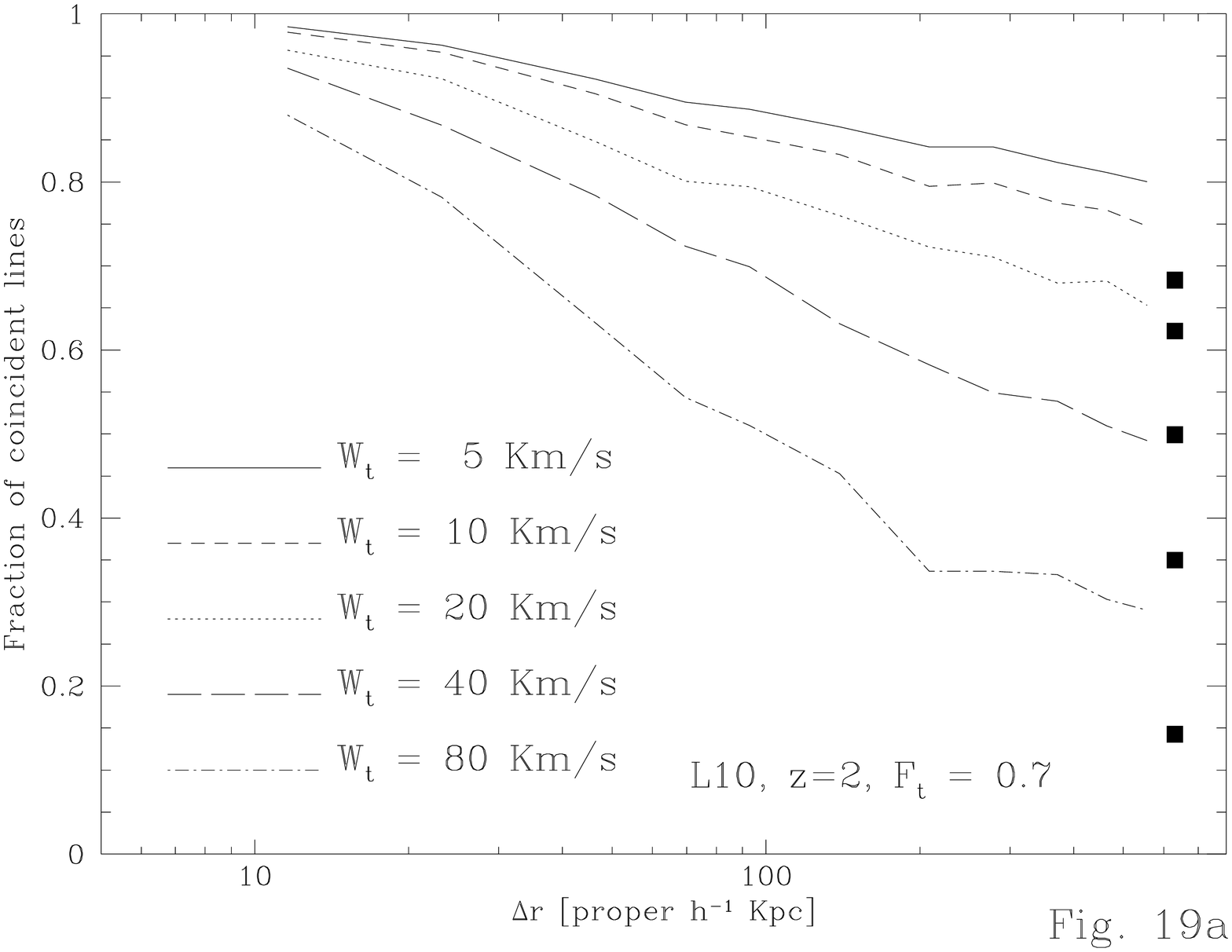} \hskip 0.3truecm
\epsfxsize=3.0in \epsfbox{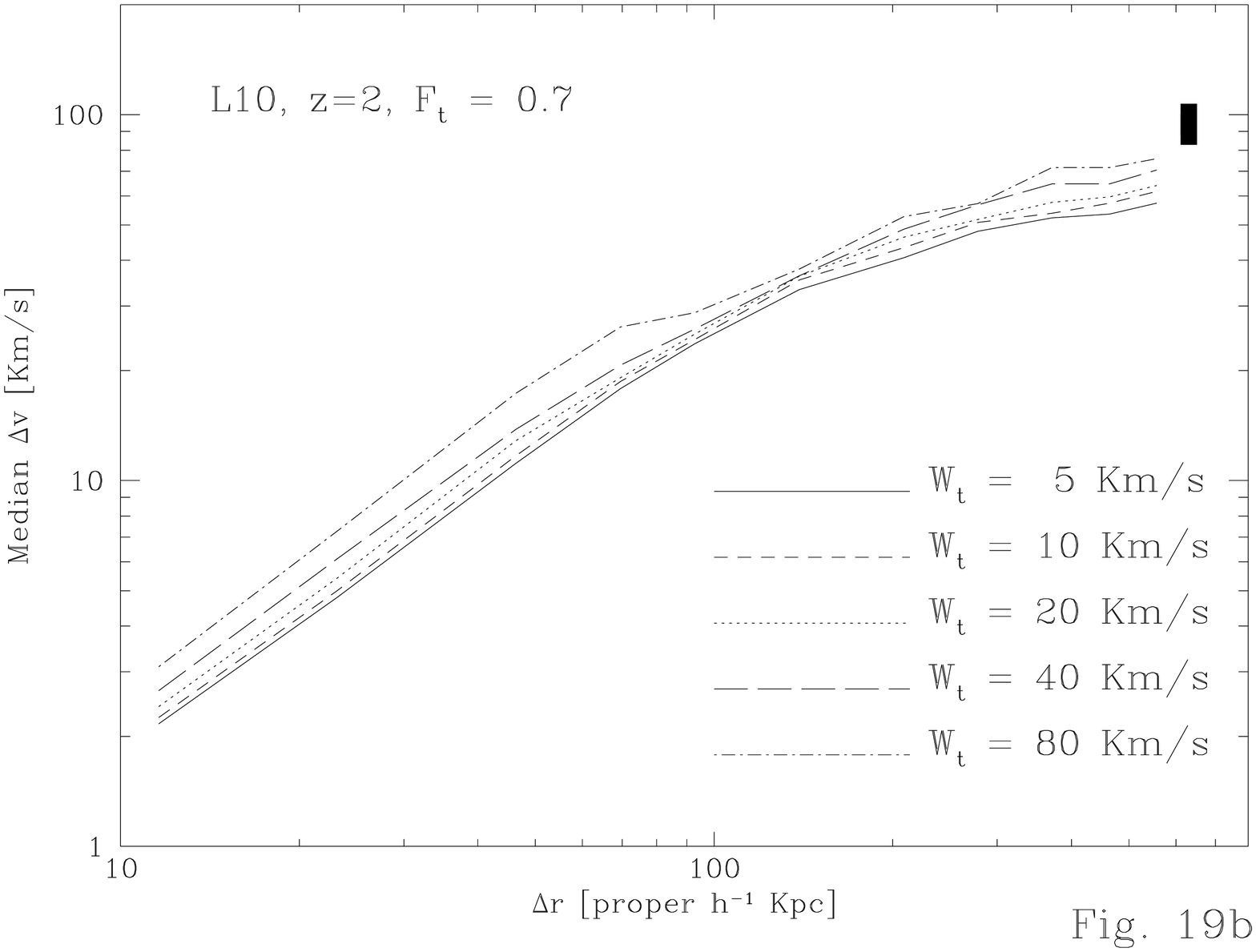}
}
\baselineskip=12truept \leftskip=3truepc \rightskip=3truepc
\noindent{\bf Fig. 19:} (a) Fraction of lines in a spectrum above the
equivalent width threshold $W_t$ which have a coincident line in the
spectrum from a parallel row at transverse distance $\Delta r$, also
with $W > W_t$, with a central velocity differing by less than $1/4$ of
the box size ($281 \kms$). Black squares indicate fraction of coincident
lines if they are placed randomly. Results are shown at z=2 to reduce
the number of chance coincidences. (b) Median of the velocity separation
$\Delta v$ between coincident lines. Black squares indicate the median
of randomly placed lines; these were computed by taking random pairs of
rows from the simulation and shifting them in velocity.
\endinsert

  Next, we compute the fraction of coincident lines above a given
equivalent width threshold on parallel lines of sight (i.e., the
fraction of the time that a line with $W > W_t$ is detected on one
line of sight when another line with $W>W_t$ has been detected on a
parallel line of sight at a transverse distance $\Delta r$). This is
shown in Figure 19a, again using $F_t=0.7$ for line detection, and at
$z=2$. In addition to the equivalent width thresholds, we require that
the lines are separated in velocity by less than 1/4 of the box size
($281 \kms$). We choose here the lower redshift because lines are much
more crowded at higher redshift (increasing the fraction of chance
coincidences), and because the observations have been done at low
redshift. The black squares on the right show the number of chance line
coincidences expected (obtained by computing the number of coincident
lines with any two random rows of the simulation, and shifting one of
two spectra by a random velocity). Most of the lines are coincident on
small scales, and the coincidences are reduced to about half the lines
(when chance coincidences are subtracted) on scales $\sim 100 h^{-1}
\kpc$. This is in qualitative agreement with the results of Bechtold
\etal (1994) and Dinshaw \etal (1994, 1995).

  We also plot in Figure 19b the median velocity separation of the
coincident lines as a function of $\Delta r$. The velocity of a line
is defined as the center of the interval between the two points where
$F=F_t$. Line pairs are identified
using the same conditions as above (if there is more than one line
satisfying these conditions, the one closest in velocity is chosen).
Black squares show the median velocity separation of the chance
coincident lines, obtained from random pairs of spectra as before.
The median velocity differences are unobservably small for small
separations (in agreement with the results of Smette \etal 1993), and
they grow to $\sim 30 \kms$ at $\Delta r = 100 h^{-1} \kpc$.

   Here again, we emphasize that measuring directly the
cross-correlation function of the flux is also a powerful method of
analysing the observations, in particular for comparison to these
simulations. As we have already seen, our simulations do not predict
correctly the number of high column density systems, which
are effected by optical depth considerations not treated
in this paper and which may be
associated with dense regions in halos and are responsible for the
large equivalent width lines. The weaker absorption lines tend to be
blended, and their cross-correlation depends on how they are
identified. Moreover, observations of the spectra of quasar pairs are
often of relatively low signal-to-noise, so that only strong lines can
be detected individually, but much information is present in the
flux cross-correlation in regions of low absorption, which can be
above the noise once it is averaged over a large region in the spectrum.

\subsec{4.9}{The HeII Gunn-Peterson Effect}

  The spectrum of a quasar below the rest-frame wavelength of the \heii
$\lya$ line was recently observed for the first time. The first
observation (Jakobsen \etal 1994) was consistent with no transmitted
flux, and could put an upper limit of 0.2 to the fraction of transmitted
flux. Later observations (Tytler \etal 1995; Davidsen \etal 1995) have
found a fraction of transmitted flux closer to $\sim 0.3$ at $z=3$.

  Several calculations have been made of the expected \heii flux
decrement from a population of clouds causing absorption lines with
Voigt profiles, given certain fits to the observed distributions of
column densities and b-parameters (Miralda-Escud\'e 1993; Jakobsen
\etal 1994; Madau \& Meiksin 1994). These calculations need to assume
an extrapolation of the column density distribution to very low column
densities, since the \heii decrement is dominated by the weakest lines
which can cover most of the spectrum; furthermore, they all assume that
the positions of the lines in the spectra are uncorrelated. In
addition, there is an uncertainty in the relation of the \hi and \heii
b-parameters, depending on whether the velocity dispersion is of
thermal or hydrodynamic origin.

  The simulations of the gas causing the $\lya$ forest allow us to make
a direct prediction of the \heii decrement, as a function of the ratio
of the ionization rate of \hi and \heii atoms, given by the spectrum of
the photoionizing background, circumventing all of the above
uncertainties. We assume that collisional ionization is negligible
for both the neutral hydrogen and the \heii, and that the clouds are
optically thin to a spatially uniform background of ionizing photons. In
that case, the ratio of \heii to \hi column densities is uniform through
the intergalactic medium and all the clouds, and is given by $N_{\heii}/
N_{\hi} = 1.7 (J_{\hi}/J_{\heii})$, and the ratio of the optical depths
given by any gas element at any position of the spectrum, without
including thermal broadening, is $\tau_{\hi}/\tau_{\heii} = 0.43\,
(J_{\hi}/J_{\heii})$ (Miralda-Escud\'e 1993). The spectra then need to
be convolved with thermal broadening with a width $b_{\heii} = 0.5\,
b_{\hi}$.

\topinsert
\centerline{
\epsfxsize=4.0in \epsfbox{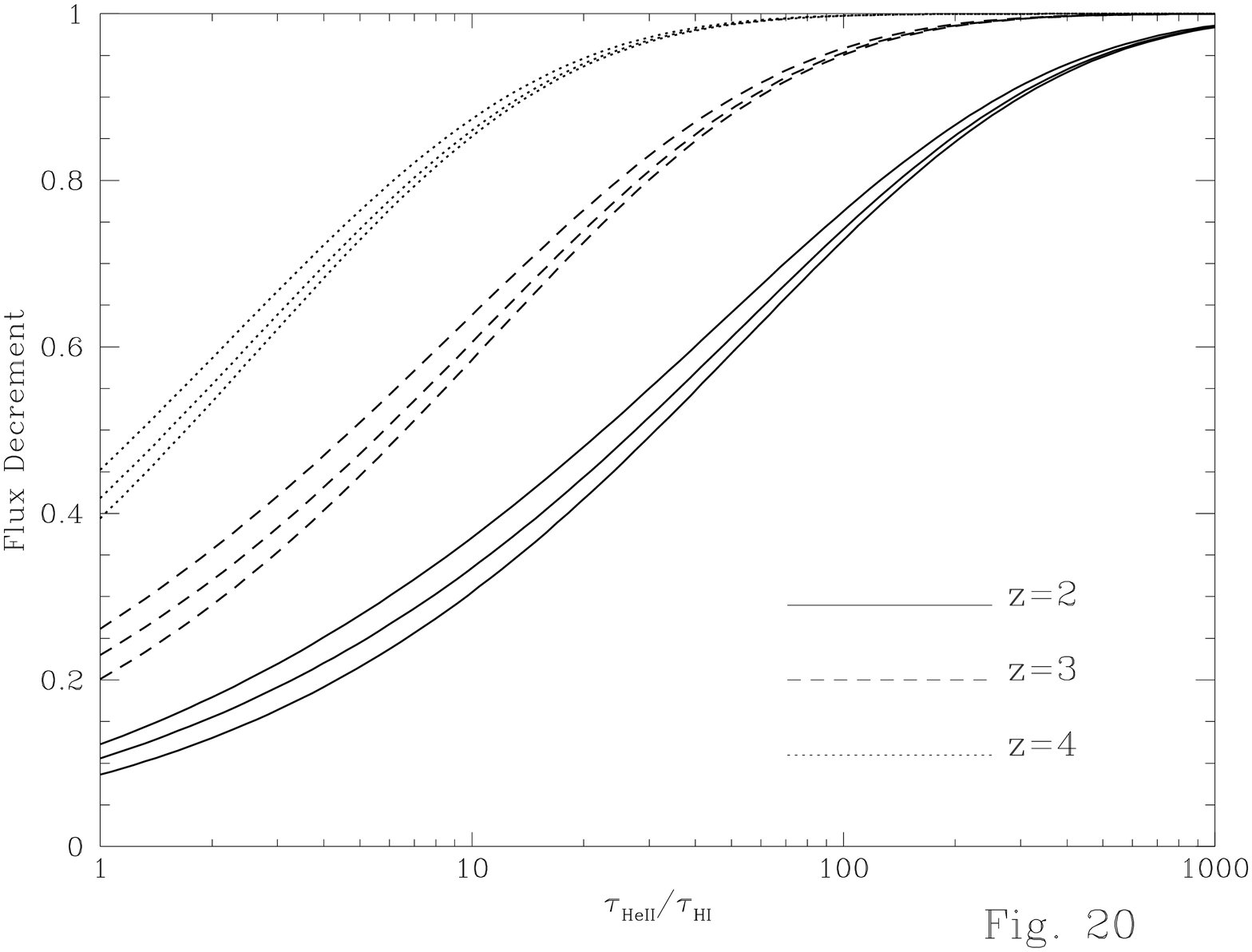}
}
\baselineskip=12truept \leftskip=3truepc \rightskip=3truepc
\noindent{\bf Fig. 20:} Average flux decrement as a function of a
factor $\tau_{\heii}/\tau_{\hi}$ by which the optical depth of the
spectra are multiplied. For each redshift, the upper line gives the
flux decrement obtained from spectra without including thermal
broadening, the upper line included thermal broadening for $\hi$, and
the middle line is for thermal broadening with half the velocity width,
appropriate for \heii. For fully ionized, optically thin clouds, and
neglecting collisional ionization (which are good approximations for
the low column density absorption lines dominating the contribution to
the average flux decrement), the middle lines predict the $\heii$ flux
decrement in terms of the spectral shape of the ionizing background,
where $\tau_{\heii}/\tau_{\hi} = 0.43 (\jhi/\jheii)$.
\endinsert

  The predicted \heii flux decrement from our L10 simulation is shown in
Figure 20, as a function of the ratio $\tau_{\heii}/\tau_{\hi}$. At each
redshift we plot three lines: the upper one is for spectra with the
thermal broadening of \hi (and therefore they are the same curves as in
Fig. 10), the middle curve is for the thermal broadening of \heii, and
the lower curve is for no thermal broadening. Reducing the thermal
broadening reduces the flux decrement, since more of the absorption is
concentrated into saturated lines, and any gaps between the lines are
left with less absorption. The small difference between the three curves
denotes the importance of hydrodynamic motions, especially in the weak
features of the spectrum. In particular, it implies that at the highest
flux decrements the absorption is being produced by intergalactic gas in
the voids, covering all the range of velocities in a spectrum (notice
from \Sec 3, Figures 7, that the range of velocities within systems
producing absorption lines is small).

  If the observed flux decrement is $0.7$ at $z=3$, then we need
$\tau_{\heii}/\tau_{\hi} \simeq 20$, which corresponds to
$J_{\hi}/J_{\heii} \simeq 50$.
{}From Figures 4a,b of Miralda-Escud\'e \& Ostriker (1992),
we see that this ratio is slightly higher than what is
predicted for a model of the ionizing background where the sources are
quasars with a power-law spectrum $F_{\nu} \propto \nu^{-1.4}$, by
about a factor of 2.5 . The
larger ratio of $J_{\hi}/J_{\heii}$ can easily be explained if
the intrinsic spectrum of quasars is slightly softer between the
ionization edges of \hi and \heii or if hot stars (which have a soft
spectrum) make a significant contribution.
In fact, because of the effects of
absorption by the $\lya$ clouds, the intrinsic spectrum of quasars
needs to change by less than a factor 2.5 in the ratio
$J_{\hi}/J_{\heii}$.
This is still consistent with observations of the
far UV and X-ray spectrum of quasars (Laor \etal 1994). The background of
the spectrum produced in our simulation has too large a ratio
$J_{\hi}/J_{\heii}$ (see Fig. 1), indicating that the chosen sources
described in \Sec 2 were too soft
and the reionization of \heii was not sufficiently advanced.
In any case, a \heii flux
decrement near 0.7 at $z=3$ is entirely consistent with the \heii having
been reionized by the expected sources of the ionizing background.


\sect{5. DISCUSSION}

  In \Sec 4, we have presented the predictions of the CDM$+\Lambda$
model we are examining for the $\lya$ forest, according to our
simulations, and we have seen that
they generally agree with the observed characteristics, although a
much more precise and exact comparison with observations should be done
in the future to establish quantitatively any differences of the
observed spectra with the predicted ones by this and other theories.
If it is indeed true that the gravitational collapse of structure from
primordial fluctations at high redshift is the origin of the $\lya$
forest, then we can derive several consequences for the average
baryon density of the universe and the distribution of the baryons in
the $\lya$ clouds. We analyze such implications in this Section.

\subsec{5.1}{Constraints on $\Omega_b$}

  In \Sec 4.1, we have defined the quantity
$$\mu^2 \equiv (\Omega_bh^2)^2/(h \jhi) ~, \eqno(\new) $$
which we have chosen to fit our predicted value of the average flux
decrement in $\lya$ spectra to the observed one.
We have then seen that this value leads to a correct prediction of the
number of absorption lines near $\nhi = 10^{14}\cm^{-2}$, and gives a
shape for the column density distribution that is similar to the
observed one. We have also examined the dependence of our predicted
average flux decrement and number of lines with the resolution and the
size of the periodic box we use in the simulation, and have found that
for fixed $\mu^2$, such effects can only be causing us to overestimate the
average flux decrement and number of lines. Thus, in the model we assume
in this paper the parameter $\mu^2$ should probably not be lower than we
have assumed because, as we see from Figure 10,
the predicted flux decrement is already on the low side
compared to observations, and it could only be made lower by the
limitations in dynamic range of the simulation. There is still the
possibility that a simulation evolved with a low value of $\jhi$ as
needed to obtain the correct flux decrement would contain cooler gas
that would lead to more absorption; unfortunately, we only have a
simulation for a high value of $\jhi$, and we have had to assume that
the only effect of lowering $\jhi$ is to increase the neutral fractions
as $\jhi^{-1}$.

  Determinations of the intensity of the ionizing background caused by
quasars (which include the observationally known absorption by $\lya$
clouds) give values of $\jhi \simeq 0.3$ at the redshifts we are
considering (e.g., Haardt \& Madau 1995), and the value could be much
larger owing to obscuration of quasars (see Fall \& Pei 1995) and to the
ionizing photons emitted by high-redshift galaxies.
At the same time, observations of the proximity effect favor an
intensity of the ionizing background of $\jhiu \simeq 0.3 - 1$
(Bechtold 1994 and references therein). If we require $\jhiu > 0.2$,
then we need $\Omega_b > 0.05$ for $h=0.65$, which is on the high side
of the estimates from primordial nucleosynthesis constraints.

  The parameter $\mu^2$ cannot be obtained from the observations of the
$\lya$ forest alone; the reason we can infer its value is that we use
the predictions of our simulation for the nature of the clouds. We now
ask what property of the absorbing systems is needed to infer $\mu^2$.
Let us consider that the absorption systems over some column
density range contain a fraction of the baryons $\Omega_c/\Omega_b$,
and have a filling factor $f$ of the volume in the universe. The gas
density in the systems is then proportional to $\Omega_c/f$, and the
density of neutrals is proportional to $(\Omega_c/f)^2/\jhi$. The
average distance between two absorbing systems along a line-of-sight,
$d$, is observationally determined, and the pathlength through them is
$d\, f$. Thus, the column density, which is also observed, is
proportional to $(\Omega_c^2 d)/(\jhi f)$, so the quantity that is
determined from observations alone is $\Omega_c^2/ (\jhi f)$. To infer
$\Omega_b^2/\jhi$, we need both the volume filling factor and the fraction
of baryons in the clouds. The reason why we infer a high value for
$\mu^2$ is that the filling factors in our theory are very large
(essentially, the weakest lines are caused by a fluctuating
Gunn-Peterson due to the intergalactic medium, which fills all the
volume), even though the fraction of baryons contained in the clouds is
very high, as we shall see below.

  The high value of $\mu^2$ that we require
could be changed by the effects of faster cooling with a low $\jhi$
mentioned above. Another possibility is that very thin slabs of cool
gas are generically formed between shocks, and they are not resolved in
our simulation. Thin slabs are always produced in one-dimensional
calculations of pancake-collapse (e.g., Shapiro \& Struck-Marcell 1985);
the question is whether they could still be important (and provide a
significant contribution to the neutral column densities) in
three-dimensional systems with small-scale structure in the dark matter.
Another possible effect is the presence of explosions, which could
reduce the filling factor of the photoionized gas owing to the presence
of hotter gas in the intergalactic medium. Basically, anything that
can increase the density of the gas contributing most of the observed
column density will reduce the required baryon content.

  A similar line of reasoning has been employed by Rauch \& Haehnelt
(1995) to derive an upper limit to the thickness of the clouds given an
upper limit to $\Omega_b\, h^2$ (using the reverse argument), and infer
a large axis ratio given the observation of the line coincidences
accross the line of sight. The clouds in our simulations have thicknesses
of the order of this upper limit; they certainly have relatively large
axis ratios, since for column densities typical of the $\lya$ forest the
absorbers originate in large-scale structures forming sheets and
filaments.


\subsec{5.2}{The Fraction of Baryons in the Intergalactic Medium}

  There is no absolute distinction in our model between the
intergalactic medium and the $\lya$ clouds. They both form a
continuous gas density field in space over the whole universe. Thus,
any separation of the intergalactic medium and the clouds necessarily
involves some arbitrariness, depending on some parameter which can be
varied. If we identify clouds as being the regions where the neutral
hydrogen density in real space is above a certain threshold (as in \Sec
3), then we can identify the gas outside this region as the
intergalactic medium. The cumulative distribution of the gas neutral
density then gives the fraction of gas in the intergalactic medium at
each isodensity contour. We do this in terms of the neutral density,
since this is what is observed in the $\lya$ spectra; if contours of the
total gas density were used, the results would not change too much,
because the neutral fraction depends mostly on density, and the
temperature tends to correlate well with density.

\topinsert
\centerline{
\hskip -0.5truecm
\epsfxsize=3.0in \epsfbox{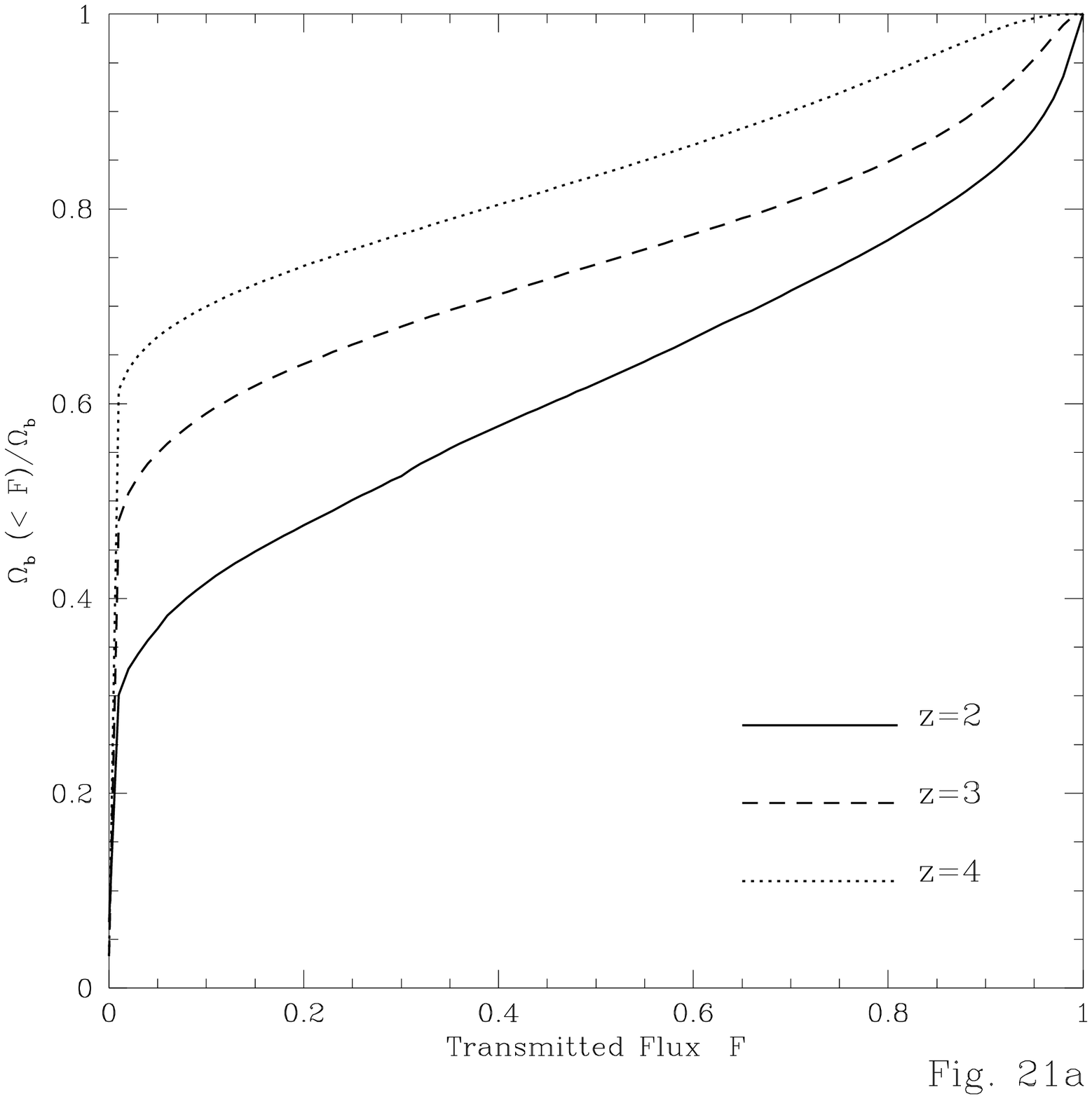} \hskip 0.3truecm
\epsfxsize=3.0in \epsfbox{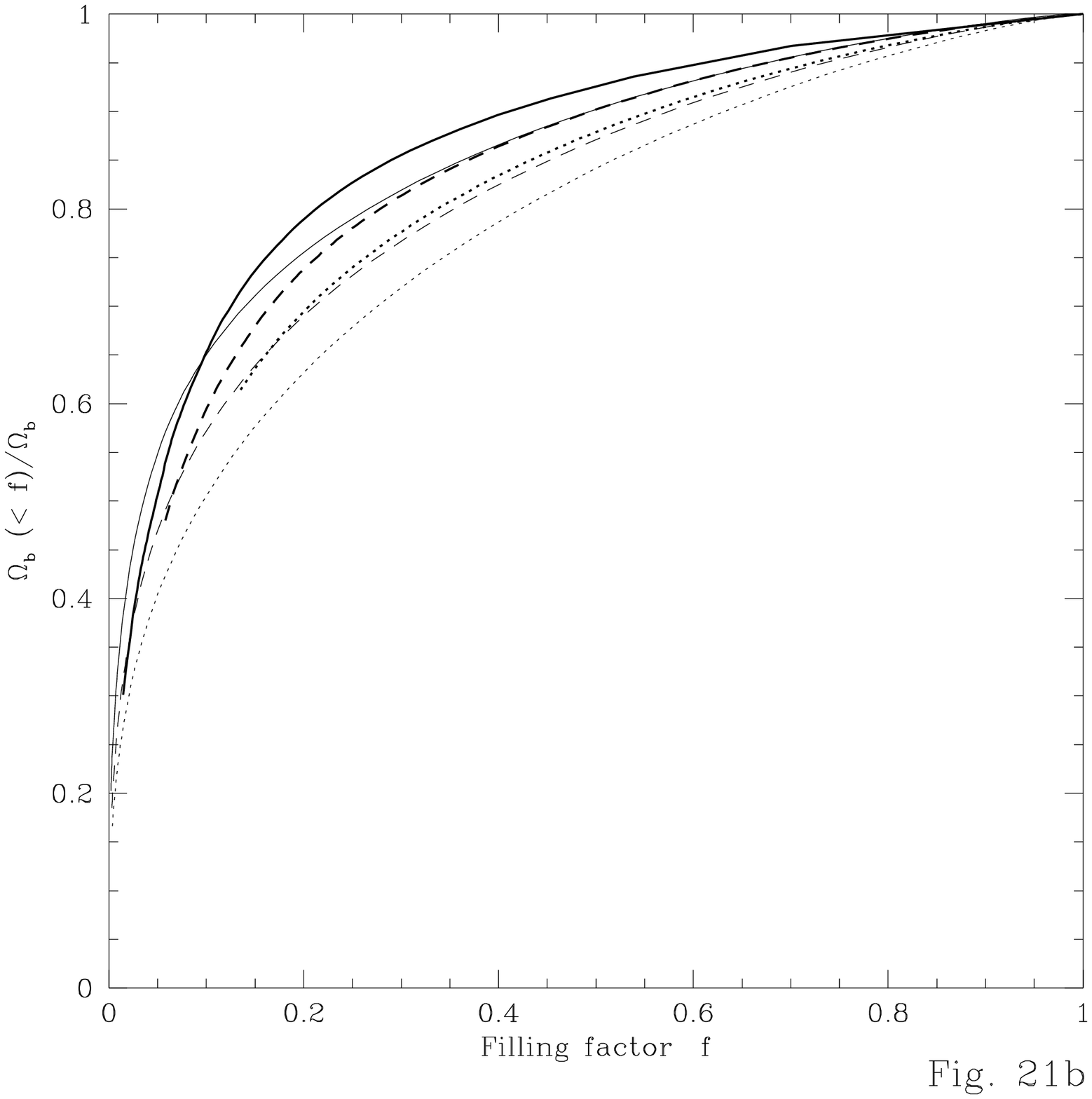}
}
\baselineskip=12truept \leftskip=3truepc \rightskip=3truepc
\noindent{\bf Fig. 21:} (a) Fraction of baryons located at a velocity
in the spectrum along random lines of sight where the transmitted flux
is below F, in the L10 simulation. (b) Same curves as in (a), as a
function of the filling factor in the spectra of regions below a fixed
transmitted flux threshold (thick lines). The thin lines give the
fraction of baryons in regions above a fixed density of neutral gas in
real space, as a function of the volume-filling factor of these regions.
\endinsert

  We can also identify clouds in the spectrum as in \Sec 4, defining
a fixed optical depth as a threshold to separate absorption lines and
the intergalactic medium. In this case, we can define the intergalactic
medium as the gas having a hydrodynamic velocity along
the line-of-sight which is not within a line (i.e., where the optical
depth is above the threshold value). While this concept is less related
to the physical properties of the gas (and, generally, any element of gas
may be considered as part of the intergalactic medium or part of an
absorption line depending on the line-of-sight that is chosen), it is
more related to what can be observed in a spectrum. Shown in Figure 21a
is the fraction of baryons outside absorption lines as a function of
the threshold in transmitted flux, at three different redshifts (all
quantities in this Section are for the L10 simulation and are obtained
from the same 18000 rows described in \Sec 4.4, with the same value
of $\mu^2$ used throughout the paper; the same
rows are used to calculate quantities from real space and from the
spectra). In Figure 21b, the thick lines are the same quantity plotted
in Figure 21a, but shown instead as a function of the filling factor in
the $\lya$ spectra of regions with transmitted flux below different
thresholds. The thin lines show the same fraction of baryons computed in
real space, as explained above.

  The evolution with redshift seen in Figure 21b is as expected: at
lower redshift, structures have collapsed to a more evolved stage, with
more baryons having moved into high density regions. Thus, the fraction
of baryons in the intergalactic medium, defined to have any constant
filling factor with redshift, decreases with time. However, we see from
Figure 21a that the fraction of baryons which are in regions of the
spectrum above a fixed transmitted flux is {\it increasing} remarkably
fast with time. For example, taking a flux threshold $F_t = 0.5$, the
fraction of baryons at velocities having less absorption is only $17\%$
at $z=2$, and it rises to $38\%$ at $z=4$. Of course, the filling
factor for such regions at $z=2$ is much larger than at $z=4$ (see
Fig. 11b); this is in fact the explanation why the gas appears to be
``moving'' to regions with less absorption. The gas actually moves, on
average, to regions of higher overdensity, but this evolution is
relatively slow and the more important effect is the expansion of the
universe (and the gas structures that are forming), which causes the
Gunn-Peterson optical depth to increase rapidly with redshift.

  We also see from Figure 21b that the fraction of baryons in the
$\lya$ spectra in regions of low absorption is lower than in real space.
This is simply due to the faster expansion of the voids compared to the
Hubble expansion. At low filling factors, the thick curves for the
$\lya$ spectra drop below the ones for real space owing to the effects
of thermal broadening and the velocity dispersion in collapsing systems.
This figure shows clearly that when weak absorption features are
detected in the $\lya$ spectra covering more than $\sim 50\%$ of the
spectrum, then such absorption must be attributed to the fluctuating
Gunn-Peterson effect from the intergalactic medium, since the gas
causing this absorption occupies an even larger fraction of the real
space volume. This must be true in any gravitational collapse theory
for the $\lya$ clouds. Of course, in alternative theories of
pressure-confined clouds the absorption lines could overlap over the
whole $\lya$ spectrum while filling a small fraction of the volume of
the universe, but such theories are in increasing conflict with
observations of the transverse sizes of the clouds (Bechtold \etal 1994,
Dinshaw \etal 1995).

  Our conclusions then are that the weakest absorption features
in the \hi $\lya$ spectra (e.g., Hu \etal 1995), as well as the observed
\heii decrement (Jakobsen \etal 1994; Tytler \etal 1995; Davidsen \etal
1995), and the \hi decrement in the highest redshift quasars (Schneider
\etal 1991) are most likely to be our first detections of the
intergalactic medium (since when the overall flux decrement is high,
then the lines contributing to the decrement must have a large filling
factor). These observations can still be interpreted alternatively in
other theories where the clouds are pressure-confined and still fill a
small fraction of the volume in real space when their absorption
profiles overlap to fill most of the spectrum. However, such a model
could not account for a correlation of the absorption fluctuations on
large scales. If the absorption fluctuations between lines are found to
be correlated on parallel lines of sight in a similar way as the strong
lines [see Fig.  13(a,b)], irrefutable evidence will have been obtained
for the detection of a photoionized intergalactic medium with density
fluctuations originating, at least partially, in gravitational collapse.

\subsec{5.3}{The Fraction of Baryons in the Absorption Lines}

\topinsert
\centerline{
\epsfxsize=4.0in \epsfbox{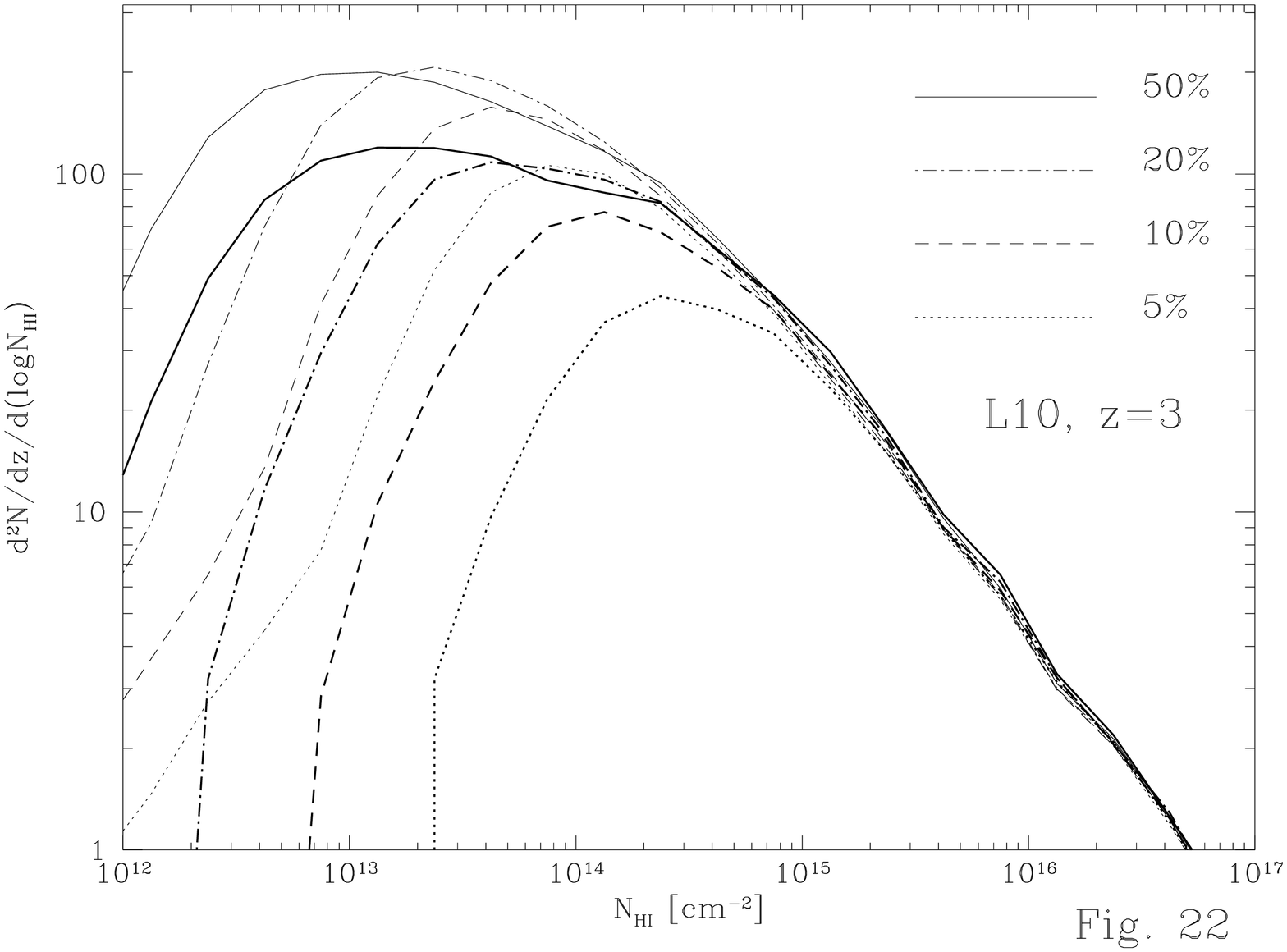}
}
\baselineskip=12truept \leftskip=3truepc \rightskip=3truepc
\noindent{\bf Fig. 22:} Column density distribution obtained from the
spectra and from real space. Thick lines are the same as in Fig. 15a,
thin lines are obtained from real space identifying clouds as intervals
along a line of sight above a fixed neutral density, such that the
volume-filling factor of these regions is the one specified in the figure.
\endinsert

  We now address the fraction of baryons which is in absorption lines of
different column densities in our simulation. All the baryons which are
not in the intergalactic medium for any fixed flux threshold, as defined
above, and have not been transformed into stars, have to be part of
absorption lines. First, we show in Figure 22 the \hi column density
distribution at different filling factors. The thick lines are the same
as shown in Figure 15a, and the thin lines is the column density
distribution obtained in real space, as in \Sec 3. The curves differ
only at low column densities, with more weak lines being found in real
space which are caused by small regions with the neutral gas density
above the threshold; such regions tend to merge with bigger lines in the
spectra, mostly due to the thermal broadening.

\topinsert
\centerline{
\hskip -0.5truecm
\epsfxsize=3.0in \epsfbox{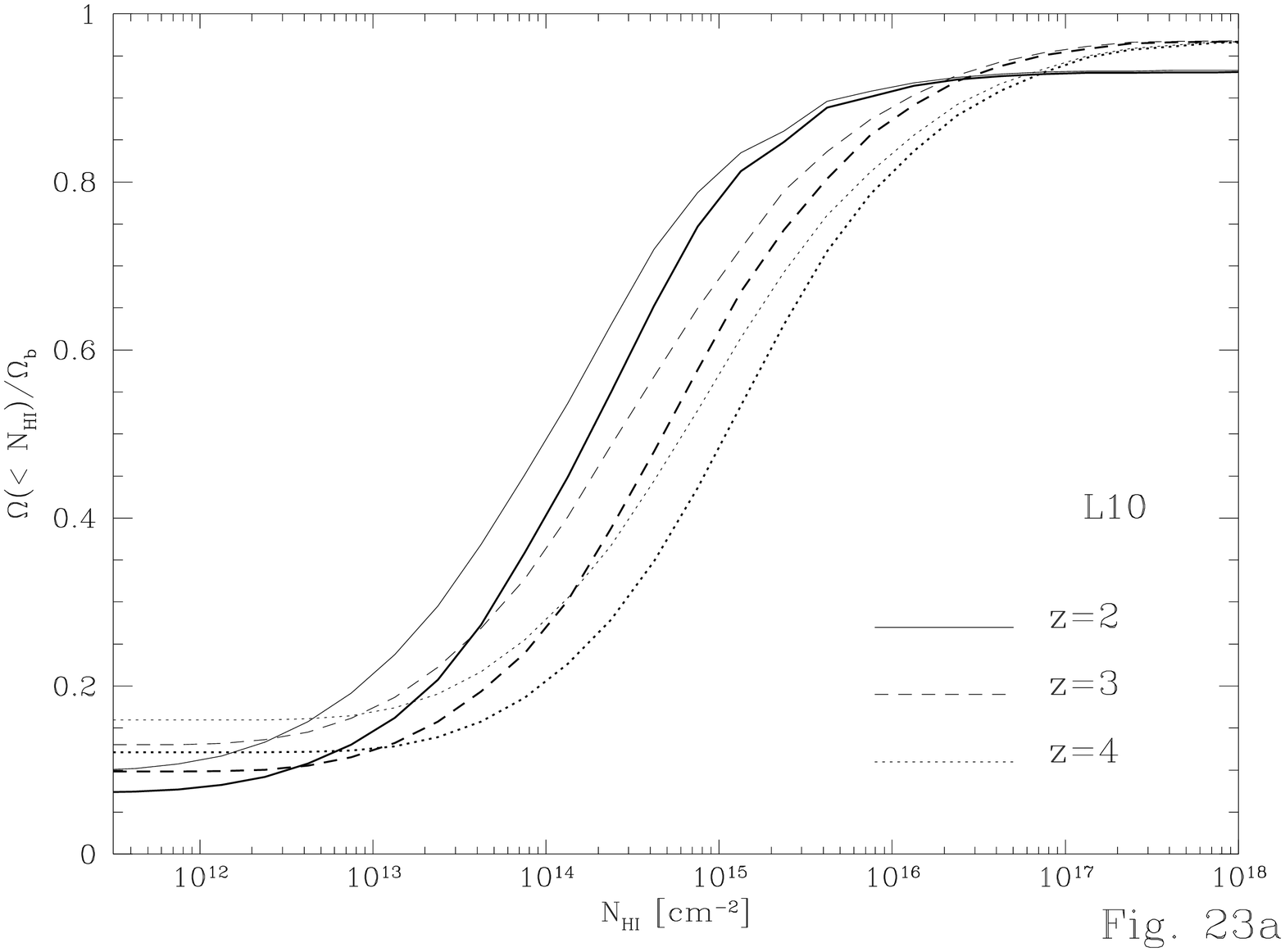} \hskip 0.3truecm
\epsfxsize=3.0in \epsfbox{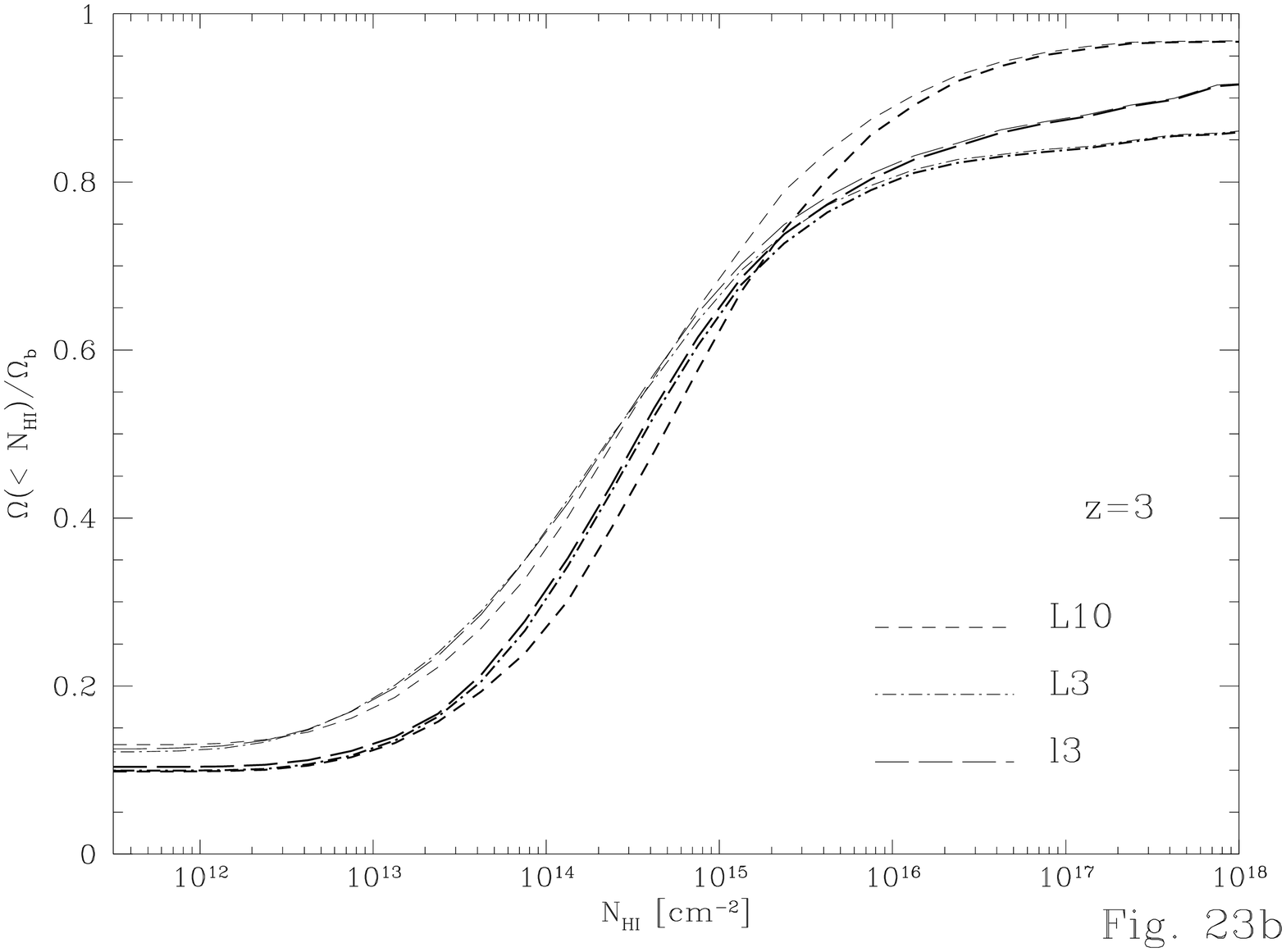}
}
\baselineskip=12truept \leftskip=3truepc \rightskip=3truepc
\noindent{\bf Fig. 23:} Fraction of baryons contained in lines below
a column density $\nhi$, for clouds identified in real space (thin
lines) and in the spectra (thick lines), using in both cases a filling
factor of 50\%.  At low column densities, the curves approach the
fraction of baryons which are left in the remaining 50\% of the volume
in real space, or in the spectra, while at high column densities the
difference of $\Omega/\Omega_b$ from unity is the fraction of baryons
turned into stars (however, there is also a Poisson error due to the
limited number of high column density systems in the set of 18000
spectra used to do the calculation).

\endinsert

  The cumulative fraction of baryons in absorption lines of different
column density is shown in Figure 23a. Again, thick lines show the
results obtained from the simulated spectra, whereas thin lines are for
clouds defined in real space. We fix the filling factor of the
``clouds'' to 50\% in both cases. The results are shown at three
different redshifts, with the same line types as in Figure 21a. At the
lowest column densities, the curves start at a value of
$\Omega/\Omega_b$ which indicates the fraction of baryons in the
``intergalactic medium'', the 50\% of the spectra (or real space) which
is left outside the clouds. At the high end, the value of
$\Omega/\Omega_b$ does not reach unity because of the fraction of
baryons that have turned into stars according to our algorithm (see \Sec
2), although there is also Poisson noise due to the small number of high
column density systems found in the 18000 spectra we have used. In
practice, this fraction of the baryons could also have been stabilized
in rotating disks and be observed as a larger number of Lyman limit and
damped absorption systems (see Katz \etal 1995), or could be in the
form of molecular clouds in such galactic gaseous disks.

  We also show in Figure 23b the same curves for the L3 and l3
simulations ({\it dash-dot and long dashed lines}), all at $z=3$. More
gas resides in low column density lines in the small box simulations,
due to the absent large-scale power. However, the biggest difference
is at high column densities: much of the gas in lines with $\nhi \sim
10^{16} \cm^{-2}$ in the L10 simulation has moved to higher column
density systems or been turned into stars in the L3 and l3 simulations.
This should be related to the higher resolution of the small box
simulations (notice the difference between L3 and l3), although the
large-scale power could also be important: it may be more difficult for
the gas to move from filamentary structures to the higher density halos
when it continues to be shocked as the filaments merge with each other
(a process which has almost stopped in the small box simulations by
$z=3$).

  In our model, {\it about 80\% to 90\% of the baryons in the universe
are in the $\lya$ clouds}, with practically no change from $z=4$ to
$z=2$. The majority of them are in clouds with column densities over
the range $10^{14}$ to $10^{15.5}\cm^{-2}$. The typical column
densities decrease by a factor $\sim 5$ from $z=4$ to $z=2$, as a
consequence of the expansion of the clouds and the progressive heating
of the gas as larger scale structures collapse. The higher temperatures
(see Fig. 9b) reduce the recombination coefficient and increase
collisional ionization, decreasing therefore the neutral column
densities.

This result, that most of the baryonic mass in the universe is in
features giving column densities of $\nhi = 10^{14}-10^{15}\cm^{-2}$
is in apparent contrast to the well known result that most of the
{\it neutral} gas is observed to be in very high column density systems
(the damped Lyman alpha systems associated with proto galaxies)
having  $\nhi > 10^{20}\cm^{-2}$.
The reason is simply the high state of ionization
of most of the gas in the low column density systems
and the increasing neutral fraction as a function of column density
shown in Figure 9a.

\subsec{5.4}{Origin of the Column Density Distribution}

  Having seen the fraction of baryons in clouds of different column
density, we shall now give an interpretation of the form of the column
density distribution we have found from the simulation. As we see from
Figure 22, the column density distribution in real space is very close
to the one obtained from the simulated spectra.

  The neutral column density of a region of overdensity $\delta$ and
length $L$ is proportional to $\delta^2\, L\, T^{-0.7}$, where $T$ is
the gas temperature and is included to reflect the variation of the
recombination coefficient. If $\Omega_c/\Omega_b$ is
the fraction of baryons in such ``clouds'', then their filling factor
should be $f=(\Omega_c/\Omega_b)\, \delta^{-1}$, and the number of
clouds found per unit length along a line of sight is
$$ dN / d\log\nhi = f/L = {\Omega_c \over \Omega_b}\, (\delta\, L)^{-1}
\propto (\Omega_c / \Omega_b) (\nhi\, L\, T^{0.7})^{-1/2} ~. \eqno(\new) $$
Let us apply this to some special cases. In the case of an isothermal
halo, where the gas density varies as $L^{-2}$ and the neutral column
density as $L^{-3}$, and $\Omega_c(\nhi)/\Omega_b \propto L \propto
\nhi^{-1/3}$, we obtain the well-known result that
$dN/d\log\nhi \propto \nhi^{-2/3}$ (e.g., Rees 1988); in general, for
spherical halos with the gas density $\rho \propto L^{-\alpha}$, we have
$dN/d\log\nhi \propto \nhi^{2/(1-2\alpha)}$.

  At column densities where the typical structures are filamentary in
nature, the scales accross the systems, $L$, are approximately
independent of the column density accross them. The filaments have
column densities up to $\nhi \sim 10^{15}\cm^{-2}$, and we see from
Fig. 23a that the baryon content is approximately constant in the
range $10^{14}\cm^{-2}$ to $10^{15}\cm^{-2}$. We then expect
$dN/d\log\nhi \propto \nhi^{-0.5}$, as is indeed found.
The pathlength $L$ starts decreasing at higher column densities, when
the systems have a more spheroidal geometry. In addition, at column
densities above $10^{15} \cm^{-2}$,
the fraction of baryons in clouds per unit $\log\nhi$ starts
declining, and the temperatures increase (Fig. 9c). All these effects
result in the steepening of the column density distribution above
$10^{15}\cm^{-2}$.
At low column densities the distribution becomes shallower, due to
the small baryon content of low-column density sheets.
Although a large number of weak lines are inferred when deblending
methods are used, most of these lines are in fact not real systems
but are only needed to fit the outer profiles of stronger lines.

\subsec{5.5}{Relation to Galaxy Formation}

The densest regions (overdensity $\ge100$)
of the inter-connecting network of baryonic
matter distribution are cooling and collapsing,
with velocity dispersions typical
of dwarf galaxies at present ($v \sim 50 - 100 \kms$).
The gas in these regions may either start forming stars,
or form relatively stable gaseous disks in the more massive regions
that could be responsible for the damped absorption systems,
and could later merge to form
present-day galaxies (see Katz \etal 1995).

It is unavoidable that, during the galaxy formation process,
absorption systems should be produced. Present galaxies are surrounded
by dark matter halos, and in order for the baryons to have
separated out and accumulated
in the central luminous parts, gas must have dissipated and flown in
through the halos at some point in the past.
During this process, gas
is shock-heated to the galactic virial temperatures, and when radiative
cooling is effective in free-falling gas temperature perturbations
are amplified (Field 1965) and one will naturally form
pressure-confined, photoionized clouds embedded in hotter gas (e.g.,
Fall \& Rees 1985).
Such clouds may be responsible for some of the
high column density systems that we do not reproduce in our simulation.
In addition, our treatment of the ionizing radiation does not
allow proper radiative transfer (rather, we assume
that the radiation field is uniform across the simulation box), so
we must have underestimated column densities of systems whose optical
depths approach unity ($\nhi \sim 10^{17}$).
These two effects can probably explain the observed number of
Lyman limit systems, and they might also contribute significantly to the
observed damped $\lya$ systems.

Regardless of whatever uncertainties might exist in the simulations,
what is clear from the results of this paper is that
the history of galaxy formation is closely related to the $\lya$
absorption systems. Any gas
which has not been incorporated into galaxies at high redshift should
be in the $\lya$ clouds, and different theories on the rate at which
gas can cool and form galaxies would lead to different predictions on
the number of lines of different column densities.
The agreement of our simulation with the observed absorption lines
at the column densities where most of the baryonic material is contained
suggests that, indeed, most of the baryons had not collapsed into
galaxies at $z\sim 3$. In fact, if more baryons had collapsed in very
dense regions the problem of accounting for the number of $\lya$ forest
lines given the lower limits on $\jhi$ would be made much more severe.

In several studies of galaxy formation, it was found that small
galaxies at high redshift accreted a large fraction of the gas, implying
that present galaxies could only have formed from mergers of a large
number of small galaxies (see White \& Frenk 1991; Kauffmann, White,
\& Guiderdoni 1993; Lacey \etal 1993).
This has been referred to as the ``overcooling problem''.
Part of the problem may be related to the use of simple Press-Schechter
models, where it is assumed that much of the gas collapses into dense
clumps after the first generation of objects where the gas can collapse
is formed. This is not borne out by our simulations, which imply a
slow depletion of the gas from the intergalactic medium (see Fig. 21b),
and suggest a picture where much of the gas stays in filamentary
structures which merge with each other to form larger-scale filaments
before much of the gas has a chance to collapse to very dense clumps.

A solution that has been investigated is the effect of
photoionization on slowing or reversing
the cooling rate and hence stablizing the collapse on small scales
(Efstathiou 1992; Quinn, Katz, \& Efstathiou 1995; Steinmetz 1995).
In all the simulations we have performed for a variety of models
(see Ostriker \& Cen 1995 and references therein),
we find that only a small fraction (2-10\%)
of all the baryons has collapsed into compact objects (stars/galaxies)
at present. In fact, we found that most of the gas resides in two major
components with comparable masses: ``Voids" and ``Hot IGM".
The former corresponds to the $\lya$ clouds that we are finding
here but the latter (with temperatures in the range
$10^5-10^7\kelvin$ and intermediate densities) is a
totally separate and new component.
This may be affected by the relatively poor
resolution of our previous simulations,
but our new simulations with higher resolution,
including those presented in this paper,
yield results in terms of collapsed fraction
consistent with our previous findings.

We suspect that the current popular variants of the CDM model
will yield the collapsed baryonic fraction
consistent with observations,
while HDM-like models and PBI-like models will perhaps
produce too small and too large a fraction of collapsed baryons, respectively.
But higher resolution simulations of these models
with proper handling of the radiative processes
are the only means to give us a definitive answer.

\sect{6. CONCLUSIONS}

  The model we have investigated here is one of several variants
of the basic Cold Dark Matter scenario
spawned after the COBE observations showed that the simplest CDM model,
while close in many respects to reality,
could not be correct in detail
(\cf Efstathiou 1991, Ostriker 1993).
This particular variant is not unique in its ability to fit all known
observational constraints (\cf Ostriker \& Steinhardt 1995),
but, we stress it was adopted for reasons having nothing to do with
observations of the Lyman alpha forest.
Bond \etal (1988)
argued at a very early time, on the basis of quite general
considerations, that this type of model should produce absorption
line features roughly like those
observed in the Lyman alpha forest, and
we find that contention to be correct.

  In massive and detailed simulations of the evolution of a CDM$+\Lambda$
model we find that the normal physics of self-gravitating fluids
containing dark matter, baryons with a primeval composition
and ionizing radiation will generate an interlocking structure
of ``sheets", ``filaments" and ``clumps" surrounding
``voids" of growing size.
Once the gas is photoionized and photoheated, the thermal energy
(as symbolized by the Jeans length and Jeans mass)
smoothes out fluctuations on small scales in the gaseous component
(but not of course in the dark matter component).

The weak shocks induced by the flows in gas are regions of moderate
overdensity $10^{0.5}-10^{1.5}$
which correspond to overdensities in neutral hydrogen of $10^{1.0}-10^{3.0}$
and column densities
of $10^{14}-10^{15}$.
Such systems contain most of the baryons at redshifts $z=4$ to $z=2$.
The somewhat smaller fraction of gas
in the voids produces a fluctuating Gunn-Peterson effect.

Gas drains along the filaments, accumulating in higher density
systems (where presumably galaxies form).
The relatively small fraction of the gas ($\sim 10$ \%)
associated with these systems is not computed accurately by our
simulations due to a lack of resolution at high density (as well as
other physical effects we do not include, such as self-shielding).
Consequently, we do not expect our predictions for absorbers with
$\nhi \gta 10^{16}\cm^{-2}$ to be accurate, and indeed the
number of them obtained in the simulations is significantly lower than
observed.
But, for the bulk of the gas in the voids and in the low and moderate
column density systems, our numerical methods
are probably accurate enough to fairly predict what our adopted model would
imply.

First we must adjust the parameter $\mu \equiv \Omega h^2/(\jhi h)^{1/2}$
to match observations. This is the only parameter we have adjusted
on a {\it post hoc} fashion.
If we take $\Omega_b h^2$ to be the value given
by standard nucleosysthesis arguments
($\Omega h^2 = 0.0125\pm 0.0025$; Walker \etal 1991),
then we require
$\jhiu h = 0.1\times 0.65=0.065$.
As we have discussed, this is too low compared to the lower limit of
$\jhiu$ obtained from the abundances of the observed quasars, indicating
that our model requires a slightly larger $\Omega_b$ than the upper limit
from primordial nucleosynthesis.

  With the parameter $\mu$ adjusted as described,
we find that there is an excellent agreement between the
computations and the observations.
The typical feature that produces a Lyman alpha cloud
is gas that collapsed to form mini-pancakes and filaments with collapse
velocities of several tens of km/s.
It is observed in its post-shock phase having an overdensity of
$\sim 10^1$, a neutral fraction of $\sim 10^{-5}$,
a Doppler $b$ parameter of $\sim 25$km/s
(produced by a similar contribution of hydrodynamic motions and thermal
broadening)
and an extent along the line of sight of $\sim 100$kpc.
The gas is typically expanding close to the Hubble rate for systems
with $\nhi \lta 10^{15} \cm^{-2}$.
Neutral fraction increases from
$10^{-5.3}$ at $\nhi=10^{13}\cm^{-2}$
to
$10^{-4.0}$ to at $\nhi=10^{16.0}\cm^{-2}$,
with the characteristic gas temperature increasing from $10^{4.3}$
to
$10^5$ in that same range of column densities. The temperatures increase
with time as
shocks from longer waves produce higher velocities.

We find approximately the correct redshift dependence of Lyman alpha
forest and attribute it very simply to the same redshift dependence
of the classical Gunn-Peterson effect.
Roughly speaking, each feature in the low column density $\lya$ forest
preserves its characteristics, but its opacity changes by approximately
$(1+z)^{9/2}$, as expected for comoving, photoionized gas in
a nearly constant radiation field.

The correlations in spectral features both along and perpendicular
to the line of sight are consistent with observations, with
detailed predictions testable only when simulations and real
data are analyzed in the same way.
The distribution of equivalent widths is also as observed,
with the same caveats as mentioned above.
We do predict a broad distribution of $b$ values and a weak
correlation of $b$ with $\nhi$, which should allow a detailed check
of the overall scenario. The ratio of the \heii to \hi flux decrements
at different redshifts is predicted in terms of the spectral shape of
the background, and we find that a model where quasars are the emitting
sources is consistent, although softer sources making a contribution to
the background similar to quasars are also possible.

  These results, and the fact that other models of structure
formation by hierarchical clustering (HKWM and Zhang \etal 1995)
give results that are similar to ours suggest that the value required
for the parameter $\mu$ could provide a reliable measurement.
The main reason why we obtain a very large baryon content in the $\lya$
forest, even when choosing the lowest reasonable value for $\jhiu$,
is the large volume-filling factor of the clouds
implied by our models.
The only way to decrease our estimate of $\Omega_b^2/\jhi$ from the $\lya$
forest observations is to reduce the volume-filling factor (or increase
the cloud overdensities). This could be the case if gas was able to cool
to lower temperatures and contract to higher densities than in our
simulation, forming thin slabs in the center of the
filamentary and sheet-like structures which we might not resolve.
This will be investigated in more detail in Gnedin \etal (1995).
If such thin slabs were important, they would probably produce narrow
components in the line profiles and, at the same time, any tendency to
fragmentation on small scales should cause small-scale differences in the
line profiles on parallel lines of sight.
In fact, it seems difficult
for this interpretation that clouds with high densities could be
correlated over large transverse scales, as implied by the
observations of Bechtold \etal and Dinshaw \etal, and practically not
change their line profiles
on scales of a few $\kpc$ (Smette \etal 1992). Further observations of
the degree of similarity of the absorption profiles in lensed quasars
are essential to make a strong case for the large baryon content of the
clouds.

  A value of $\Omega_b$ as large as required by our models would be
consistent with a baryon fraction $\sim 15\%$ in clusters (e.g., White
\& Fabian 1995) for the values of our adopted model
$\Omega=0.4$ and $h=0.65$. This would then exacerbate the problem
of the missing baryons at the present time. One possibility is that
a large fraction of baryons are still in the intergalactic medium,
forming a present-time ``$\lya$ forest'' on much larger scales and
much higher temperatures ($T\sim 10^5-10^6\kelvin$) than at high redshift.
This would yield very weak neutral hydrogen absorption lines owing to
the high degree of collisional ionization. Only the low temperature
tail of the gas would be seen in the low-redshift $\lya$ spectra.
Given the small density of baryons observed in galaxies and the X-ray
emitting gas ($\Omega_b \simeq 0.003$; Persic \& Salucci 1992),
a very large (perhaps implausible)
fraction of the baryons would have to be in such low-density structures.
The missing baryons might also have formed low-mass stars in galaxy
halos (e.g., Fabian, Nulsen, \& Canizares 1991),
might be in low surface brightness, undetected galaxies
(e.g., McGaugh 1994), or might reside in low surface brightness stellar
halos around normal galaxies and in clusters of galaxies (Sackett \etal
1994).

%

\bigskip
\bigskip

We thank John Bahcall, Mike Fall, Piero Madau, Martin Rees, and
David Weinberg for stimulating discussions, and Bob Carswell,
John Webb, and Andrew Cooke for providing us with the latest version of
their profile fitting software and very helpful discussions and advice
on its usage.
The simulation was performed on Convex-3880 in NCSA with
time allocation provided by M. Norman.
RC and JPO are supported by NASA grant NAGW-2448 and
and NSF grants AST91-08103 and ASC-9318185;
JM gratefully acknowledges support from the W. M. Keck Foundation, and
from NASA grant NAG-51618. MR is supported by NASA through grant
HF-01075.01-94A from the Space Telescope Science Institute, which is
operated by the Association of Universities for Research in Astronomy, Inc.,
under NASA contract NAS5-26555.

Finally, we would like to add a note
on the availability of the simulations presented in this paper
to the larger cosmological/astrophysical community.
We are making the synthetic spectra generated
from these simulations
available freely to users in the community.
The simulated quasar spectra can be found on the Web page
"http://astro.princeton.edu/~cen/LYA/entry.html".
\vfill\eject

\parskip .0cm plus .03cm
\sect{REFERENCES}

\apj{Arons, J., 1972}{172}{553}
\apj{Arons, J., \& Wingert, D. W. 1972}{177}{1}
\apj{Bahcall, J.~N., \& Spitzer, L. 1969}{156}{L63}
\apjsup{Bechtold, J. 1994}{91}{1}
\apj{Bechtold, J., Crotts, A. P. S., Duncan, R. C., \& Fang, Y. 1994}
{437}{L83}
\apj{Bi, G. 1993}{405}{479}
\apj{Bond, J. R., Szalay, A. S., \& Silk, J., 1988}{324}{627}
\apj{Carswell, R. F., Lanzetta, K. M., Parnell, H. C., \& Webb, J. K.,
1991}{371}{36}
\apjsup{Cen, R. 1992}{78}{341}
\apj{Cen, R., Gnedin, N.Y., \& Ostriker, J.P, 1993}{417}{387}
\apj{Cen, R., Miralda-Escud\'e, J., Ostriker, J.~P., \& Rauch, M. 1994}
{437}{L9({\bf Paper I})}
\apj{Cen, R., \& Ostriker, J. P, 1992}{399}{L113}
\apj{\refrule~ 1993a}{404}{415}
\apj{\refrule~ 1993b}{417}{415}
\apj{Charlton, J. C., Salpeter, E. E., \& Hogan, C. J. 1993}{402}{493}
\mnras{Cristiani, S., D'Odorico, S., Fontana, A., Giallongo, E.,
\& Savaglio, S. 1995}{273}{1016}
\refbook{Davidsen, A., et al. 1995, preprint}
\apj{Dinshaw, N., Impey, C. D., Foltz, C. B., Weymann, R. J., \& Chaffee,
F. H. 1994}{437}{L87}
\nature{Dinshaw, N., Foltz, C. B., Impey, C. D., Weymann, R. J., \& Morris,
S. L. 1995}{373}{223}
\apj{Edelson, \& Malkan 1986}{308}{59}
\refbook{Efstathiou, G. 1991, Physica Scripta, T36, 88}
\mnras{\refrule~ 1992}{256}{43p}
\mnras{Efstathiou, G., Bond, J. R., \& White, S. D. M. 1992}{258}{1p}
\aarev{Fabian, A. C., Nulsen, P. E. J., \& Canizares, C. R. 1991}{2}{191}
\apj{Fall, S.~M., \& Pei, 1993}{402}{479}
\refbook{\refrule~ 1995, in
{\it QSO Absorption Lines}, Proc. ESO
Workshop, ed. G. Meylan (Heidelberg: Springer), p. 23}
\apj{Fall, S.~M., \& Rees, M.~J. 1985}{298}{18}
\apj{Fardal, M.~A., \& Shull, J.~M. 1993}{415}{524}
\apj{Field, G. 1965}{142}{153}
\mnras{Fransson, C., \& Epstein, R. 1982}{198}{1127}
\refbook{Gnedin, N., et al. 1995, in preparation}
\apj{Gunn, J.~E., \& Peterson, B.~A. 1965}{142}{1633}
\refbook{Haardt, F., \& Madau, P. 1995, submitted to ApJ (astro-ph 9509093)}
\refbook{Harten, 1984, J. Comp. Phys., 49, 357}
\refbook{Hernquist, L., Katz, N., Weinberg, D.~H., \& Miralda-Escud\'e,
J. 1995, ApJLet, submitted ({\bf HKWM})}
\aj{Hu, E. M., Kim, T.-S., Cowie, L. L., Songaila, A., \&
Rauch, M. 1995}{110}{1526}
\asps{Ikeuchi, S. 1986}{118}{509}
\apj{Ikeuchi, S., \& Ostriker, J. P. 1986}{301}{522}
\nature{Jakobsen, P., Boksenberg, A., Deharveng, J.~M., Greenfield, P.,
Jedrzejewski, R., \& Paresce, F. 1994}{370}{35}
\apj{Jenkins, E.~B., \& Ostriker, J.~P. 1991}{376}{33}
\refbook{Katz, N., Weinberg, D.~H., Hernquist, L., \& Miralda-Escud\'e, J.
1995, ApJLet, submitted}
\mnras{Kauffmann, G., White, S. D. M., \& Guiderdoni, B. 1993}{264}{201}
\apj{Kofman, L., Gnedin, N., \& Bahcall, N. 1993}{413}{1}
\mnras{Lacey, C. G., Guiderdoni, B., Rocca-Volmerange, B., \& Silk, J.
1993}{402}{15}
\apj{Laor, A., Fiore, F., Elvis, M., Wilkes, B.~J., \& McDowell, J. C. 1994}
{435}{611}
\apj{Lu, L., Wolfe, A. M., \& Turnshek, D. A. 1991}{367}{19}
\apj{Madau, P., \& Meiksin, A. 1994}{433}{L53}
\apj{McGaugh, S. S. 1994}{462}{135}
\apj{Miralda-Escud\'e, J., \& Ostriker, J.~P. 1992}{392}{15}
\mnras{Miralda-Escud\'e, J. 1993}{262}{273}
\mnras{Miralda-Escud\'e, J., \& Rees, M. J. 1993}{260}{624}
\mnras{\refrule~ 1994}{266}{343}
\apj{Morris, S. L., \& van den Bergh, S. 1994}{427}{696}
\apj{Murdoch, H. S., Hunstead, R. W., Pettini, M., \& Blades, J. C.
1986}{309}{19}
\araa{Ostriker, J. P. 1993}{31}{689}
\refbook{Ostriker, J. P., \& Cen, R. 1995, submitted to ApJ}
\apj{Ostriker, J. P., \& Ikeuchi, S. 1983}{268}{L63}
\refbook{Ostriker, J. P., \& Steinhardt, P. 1995, Nature, in press}
\refbook{Peebles, P.~J.~E. 1980, {\it The Large-Scale Structure of
the Universe} (Princeton Univ. Press: Princeton)}
\aa{Petitjean, P., M\"ucket, J. P., \& Kates, R. E., 1995}{295}{L9}
\mnras{Petitjean, P., Webb, J. K., Rauch. M., Carswell, R. F., \& Lanzetta, K.,
1993}{262}{499}
\mnras{Persic, M., \& Salucci, P. 1992}{258}{14p}
\apj{Press, W.~H., \& Rybicki, G.~B. 1993}{418}{585}
\apj{Press, W.~H., Rybicki, G.~B., \& Schneider, D.~P. 1993}{414}{64}
\refbook{Quinn, T., Katz, N., \& Efstathiou, G. 1995, preprint}
\apj{Rauch, M., Carswell, R. F, Chaffee, F. H., Foltz, C. B., Webb, J. K.,
Weymann, R. J., Bechtold, J., \& Green, R. F. 1992}{390}{387}
\mnras{Rauch, M., \& Haehnelt, M.~G. 1995}{275}{L76}
\refbook{Rauch, M., et al. 1995, in preparation}
\mnras{Rees, M.~J. 1986}{218}{25p}
\refbook{\refrule~ 1988. In: {\it QSO Absorption Lines}, p. 107, ed.
J.~C. Blades \etal , Cambridge Univ. Press, Cambridge.}
\apj{Reisenegger, A., \& Miralda-Escud\'e, J. 1995}{449}{476}
\apj{Ryu, D., Ostriker, J. P., Kang, H., \& Cen, R. Y. 1993}{414}{1}
\nature{Sackett, P. D., Morrison, H. L., Harding, P., \& Boroson, T. A.
1994}{370}{441}
\apjsup{Sargent, W. L. W., Steidel, C. C., \& Boksenberg, A. 1989}{69}{703}
\apjsup{Sargent, W. L. W., Young, P. J., Boksenberg, A., \& Tytler, D.
1980}{42}{41}
\refbook{Scalo, J.M. 1986, Fund. Comis Phys., 11, 1}
\aj{Schneider, D.~P., Schmidt, M., \& Gunn, J.~E. 1991}{101}{2004}
\apjsup{Schneider, D. P., et al. 1993}{87}{45}
\apjsup{Shapiro, P., \& Struck-Marcel, P. 1985}{57}{205}
\apj{Smette, A., Surdej, J., Shaver, P. A., Foltz, C. B., Chaffee, F. H.,
Weymann, R. J., Williams, R. E., \& Magain, P. 1992}{389}{39}
\refbook{Smette, A., Surdej, J., Shaver, P. A., Reimers, D.,
Wisotzki, L., \& K\"ohler, T. 1995, A\&A, in press}
\refbook{Steinmetz, M. 1995, preprint}
\apj{Stengler-Larrea, E., et al.\ 1995}{444}{64}
\apj{Stompor, R., \& Gorski, K. M. 1994}{422}{L41}
\apj{Storrie-Lombardi, L., et al.\ 1994}{427}{L13}
\aa{Sunyaev, R.~A., \& Zel'dovich, Ya.~B. 1972}{20}{189}
\refbook{Tytler, D., Fan, X.-M., Burles, S., Cottrell, L., Davis, C.,
Kirkman, D., \& Zuo, L. 1995, in {\it QSO Absorption Lines}, Proc. ESO
Workshop, ed. G. Meylan (Heidelberg: Springer), p. 289}
\apj{Walker, T. P., Steigman, G., Schramm, D. N., Olive, K. A.,
\& Kang, H. S. 1991}{376}{51}
\apj{Wang, B. 1993}{415}{174}
\apj{\refrule~ 1995}{444}{L17}
\mnras{Webb, J.~K., Barcons, X., Carswell, R.~F., \& Parnell, H.~C.
1992}{255}{319}
\refbook{White, M., \& Bunn. E 1995, preprint}
\mnras{White, D., \& Fabian, A. C. 1995}{273}{72}
\apj{White, S. D. M., \& Frenk, C. S. 1991}{379}{25}
\apj{Williger, G. M., Baldwin, J. A., Carswell, R. F., Cooke, A. J.,
Hazard, C., Irwin, M. J., McMahon, R. G., \& Storrie-Lombardi, L. J.
1994}{428}{574}
\refbook{Zhang, Y., Anninos, \& Norman, M. 1995, submitted to ApJ}
\mnras{Zuo, L. 1992a}{258}{36}
\mnras{\refrule~ 1992b}{258}{45}

\vfill
\eject
\bye